\documentclass[a4paper, 11pt]{article}
\pdfoutput=1
\usepackage{jheppubnew}
\usepackage{graphicx}
\usepackage{bm}
\usepackage{caption, subcaption}
\usepackage{ifthen}
\usepackage{amsmath}
\usepackage{amsfonts}
\usepackage{amssymb}
\usepackage{amsthm}
\usepackage{mathrsfs}
\usepackage[all]{xy}
\usepackage{xcolor}
\usepackage{comment}
\usepackage[numbers,sort&compress]{natbib}

\newcommand{\be}{\begin{equation}}
\newcommand{\ee}{\end{equation}}
\renewcommand{\[}{\begin{equation}}
\renewcommand{\]}{\end{equation}}

\newcommand{\D}{\mathrm{d}}

\renewcommand{\O}{\mathcal{O}}
 
\newcommand{\F}{\mathcal{F}}
\newcommand{\ep}{\epsilon}

\newcommand{\g}{\gamma}
\newcommand{\<}{\langle}
\renewcommand{\>}{\rangle}

\newcommand{\nn}{\nonumber}
\newcommand{\lla}{\langle \! \langle}
\newcommand{\rra}{\rangle \! \rangle}

\newcommand{\bs}[1]{\boldsymbol{#1}}
\newcommand{\x}{{\bs{x}}}

\newcommand{\p}{\partial}

\DeclareMathOperator{\K}{K}

\DeclareMathOperator{\Lo}{L}
\DeclareMathOperator{\Ro}{R}

\renewcommand{\2}{C}
\renewcommand{\sl}{\2^{(0)}}
\newcommand{\sdt}{D}
\newcommand{\ThreePt}{\empty}
\newcommand{\3}[1]{C_{
		\ifthenelse{\equal{\ThreePt}{\empty}}{#1}{
			\ifthenelse{\equal{#1}{\empty}}{\ThreePt}{\ThreePt,#1}}}}
\newcommand{\redef}[1]{{C'}_{
		\ifthenelse{\equal{\ThreePt}{\empty}}{#1}{
			\ifthenelse{\equal{#1}{\empty}}{\ThreePt}{\ThreePt,#1}}}}
\newcommand{\ren}[1]{C_{
		\ifthenelse{\equal{\ThreePt}{\empty}}{#1}{
			\ifthenelse{\equal{#1}{\empty}}{\ThreePt}{\ThreePt,#1}}}}
\newcommand{\sd}[1]{D_{
		\ifthenelse{\equal{\ThreePt}{\empty}}{#1}{
			\ifthenelse{\equal{#1}{\empty}}{\ThreePt}{\ThreePt,#1}}}}

\newcommand{\ct}{\mathfrak{c}}

\title{
Renormalised CFT 3-point functions of scalars, currents and stress tensors}
\author[a]{Adam Bzowski,}
\author[b]{Paul McFadden}
\author[c]{and Kostas Skenderis.}
\affiliation[a]{Institut de Physique Th\'{e}orique, CEA Saclay, Gif-sur-Yvette, France.}
\affiliation[b]{Centre for Particle Theory, Department of Mathematical Sciences, Durham University, U.K.}
\affiliation[c]{STAG Research Centre and Mathematical Sciences, University of Southampton, U.K.}
\emailAdd{adam.bzowski@ipht.fr} 
\emailAdd{paul.l.mcfadden@durham.ac.uk} 
\emailAdd{k.skenderis@soton.ac.uk}

\begin{document}

\abstract{

We discuss the  renormalisation of mixed 3-point functions involving 
tensorial and scalar operators in conformal field theories of general dimension.
In previous work we analysed correlators of either purely scalar or purely tensorial operators, 
in each case finding new features and new complications: 
for scalar correlators, renormalisation leads to beta functions, novel conformal anomalies of type B, and unexpected analytic structure in momentum space; for correlators of stress tensors and/or conserved currents, beta functions vanish but anomalies of both type B and type A (associated with a 0/0 structure) are present. Mixed correlators combine all these features:
beta functions and anomalies of type B, plus the possibility of new type A anomalies. 
Following a  non-perturbative and general  momentum-space analysis,
we present explicit results in  dimensions 
$d=3,4$ for all renormalised 3-point functions of stress tensors, conserved currents and scalars of dimensions $\Delta=d$ and $\Delta=d-2$. 
We identify all anomalies and beta functions, and explain the form of the anomalous conformal Ward identities.
In $d=3$, we find  a 0/0 structure but  
the corresponding type A  anomaly  
turns out to be trivial.
In addition, the correlators 
of two currents and a scalar, and of two stress tensors and a scalar, both feature 
universal tensor structures  that are independent of the scalar dimension and vanish for opposite helicities.

}

\maketitle

\section{Introduction}

Momentum-space methods, ubiquitous 
in quantum field theory, 
are not yet widely 
available for conformal field theories.
Even 
elementary results such as the form of 3-point functions, known in position space for decades \cite{Polyakov:1970xd, Schreier:1971um, Osborn:1993cr, Erdmenger:1996yc}, have only recently 
been studied in momentum space \cite{Cappelli:2001pz, Giannotti:2008cv, Armillis:2009pq, Coriano:2012wp,  Coriano:2013jba, Bzowski:2013sza, Bzowski:2015pba, Bzowski:2015yxv, Bzowski:2017poo, Coriano:2017mux, Coriano:2018zdo, Coriano:2018bbe}. 
With new applications in 
cosmology \cite{ Antoniadis:2011ib, Maldacena:2011nz, Schalm:2012pi,  Bzowski:2012ih, Mata:2012bx, McFadden:2013ria, Ghosh:2014kba, Anninos:2014lwa, Kundu:2014gxa,  Arkani-Hamed:2015bza,  Baumann:2015xxa, Isono:2016yyj} 
and  condensed matter  \cite{Chowdhury:2012km, Huh:2013vga, Jacobs:2015fiv, Myers:2016wsu, Lucas:2016fju, Lucas:2017dqa} for which momentum space is ideally suited, now is the time to close this surprising gap.

General results for  
3-point functions, valid for generic values of the operator and spacetime dimensions, were presented 
in \cite{Bzowski:2013sza}.
In certain special cases, however,
these results are invalidated by the presence of divergences. 
Starting with 3-point functions of scalar operators, we formulated a corresponding renormalisation  prescription in \cite{Bzowski:2015pba}.
In \cite{Bzowski:2017poo}, we extended this prescription to obtain the renormalised 3-point functions of stress tensors and conserved currents.
This paper completes 
our analysis by presenting
renormalised 3-point functions for 
mixed correlators 
involving scalars, stress tensors and conserved currents.
Our entire approach is fully non-perturbative, 
making use only of conformal symmetry.

The standard position-space results for CFT correlators are valid only at separated points, with coincident insertions leading to singularities.  Correlators should nevertheless be well-defined distributions, and in particular should have a well-defined Fourier transform.  In certain cases, to achieve this we must first regulate then supplement the standard position space expressions with suitable contact terms. 
The dependence on the renormalisation scale thus introduced then leads to conformal anomalies and/or beta functions.
From a purely position-space perspective, while it might appear that these contact terms can be neglected, this is by no means universally true:
certain correlators -- for example, the 3-point function of a stress tensor, conserved current and a scalar we study here -- appear to be nonzero at separated points but vanish when the constraints at coincident points are taken into account. 
Examples also exist where an analysis at separated points implies the correlator vanishes, whereas in fact it is nonzero due to contact terms \cite{Bonora:2015nqa}.
Contact contributions are naturally also important in applications where operator insertions are integrated over the spacetime coordinates, 
such as conformal perturbation theory. 

For our present purposes, the momentum-space 3-point functions are most easily found by solving the relevant momentum-space conformal Ward identities.  (Certainly, this is far easier than attempting to Fourier transform from position space.)
Using the minimal decomposition for tensorial structure introduced in \cite{Bzowski:2013sza}, these Ward identities take a simple form and can  be solved through surprisingly elementary means: in fact, just separation of variables followed by a Mellin transform to extract the component of appropriate scaling dimension.  The resulting momentum-space 3-point functions can then be expressed in terms of {\it triple-$K$ integrals}, a general class of parametric integrals involving three modified Bessel functions and a power. 

For special values of the operator and spacetime dimensions,  divergences now arise from the lower limit of these triple-$K$ integrals, corresponding to the divergences one would have obtained from naively Fourier-transforming the position-space expressions over coincident configurations.  
These divergences can be dimensionally regulated 
through infinitesimal shifts of the operator and spacetime dimensions, producing corresponding shifts in the parameters of the triple-$K$ integrals. A major advantage of this regularisation is that it respects conformal symmetry. Moreover, it allows 
the form of divergences to be read off from a simple series expansion of the triple-$K$ integrand.   As shown in \cite{Bzowski:2015pba}, this result is a consequence of the Mellin mapping theorem which relates the singularities of a triple-$K$ integral  
to the poles of its integrand.

The nature of the singularities present now dictates the type of 
counterterms required for their removal.  
In general, singularities in triple-$K$ integrals arise whenever a certain singularity condition 
is satisfied.  This condition involves a choice of three independent $\pm$ signs, according to which singularity can be classified. 
Here, we will encounter 
singularities of type $(---)$ and $(+--)$, along with permutations, 
whose implications are as follows.

Singularities of type $(---)$ correspond to triple-$K$ integrals whose divergences are {\it ultralocal}, meaning they are purely analytic functions of the squared momenta.  In position space, they represent contact terms contributing only when all three insertion points are coincident.  Singularities of this type can be removed by local counterterms that are cubic in the sources, giving rise to conformal anomalies. 
Singularities of type $(+--)$ (and permutations) correspond instead to triple-$K$ integrals whose divergences are  
{\it semilocal}, meaning they are non-analytic in only one of the squared momenta, or else a sum of such terms.  In position space, they represent contact terms contributing when only two of the three insertions are coincident.  Singularities of this type can be removed by cubic counterterms involving two sources and one operator.  The  3-point contribution from such counterterms involves a 2-point function, generating the necessary semilocal momentum dependence required to cancel the divergence.
Counterterms of this nature effectively reparametrise the source of the operator involved, and hence give rise to a nontrivial beta function.
Being quadratic in the sources, however, this beta function vanishes for the CFT itself where the sources are set to  zero. 

For correlators featuring only stress tensors and/or conserved currents, as studied in \cite{Bzowski:2017poo}, only singularities of type $(---)$  arise consistent with the fact that only counterterms cubic in the sources are available.
With the introduction of scalar operators, this is no longer the case and hence we now find both $(---)$ and $(+--)$ type singularities  arising.
The elimination of these singularities can nevertheless be surprisingly  intricate.  Due to the restrictions imposed by gauge and Lorentz invariance, we sometimes encounter cases where either {\it no counterterms} are available, or else only counterterms of the wrong type, despite  the presence of singularities.   
In such cases, as we will see, the singularities are instead eliminated through two additional mechanisms, 
acting either singly or in combination. 

The first 
is that cancellations can occur between the singularities of  {\it different} triple-$K$ integrals contributing to the same correlator.  More precisely, we decompose the tensorial structure of correlators  into a minimal set of basis tensors, each of which is multiplied by a scalar form factor depending only on the momentum magnitudes.  The conformal Ward identities then impose that each of these form factors is given by a specific linear combination of triple-$K$ integrals.  
 We can then obtain cancellations between the singularities of the different triple-$K$ integrals appearing in these sums.
Alternatively, as a second mechanism, the relevant linear combination can be such that a singular triple-$K$ integral is multiplied by a coefficient that vanishes as some appropriate power of the regulator.  

Remarkably, 
 no arbitrariness is involved in 
 either of these cancellation mechanisms. The set of constants determining these linear combinations of triple-$K$ integrals, which we refer to as {\it primary constants}, are themselves constrained by a subset of the conformal Ward identities.
In general, these can be split into two sets, the {\it primary} and {\it secondary} Ward identities, the first of which can be solved in terms of triple-$K$ integrals up to constants of integration which are the primary constants.  The secondary conformal Ward identities 
then act simply to constrain these primary constants.  (As such, they can most easily be analysed in special kinematic configurations for which the triple-$K$ integrals simplify, such as the soft limit where one momentum vanishes.) Their action not only reduces the number of undetermined parameters to their final physical values, but moreover selects primary constants such that all singularities not removed by counterterms are automatically cancelled through the mechanisms 
described above.

With all singularities eliminated, the regulator can now be removed to obtain the renormalised correlators.  Where  
anomalies and/or beta functions are present, these renormalised correlators obey modified conformal Ward identities now featuring additional inhomogeneous (or `anomalous') terms.   The form of these additional terms can  easily be found by inserting the renormalised correlators back into the original homogeneous Ward identities, but can also  be understood from theoretical considerations.  
This is particularly simple for the dilatation Ward identity, since a dilatation is equivalent to changing the renormalisation scale $\mu$ while holding the renormalised couplings fixed.  
On general grounds,  the renormalised generating functional $W$ satisfies
\[\label{dmuW}
A = \mu \frac{\D}{\D \mu} W
 = \Big[\mu\frac{\p}{\p \mu} +  \int\D^d\x\, \sum_I\beta_{\Phi^I}\frac{\delta}{\delta\Phi^I} \Big] W,
\]
where $A$ is the anomaly action and the $\beta_{\Phi^I}$ are beta functions for
some 
set of couplings which we schematically denote  
$\Phi^I$.  Note that anomalous dimensions are absent, as we work non-perturbatively assuming a CFT with some specific set of dimensions is given.
The anomaly action and beta functions can be read off from the counterterm action (see section \ref{sec:anomaliesandbetafns}), 
after which the anomalous dilatation Ward identities for  correlators follow by functional differentiation.

Understanding the anomalous terms entering the {\it special} conformal Ward identities, namely the primary and secondary Ward identities above, requires more effort.  
One way, 
as we show in section \ref{sec:AnomalousCWIs}, is to view the CFT as the flat-space limit of a Weyl-invariant theory on curved spacetime \cite{Osborn:1991gm}.
This perspective is also natural since we use the metric as a source for the stress tensor.
The anomaly 
then derives from the Weyl variation of the renormalised generating functional,\footnote{Derivatives of $\sigma$ can be removed through integration by parts.  The resulting $\nabla_\mu \mathcal{J}^\mu$ contributions to the anomaly correspond to finite local counterterms (see, {\it e.g.}, \cite{Henningson:1998gx}), and are thus scheme-dependent.}
\[\label{WeylofW}
\delta_\sigma W = \int\D^d\x\sqrt{g}\,\mathcal{A}\,\sigma(\x), \qquad  A = \int\D^d\x\sqrt{g}\,\mathcal{A},
\]
with the variation of the sources given by the beta functions.  Since conformal transformations are equivalent to a diffeomorphism followed by a Weyl transformation, their action on the renormalised correlators can now be evaluated.

Denoting the generating functional as
\[\label{Wlim}
W = \lim_{\ep\rightarrow 0}\,\ln \, \<\,e^{-S_{\mathrm{ct}}-S_{\mathrm{source}}}\,\>,
\]
we further require the regulated counterterm action satisfies the Weyl covariance condition
\begin{align}\label{fullWeylcovcond}
\delta_\sigma(S_{\mathrm{ct}}+S_{\mathrm{source}})
=\int\D^d\x\sqrt{g}\,\sigma\mu\frac{\D}{\D\mu}(\mathcal{L}_{\mathrm{ct}}+\mathcal{L}_{\mathrm{source}}).
\end{align}
In fact,  this relation is automatically satisfied by the  divergent part of the counterterm action.  As we will see, all divergences  in the regulated theory satisfy non-anomalous Ward identities meaning their Weyl variation follows from their scaling dimension.  (Anomalies arise only after adding counterterms and removing the regulator.)  The relation \eqref{fullWeylcovcond} is {\it not} however automatically satisfied by the finite ({\it i.e.,} scheme-dependent) part of counterterms.  Imposing  \eqref{fullWeylcovcond}  thus restricts us to a narrow class of (conformal) renormalisation schemes, reducing the number of scheme-dependent terms entering the renormalised correlators.  
This restriction to conformal schemes is especially useful for the mixed scalar/tensor correlators we study here, for which a large number of potential counterterms exist even after gauge- and Lorentz-invariance have been imposed.

Finally, we wish to highlight two features of our results in three spacetime dimensions:
\begin{itemize}
\item For the 3-point function $\<T_{\mu_1\nu_2}T_{\mu_2\nu_2}\O\>$, the tensorial basis in which correlators are decomposed has a degeneracy due to the appearance of an evanescent tensor \cite{Bzowski:2013sza, Bzowski:2017poo}.  
As a contraction of a 4-form, this tensor vanishes in three spacetime dimensions but not in the dimensionally regulated theory.  
When such tensors appear with divergent coefficients, one usually finds a type A conformal anomaly which preserves scale but not special conformal invariance \cite{Deser:1993yx}. This  mechanism was recently demonstrated for the four-dimensional Euler anomaly in \cite{Bzowski:2017poo}.
Here, in section \ref{typeAex1}, we find instead a three-dimensional counterexample where a 0/0 limit
leads to a {\it trivial} anomaly that can be removed by counterterms. 
\item In three dimensions, we find that  $\<T_{\mu_1\nu_1}T_{\mu_2\nu_2}\O\>$ and $\<J^{\mu_1}J^{\mu_2}\O\>$ each involve only a {\it single} tensorial structure, 
which is {\it independent} of the dimension of the scalar operator.  
This tensorial structure is such that the  correlators vanish when the stress tensors or currents have opposite helicities.
Dependence on the scalar dimension enters only via an overall (scalar) form factor involving the momentum magnitudes.  
It would be interesting to find a deeper explanation for this curious behaviour.
Holography is perhaps relevant, as both cases involve a single four-dimensional bulk vertex whose non-scalar part is Weyl invariant. 
\end{itemize}

The layout of this paper is as follows. In section \ref{sec:overview}, we introduce our momentum-space technology summarising the relevant Ward identities; their tensorial decomposition and solution in terms of triple-$K$ integrals; regularisation and renormalisation; results for 2-point functions; and the extraction of anomalies and beta functions.  Our main results for renormalised 3-point functions are then presented in section \ref{sec:results}.  In section \ref{sec:AnomalousCWIs}, we give a detailed analysis explaining the form of the anomalous conformal Ward identities, after which we  conclude in section \ref{sec:Discussion}.  Two appendices discuss the analysis of counterterms, while a third presents general shadow relations for the correlators of interest.

We have endeavoured to make this paper self-contained, with the relevant background material summarised in section \ref{sec:overview}. 
The different subsections of section \ref{sec:results} can then be read independently of one another, so readers interested only in results for a particular correlator may head directly to the relevant subsection after reviewing our conventions in section \ref{sec:overview}.

\section{Overview of the method}
\label{sec:overview}

\label{sec:Definition}

This section summarises the notation and definitions we will use for presenting our results.  
A complementary discussion of our solution method  can be found in section 2 of \cite{Bzowski:2017poo}; here we focus principally on the new  features arising for mixed correlators of scalars and tensors.
Throughout, we assume we are working in $d \geq 3$ Euclidean dimensions.

\subsection{Transverse Ward identities}

Under a variation of the sources, 
we find the variation of the renormalised generating functional 
\[\label{variationofW}
\delta W[g_{\mu\nu},A^a_\mu,\phi^I]  = -\int\D^d\x\sqrt{g}\,
\Big(\frac{1}{2}\<T_{\mu\nu}\>_s\,\delta g^{\mu\nu}+\<J^{\mu a}\>_s\,\delta A^a_{\mu}+\<\O^I\>_s\delta\phi^I
\Big),
\]
where the subscript $s$ denotes 
a nontrivial source profile.
The gauge field $A^a_\mu$ sources a symmetry current $J^{\mu a}$ associated with some group $G$,  in general non-Abelian.  The indices $a=1,\,\ldots,\,\dim G$ and repeated indices are to be summed.
The $\phi^I$ source scalar primary operators $\O^I$,  transforming in some representation $R$ 
with generator matrices $(T_R^a)^{IJ}$.

Under a gauge transformation $\alpha^a$, 
the sources transform as
\begin{align}\label{sourcegauge}
\delta g^{\mu\nu} &=0, \\
\delta A^a_\mu &=- \nabla_{\mu}\alpha^a - g f^{abc}A^b_\mu\alpha^c, \\ 
\delta\phi^I &= -ig\alpha^a(T_R^a)^{IJ}\phi^J,
\end{align} 
where $f^{abc}$ is the structure constant and $g$ the gauge coupling.   
Under a diffeomorphism $\xi^\mu$, 
\begin{align}\label{sourcediff}
\delta g^{\mu\nu} &= -2\nabla^{(\mu}\xi^{\nu)}, \\
\delta A^a_\mu &= \xi^\nu\nabla_\nu A^a_\mu+A_\nu^a\nabla_\mu \xi^\nu, \\
\delta\phi^I &= \xi^\mu\nabla_\mu\phi^I,
\end{align} 
where we perform all symmetrisations (and antisymmetrisations) with unit strength.
The invariance of the generating functional under these transformations yields the transverse Ward identities,
\begin{align}\label{p:Jward}
0&=\nabla_\mu\<J^{\mu a}\>_s-ig\<\O^I\>_s(T^a_R)^{IJ}\phi^J+gf^{abc}A_\mu^b\<J^{\mu c}\>_s,\\[1ex]\label{p:Tward}
0&=\nabla^\mu\<T_{\mu\nu}\>_s+\<\O^I\>_s \nabla_\nu\phi^I+\<J^{\mu a}\>_s2\nabla_{[\nu}A_{\mu]}^a-A^a_\nu\nabla_\mu\<J^{\mu a}\>_s.
\end{align}
Introducing the covariant derivative and field strength 
\begin{align}
D_\mu^{IJ} &= \delta^{IJ}\nabla_\mu-i g A^a_\mu(T^a_R)^{IJ}, \\
F^a_{\mu\nu} &= 2\nabla_{[\mu}A_{\nu]}^a+gf^{abc}A^b_\mu A^c_\nu,
\end{align}
we can alternatively write \eqref{p:Jward} and  \eqref{p:Tward} as 
\begin{align}
0 &= D_{\mu}\<J^{\mu a}\>_s - ig\<O^I\>_s (T^a_R)^{IJ}\phi^J,\\
0 &=\nabla^\mu\<T_{\mu\nu}\>_s+\<\O^I\>_s D_\nu\phi^I-F^a_{\mu\nu}\<J^{\mu a}\>_s.
\end{align}
To obtain the second of these equations, we substituted \eqref{p:Jward} into \eqref{p:Tward}, and for the first we used the adjoint representation $(T^a_{A})^{bc}=-if^{abc}$ for the current.

The corresponding transverse Ward identities for 3-point functions now follow by functionally differentiating twice with respect to the sources.
In momentum space, these identities give the longitudinal components of 3-point functions in terms of 2-point functions, for example
\begin{align} 
& p_{1 \mu_1} \lla J^{\mu_1 a}(\bs{p}_1) \mathcal{O}^{I_2}(\bs{p}_2) \mathcal{O}^{I_3}(\bs{p}_3) \rra \nn\\[1ex]
& \qquad = - g (T_R^a)^{K I_3} \lla \mathcal{O}^K(\bs{p}_2) \mathcal{O}^{I_2}(-\bs{p}_2) \rra - g (T_R^a)^{K I_2} \lla \mathcal{O}^K(\bs{p}_3) \mathcal{O}^{I_3}(-\bs{p}_3) \rra.
\end{align}
Our double-bracket notation for correlators simply indicates stripping off the overall delta function of momentum conservation, thus
\begin{align}
\<J^{\mu_1 a_1}(\bs{p}_1) \O^{I_2}(\bs{p}_2) \O^{I_3}(\bs{p}_3) \> &=
 \lla J^{\mu_1 a_1}(\bs{p}_1) \O^{I_2}(\bs{p}_2) \O^{I_3}(\bs{p}_3) \rra \,(2\pi)^d\delta (\bs{p}_1+\bs{p}_2+\bs{p}_3), \\[1ex]
 \< \mathcal{O}^{I_2}(\bs{p}_2) \mathcal{O}^{I_3}(\bs{p}_3) \>&=\lla \mathcal{O}^{I_2}(\bs{p}_2) \mathcal{O}^{I_3}(-\bs{p}_2) \rra(2\pi)^d\delta (\bs{p}_2+\bs{p}_3).
\end{align}
We list the transverse Ward identities for each correlator at the beginning of every section.

\subsection{Trace Ward identities}
The trace of stress tensor correlators are determined by the trace Ward identities. These derive from Weyl transformations, under which the sources transform as
\begin{align}\label{Wvar1}
\delta g^{\mu\nu} = -2 \sigma g^{\mu\nu},
\qquad \delta A^a_\mu = \beta_{A_\mu^a}\sigma, \qquad \delta\phi^I = \Big(-(d-\Delta^I)\phi^I+\beta_{\phi^I}\Big)\sigma.
\end{align}
By definition, our beta functions  $\beta_{A_\mu^a}$ and $\beta_{\phi^I}$ begin at {\it quadratic order} in the sources, and hence arise from the renormalisation of $(+--)$ type singularities. 
In the CFT, where all sources take their background values, $A_\mu^a$ then has Weyl weight zero while $\phi^I$ has Weyl weight $d-\Delta^I$.  These weights are chosen so that under a   {\it conformal} transformation (constructed from a Weyl transformation and a diffeomorphism as discussed in section \ref{sec:AnomalousCWIs}), the 
operators $J^{\mu a}$ and $\O^I$ now
have the required dilatation weights $d-1$ and $\Delta^I$.

The renormalised generating functional is not in general invariant under a Weyl transformation and transforms anomalously as given in \eqref{WeylofW}.
From \eqref{variationofW}, we then obtain the trace Ward identity
\[\label{WeylWI}
\<T\>_s = \big[-(d-\Delta^I)\phi^I+\beta_{\phi^I}\big]\<\O^I\>_s+\beta_{A^a_\mu}\<J^{\mu a}\>_s+\mathcal{A}.
\]
The corresponding Ward identities for the renormalised 3-point functions follow by functionally differentiating this identity twice with respect to the sources, before restoring them to their background values.  As $\beta_{A_\mu^a}$ and $\beta_{\phi^I}$ are quadratic in the sources, and all 1-point functions in the CFT vanish, these beta function terms make no contribution to the 3-point trace Ward identities.  They may however contribute to the anomalous conformal Ward identities, as discussed in section \ref{sec:AnomalousCWIs}.

Where present, we list the trace Ward identities at the start of each section of our results.  
Note these identities apply only in the {\it renormalised} theory; in the dimensionally regulated theory from which our analysis begins, all stress tensor correlators are traceless.
Through identifying and eliminating the divergences that arise, we determine the anomaly and beta functions, and hence the trace Ward identities for the renormalised correlator.

\subsection{Defining the 3-point function}
\label{subsec:3ptdef}
  We define all 3-point functions through three functional derivatives of the generating functional, as in \cite{Osborn:1993cr}.  All metric factors are positioned {\it outside} the functional derivatives to preserve  symmetry under permutations, {\it e.g.,}
\[
\< T_{\mu_1\nu_1}(\x_1)T_{\mu_2\nu_2}(\x_2)\O^I(\x_3)\> = -\frac{4}{\sqrt{g(\x_1)g(\x_2)g(\x_3)}}\frac{\delta^3 W}{\delta g^{\mu_1\nu_1}(\x_1)\delta g^{\mu_2\nu_2}(\x_2)\delta \phi^I(\x_3)}\Big|_0,
\]
where the subscript zero indicates switching off the sources.
We caution this definition differs from that in \cite{Bzowski:2013sza}, where  3-point functions were defined through three insertions of the relevant operators.  
All results can easily be converted between these definitions, however, and we point out differences where they arise.  Our present convention simplifies the treatment of semilocal terms in divergent correlators, for reasons discussed in \cite{Bzowski:2017poo}.

Once the definition of the 3-point function has been fixed, the solution of the conformal Ward identities is unique.\footnote{The possibility of constructing local solutions corresponding to non-triple-$K$ integrals, as discussed in appendix A.3 of \cite{Bzowski:2015pba}, can be ruled out by explicit computation, see  \cite{Bzowski:2017poo}.}  Of course, if this definition is changed, the form of the Ward identities and hence that of the solutions can change (see {\it e.g.,} \cite{Gomis:2015yaa}).  Nevertheless, for a given definition the solution is still unique hence such modifications do not constitute an intrinsic ambiguity.

\subsection{Momentum variables}

We denote momentum vectors with bold letters, while their magnitudes are
\begin{equation}
p_j = | \bs{p}_j | = \sqrt{ \bs{p}_j^2 }, \qquad j = 1, 2, 3.
\end{equation}
For 3-point functions, momentum conservation allows any Lorentz scalar $\bs{p}_i\cdot\bs{p}_j$ to be re-expressed purely in terms of the momentum magnitudes.
To write our results in compact form, we define the following symmetric polynomials of the momentum magnitudes
\begin{align}
 a_{123} &= p_1 + p_2 + p_3, \qquad b_{123} = p_1 p_2 + p_1 p_3 + p_2 p_3, \qquad c_{123} = p_1 p_2 p_3, \nn\\
a_{ij} &= p_i + p_j, \qquad\qquad\quad b_{ij} = p_i p_j, \label{e:variables}
\end{align}
where $i,j = 1,2,3$, as well as the combination
\begin{align}
J^2 & = (p_1 + p_2 + p_3) (- p_1 + p_2 + p_3) (p_1 - p_2 + p_3) (p_1 + p_2 - p_3) \nn\\[0.5ex]
&= -p_1^4-p_2^4-p_3^4+2p_1^2p_2^2+2p_2^2p_3^2+2p_3^2p_1^2. \label{Jsqdef}
\end{align}
By Heron's formula, $\sqrt{J^2}/4$ represents the area of the triangle formed by the momenta.  Equivalently, the Gram determinant of any two such momenta is given  by $J^2/4$.

\subsection{Tensorial decomposition}
We decompose all correlators into a basis of scalar form factors multiplying tensor structures built from the metric and momenta. 
The form factors are functions of the momentum magnitudes, $A_j = A_j(p_1, p_2, p_3)$. 
When no arguments are specified, this standard ordering of momenta is assumed; otherwise the exchange of arguments is indicated with an arrow,  \textit{e.g.}, $A_j(p_1 \leftrightarrow p_2) = A_j(p_2, p_1, p_3)$.

Through momentum conservation, only two of the three momenta appearing in a 3-point function are independent.
Rather than making a global choice for these two independent momenta, which would obscure  the permutation symmetries for 3-point functions of identical operators, we will instead make a different choice of independent momenta for each operator insertion. 
Numbering all Lorentz indices according to the operator insertion they are associated with, we  choose the independent momenta according to the cyclic rule: 
\begin{equation} \label{a:momenta}
\bs{p}_1, \bs{p}_2 \text{ for } \mu_1, \nu_1; \ \bs{p}_2, \bs{p}_3 \text{ for } \mu_2, \nu_2 \text{  and  } \bs{p}_3, \bs{p}_1 \text{ for }\mu_3, \nu_3.
\end{equation}
Thus, for example, for the second operator insertion, which carries Lorentz indices labelled with a $2$ subscript, the independent momenta are $\bs{p}_2$ and $\bs{p}_3$.
By respecting the permutation symmetry of identical operators, this choice of independent momenta leads to a basis with the minimal number of scalar form factors; see section 1 of \cite{Bzowski:2013sza}.
Use of this cyclic convention is assumed whenever we refer to the ``coefficient of'' some specific tensorial structure in a 3-point function. 
Thus, before reading off this coefficient, we first replace momenta as required (using momentum conservation) so as to be consistent with \eqref{a:momenta}.

One of the main advantages of momentum space is that the transverse Ward identities are algebraic.  As we saw above, this means all the  longitudinal components of 3-point functions can  be reduced to 2-point functions.  Moreover, all trace components can be obtained from the trace Ward identities.
The remaining transverse-traceless tensor structure can then be decomposed with the aid of the transverse and transverse-traceless projectors,
\begin{align}
\pi_{\mu\nu}(\bs{p}) & = \delta_{\mu\nu} - \frac{p_{\mu} p_{\nu}}{p^2}, \\
\label{TTprojdef}
\Pi_{\mu \nu\alpha \beta}(\bs{p}) & = \frac{1}{2} \Big( \pi_{\mu\alpha}(\bs{p}) \pi_{\nu\beta}(\bs{p}) + \pi_{\mu\beta}(\bs{p}) \pi_{\nu\alpha}(\bs{p}) \Big) - \frac{1}{d - 1} \pi_{\mu \nu}(\bs{p}) \pi_{\alpha \beta}(\bs{p}).
\end{align}
We will write the transverse(-traceless) parts of the conserved current and stress tensor  as
\begin{align}
j^{\mu} \equiv \pi^{\mu}_{\alpha} J^{\alpha}, & \qquad\qquad 
t_{\mu \nu} \equiv \Pi_{\mu \nu}{}^{\alpha \beta} T_{\alpha \beta}.
\end{align}

\subsection{Conformal Ward identities}

The conformal Ward identities (or CWI) can be split into two types, which we label primary and secondary \cite{Bzowski:2013sza}. 
The primary conformal Ward identities can be expressed using the second-order differential operators
\begin{align}
\K_{ij} & = \K_i - \K_j, \qquad
\K_j  = \frac{\partial^2}{\partial p_j^2} + \frac{d + 1 - 2 \Delta_j}{p_j} \frac{\partial}{\partial p_j}, \qquad i,j=1,2,3\label{a:Kij}
\end{align}
where $\Delta_j$ denotes the dimension of the $j$-th operator in the 3-point function at hand. In $\< T_{\mu_1 \nu_1} J^{\mu_2} J^{\mu_3} \>$, for example, we thus have $\Delta_1 = d$ and $\Delta_2 = \Delta_3 = d - 1$.

The solution to these primary CWIs can be written in terms of the triple-$K$ integral $I_{\alpha\{\beta_j\}}$ and (for convenience) its reduced counterpart $J_{N\{k_j\}}$.  These integrals are defined by
\begin{align}
I_{\alpha \{ \beta_1 \beta_2 \beta_3 \}}(p_1, p_2, p_3) & = \int_0^\infty \D x \: x^\alpha \prod_{j=1}^3 p_j^{\beta_j} K_{\beta_j}(p_j x), \label{a:I} \\
J_{N \{ k_j \}} & = I_{\frac{d}{2} - 1 + N \{ \Delta_j - \frac{d}{2} + k_j \}}, \label{a:J}
\end{align}
where $K_\nu$ is a modified Bessel function of the second kind and we use the compressed notation $\{k_j\} = \{k_1 k_2 k_3\}$.
Solutions to the primary CWIs consist of  linear combinations of reduced triple-$K$ integrals multiplied by constants $C_j$ which we refer to as primary constants.  Some of these primary constants are fixed by the secondary CWIs as below; the remainder are free parameters characterising the specific CFT at hand.

The secondary CWI involve the first-order differential operators\footnote{In the original arXiv version of this paper, we used $\Lo_{s,n}$ and $\Ro_s$ where $s$ is the spin of the first operator.  As this operator is always either $T_{\mu\nu}$ or $J^\mu$, we have simplified by substituting $s=\Delta_1-d+2$ and redefining $\Lo_{s,n}\rightarrow \Lo_{N}$ with $N=s+n$ and $\Ro_s\rightarrow \Ro$.} 
\begin{align}
\Lo_{N} & = p_1 (p_1^2 + p_2^2 - p_3^2) \frac{\partial}{\partial p_1} + 2 p_1^2 p_2 \frac{\partial}{\partial p_2} \nn\\
& \qquad + \: \left[ (2d  - \Delta_1-2 \Delta_2 + N) p_1^2 + (2\Delta_1 -d) (p_3^2 - p_2^2) \right], \label{a:L} \\
\Ro & = p_1 \frac{\partial}{\partial p_1} - 2\Delta_1 +d, \label{a:R} 
\end{align}
and their permutations
\begin{align}
\Lo'_{N} & = \Lo_{N} \text{ with } (p_1 \leftrightarrow p_2) \text{ and } (\Delta_1 \leftrightarrow \Delta_2), \label{a:L2} \\
\Ro' & = \Ro \text{ with } (p_1 \rightarrow p_2) \text{ and } (\Delta_1 \rightarrow \Delta_2). \label{a:R2}
\end{align}
Substituting in our solution of the primary CWI, these secondary CWIs serve to fix a number of the undetermined primary constants $C_j$ appearing in our solution.  
To evaluate these relationships it is useful to examine the secondary CWI in the soft limit $p_3\rightarrow 0$, where the action of the differential operators \eqref{a:L}\,-\,\eqref{a:R2} on triple-$K$ integrals can be explicitly evaluated \cite{Bzowski:2013sza}.
To do this, we use the general relation
\begin{align}
\frac{\partial}{\partial p_i} I_{\alpha \{ \beta_j \}} & = - p_i I_{\alpha + 1 \{ \beta_j - \delta_{ij} \}}, \qquad i,j=1,2,3 \label{e:Jid1} 
\end{align}
to eliminate derivatives, after which the 
soft limit is given by
\begin{equation} \label{e:limJ}
\lim_{p_3 \rightarrow 0} I_{\alpha \{\beta_j\}}(p, p, p_3) = \ell_{\alpha \{\beta_j\}} p^{\beta_1+\beta_2+\beta_3 - \alpha - 1},
\end{equation}
where
\begin{equation} \label{e:l}
\ell_{\alpha \{ \beta_k \}} = \frac{2^{\alpha - 3} \Gamma(\beta_3)}{\Gamma(\alpha - \beta_3 + 1)} \prod_{\sigma_1, \sigma_2 \in \{-1, 1\}} \Gamma \left( \frac{\alpha - \beta_3 + 1 + \sigma_1 \beta_1 + \sigma_2 \beta_2}{2} \right).
\end{equation}
This formula is valid for $\beta_3>0$ and away from poles of the gamma functions.

\subsection{Regularisation}  

Triple-$K$ integrals diverge whenever \cite{Bzowski:2015pba} 
\[\label{singcond}
\alpha+1\pm \beta_1\pm\beta_2\pm\beta_3=-2n,
\]
where $n$ is any non-negative integer $n=0,1,2\ldots$, 
and any independent choice of $\pm$ signs can be made for each $\beta_j$.
The singularity type $(\pm \pm \pm)$ is then given by this set of signs.  
In such cases, we regulate using the generalised dimensional scheme 
\begin{equation} \label{a:scheme}
I_{\alpha \{\beta_1 \beta_2 \beta_3\}} \mapsto
I_{\tilde{\alpha} \{\tilde{\beta_1}\tilde{\beta_2} \tilde{\beta_3}\}}= I_{\alpha + u \epsilon \{\beta_1 + v_1 \epsilon, \beta_2 + v_2 \epsilon, \beta_3 + v_3 \epsilon\}},
\end{equation}
where the constants $\{u,v_j\}$ parametrise our freedom to shift the operator and spacetime dimensions according to
\[
d\rightarrow \tilde{d}=d+2u\ep, \qquad \Delta_j\rightarrow\tilde{\Delta}_j=\Delta_j+(u+v_j)\ep, \qquad j=1,2,3.
\]
Each choice of the constants $\{u,v_j\}$ thus defines a different regularisation scheme.  
After the divergences have been removed, physical dimensions are then restored by sending $\ep\rightarrow 0$.

For the conserved current $J^{\mu a}$ and stress tensor $T_{\mu\nu}$, our choice of scheme is restricted by the necessity for these operators to retain their canonical dimensions, {\it i.e.,}
\[
\tilde{\Delta}_{J}=\tilde{d}-1, \qquad \tilde{\Delta}_{T} = \tilde{d} \qquad \Rightarrow \qquad v=u.
\]
With this scheme, gauge- and diffeomorphism invariance are respected and the transverse Ward identities take the same form in both the regulated and the renormalised theory.

The divergences of triple-$K$ integrals can be directly read off from a series expansion of their integrand about the origin \cite{Bzowski:2015pba}.
Writing this expansion as
\[
x^{\tilde{\alpha}} \prod_{j=1}^3 p_j^{\tilde{\beta}_j} K_{\tilde{\beta}_j}(p_j x) = \sum_{\eta} c_{\eta} x^\eta,
\]
the divergences arise from terms of the form $x^{-1+w\ep}$ (for some finite nonzero $w$) that become poles in the limit as $\ep\rightarrow 0$.
Via the Mellin mapping theorem, these divergences 
are given by the formula
\[
I^{\mathrm{div}}_{\tilde{\alpha},\{\tilde{\beta}_j\}} = \sum_{w}\frac{c_{-1+w\ep}}{w\ep} +O(\ep^0).
\]
When the singularity condition \eqref{singcond} is multiply satisfied, the coefficients $c_{-1+w\ep}$  themselves contain poles in $\ep$, leading to a higher-order overall divergence.

\subsection{Reduction scheme and master integral}

For four-dimensional CFTs, all the triple-$K$ integrals we encounter 
can be derived from a single finite master integral 
$I_{1\{000\}}$.  
The corresponding reduction scheme, which we use to obtain many of our results, is derived in \cite{Bzowski:2015yxv}.
The master integral $I_{1\{000\}}$ is well-known in the literature (see \textit{e.g.}, \cite{ tHooft:1978xw, Boos:1987bg, Davydychev:1992xr, Bzowski:2015yxv}) and represents, for example,  the 3-point function of $\O_2 = {:}\,\varphi^2\,{:}$ 
for a  free conformal scalar $\varphi$ in four dimensions.  
It can be evaluated as 
\begin{align}\label{I1000}
I_{1\{000\}} & = \frac{1}{2\sqrt{-J^2}}\Big[\frac{\pi^2}{6}-2\ln\frac{p_1}{p_3}\ln\frac{p_2}{p_3}+\ln X\ln Y-\mathrm{Li}_2X-\mathrm{Li}_2Y\Big],
\end{align}
where $J^2$ is given in \eqref{Jsqdef} and 
\begin{align}\label{XYdef}
X & = \frac{-p_1^2+p_2^2+p_3^2-\sqrt{-J^2}}{2p_3^2}\,, \qquad 
Y = \frac{-p_2^2+p_1^2+p_3^2-\sqrt{-J^2}}{2p_3^2}.
\end{align}
Here, $X$ is effectively a dimensionless complex variable  with conjugate $\bar{X}=1-Y$.  

To express our results efficiently, we also define the  finite auxiliary integrals
\begin{align}\label{I2111fin0}
I_{2\{111\}}^{\text{(fin)}}
& =-\frac{1}{3}\Big[p_1^2\Big(2+p_1\frac{\p}{\p p_1}\Big)+p_2^2\Big(2+p_2\frac{\p}{\p p_2}\Big)+p_3^2\Big(2+p_3\frac{\p}{\p p_3}\Big)\Big]I_{1\{000\}},
\\[2ex]
I_{3\{222\}}^{\text{(fin)}} & = \left( 2 - p_1 \frac{\partial}{\partial p_1} \right) \left( 2 - p_2 \frac{\partial}{\partial p_2} \right) \left( 2 - p_3 \frac{\partial}{\partial p_3} \right) \left( \frac{1}{4} J^2 I_{1\{000\}} \right).\label{I3222fin}
\end{align}
Up to scheme-dependent terms, these auxiliary integrals are simply the finite parts of the  divergent triple-$K$ integrals after which they are named.\footnote{For example, we define $I_{2\{1+\ep,1+\ep,1+\ep\}}=-1/3\ep+I_{2\{111\}}^{\text{(fin)}}+O(\ep)$. From (4.20) of \cite{Bzowski:2015yxv}, this definition is equivalent to \eqref{I2111fin}.  Alternatively, from (3.16) of \cite{Bzowski:2015yxv} with $\alpha=3$ and $\beta_j=1+\ep$, we obtain \eqref{I2111fin0} after noting that $I_{1\{\ep,\ep,\ep\}}=I_{1\{000\}}+O(\ep^2)$ and using the dilatation Ward identity for $I_{1\{000\}}$.} 
Similarly, although we will not write it this way, the frequently occurring combination $(1/4)J^2 I_{1\{000\}}$ is simply the finite non-local part of the divergent integral $I_{0\{111\}}$ (see (4.6) of \cite{Bzowski:2015yxv}). 
The derivatives appearing in these formulae, and in all our later results, 
can be trivially evaluated using 
\begin{align}\label{Ireduce}
p_1\frac{\p}{\p p_1}I_{1\{000\}} &=\frac{1}{J^2}\Big[ 2p_1^2(p_1^2-p_2^2-p_3^2)I_{1\{000\}} \nn\\&\quad -
p_1^2\ln p_1^2+\frac{1}{2}(p_1^2+p_2^2-p_3^2)\ln p_2^2 +\frac{1}{2}(p_1^2-p_2^2+p_3^2)\ln p_3^2\Big],
\end{align}
with the analogous results for other momenta following by permutation.
In this fashion, for example, we can re-write \eqref{I2111fin0} as given in equation (3.48) of \cite{Bzowski:2017poo}, 
\begin{align}\label{I2111fin}
I_{2\{111\}}^{\text{(fin)}}
&=- \frac{4 p_1^2 p_2^2 p_3^2}{J^2} I_{1\{000\}} - \frac{1}{6 J^2} \left[ p_1^2 (p_2^2 + p_3^2 - p_1^2) \ln \left( \frac{p_1^4}{p_2^2 p_3^2} \right) \right.\nn\\
& \qquad\qquad \left. + \: p_2^2 (p_1^2 + p_3^2 - p_2^2) \ln \left( \frac{p_2^4}{p_1^2 p_3^2} \right) + p_3^2 (p_1^2 + p_2^2 - p_3^2) \ln \left( \frac{p_3^4}{p_1^2 p_2^2} \right) \right]. 
\end{align}
To obtain compact expressions, however, we usually leave such derivatives unevaluated.

\subsection{Renormalised 2-point functions}
\label{sec:2ptfns}

In spacetimes of general dimension $d>2$, the momentum-space 2-point functions read
\begin{align}
\lla \O^I(\bs{p}) \O^J(-\bs{p}) \rra & = \2_{\O\O}\delta^{IJ} p^{2 \Delta - d}, \label{e:2ptOO} \\[1ex]
\lla J^{\mu a}(\bs{p}) J^{\nu b}(-\bs{p}) \rra & = \2_{JJ} \delta^{ab} \pi^{\mu\nu}(\bs{p}) p^{d-2}, \label{e:2ptJJ} \\[1ex]
\lla T_{\mu\nu}(\bs{p}) T_{\rho\sigma}(-\bs{p}) \rra & = \2_{TT} \Pi_{\mu\nu\rho\sigma}(\bs{p}) p^{d}, \label{e:2ptTT}
\end{align}
while all 2-point functions of different operators vanish. 
These expressions are unsuitable however when 
the power of the momenta becomes an even non-negative integer.  When this occurs, the 2-point function is ultralocal in position space ({\it i.e.,} has support only when the operator insertions coincide) and can be removed by a local counterterm.  
The 2-point function would then vanish implying the operator has zero norm in violation of unitarity.

In reality, in all such cases the corresponding coefficient $\2_{\O\O}$, $\2_{JJ}$ or $\2_{TT}$ has a UV divergence, and renormalisation is necessary.  For the scalar 2-point function, this occurs whenever
\begin{equation}\label{2ptdivcases}
n = \Delta - \frac{d}{2}
\end{equation}
is a non-negative integer, while for 2-point functions of conserved currents and the stress tensor, renormalisation is necessary in all even spacetime dimensions $d=2N$.  

To remove these divergences we pass to the  dimensionally regulated theory with
\[
\tilde{d}= d+2u\ep, \qquad \tilde{\Delta} = \Delta+(u+v)\ep,
\]
where $u$ and $v$ are constants parametrising our choice of regularisation scheme.  (For stress tensors and  currents, conservation enforces $v=u$, however for a scalar $v$ is unrestricted.)  Provided $v\neq 0$, the 2-point functions are now finite.  The overall normalisation constant is now a divergent function of the regulator,  
\[\label{2ptnormexp}
\2_{j}(\ep) = \frac{\2_{j}}{v \ep}+\2^{(0)}_{j}+O(\ep), \qquad j\in \{\O\O, JJ, TT\}.
\]
allowing the regulated 2-point functions to be expanded as 
\begin{align}
\lla \O^I(\bs{p}) \O^J(-\bs{p}) \rra_{\text{reg}} & = \delta^{IJ} p^{2\Delta-d} \Big[ \frac{\2_{\O\O}}{v \epsilon} + \2_{\O\O} \ln p^2 + \sl_{\O\O} + O(\epsilon) \Big], \label{e:2ptOOreg} \\[1ex]
\lla J^{\mu a}(\bs{p}) J^{\nu b}(-\bs{p}) \rra_{\text{reg}} & = \delta^{ab} \pi^{\mu\nu}(\bs{p}) p^{d-2} \Big[ \frac{\2_{JJ}}{v \epsilon} + \2_{JJ} \ln p^2 + \sl_{JJ} + O(\epsilon) \Big], \label{e:2ptJJreg} \\[1ex]
\lla T_{\mu\nu}(\bs{p}) T_{\rho\sigma}(-\bs{p}) \rra_{\text{reg}} & = \Pi_{\mu\nu\rho\sigma}(\bs{p}) p^{d} \Big[ \frac{\2_{TT}}{v \epsilon} + \2_{TT} \ln p^2 + \sl_{TT} + O(\epsilon) \Big]. \label{e:2ptTTreg}
\end{align}

These expressions can be renormalised through the addition of suitable counterterms.  To quadratic order in the sources, these counterterms are
\begin{align} \label{e:SctOO}
S_{\text{ct}}  =  \int \D^{d + 2 u \epsilon} \bs{x}\, \sqrt{g}\, \mu^{2 v  \epsilon}\Big[&\ct_{\O\O} \phi^I \Box^{\Delta-d/2} \phi^I +\ct_{JJ} \delta^{ab}  F_{\mu\nu}^a \Box^{(d-4)/2} F^{\mu\nu b}\nn\\&\quad +\ct_{TT}  W_{\mu\nu\rho\sigma} \Box^{(d-4)/2} W^{\mu\nu\rho\sigma}\Big],
\end{align}
where $W_{\mu\nu\rho\sigma}$ denotes the Weyl tensor and the renormalisation scale $\mu$ enters on dimensional grounds. 
All the Laplacians are raised to positive integer powers, as required for locality, since the scalar counterterm exists only when $\Delta-d/2=n$ while those for the currents and stress tensors only when $d=2N\ge 4$. 
Beyond quadratic order in the sources, these Laplacians should be replaced by their Weyl-covariant generalisations.
 
Choosing the counterterm coefficients
\begin{align} 
\ct_{\O\O} & = \frac{(-1)^{\Delta-d/2} \2_{\O\O}}{2 v \epsilon} + \ct_{\O\O}^{(0)} + O(\epsilon), \label{e:ctOO} \\[1ex]
\ct_{JJ} & = \frac{(-1)^{d/2} \2_{JJ}}{4 v \epsilon} + \ct_{JJ}^{(0)} + O(\epsilon), \label{e:ctJJ} \\[1ex]
\ct_{TT} & = \frac{(-1)^{d/2} \2_{TT}}{4 v \epsilon} + \ct_{TT}^{(0)} + O(\epsilon), \label{e:ctTT}
\end{align}
where the finite coefficients $\ct_{\O\O}^{(0)}$, $\ct_{JJ}^{(0)}$ and $\ct_{TT}^{(0)}$  encode a particular choice of renormalisation scheme, the renormalised correlation functions can now be obtained by subtracting the counterterm contributions from the regulated correlators and taking the limit $\ep\rightarrow 0$.  This procedure yields the renormalised correlators
\begin{align}
\lla \O^I(\bs{p}) \O^J(-\bs{p}) \rra & = \delta^{IJ} p^{2\Delta-d} \Big[ \2_{\O\O} \ln \frac{p^2}{\mu^2} + \sdt_{\O\O} \Big], \label{e:2ptOOren} \\[1ex]
\lla J^{\mu a}(\bs{p}) J^{\nu b}(-\bs{p}) \rra & = \delta^{ab} \pi^{\mu\nu}(\bs{p}) p^{d-2} \Big[ \2_{JJ} \ln \frac{p^2}{\mu^2} + \sdt_{JJ} \Big], \label{e:2ptJJren} \\[1ex]
\lla T_{\mu \nu}(\bs{p}) T_{\rho \sigma}(-\bs{p}) \rra & = \Pi_{\mu\nu\rho\sigma}(\bs{p}) p^{d} \Big[ \2_{TT} \ln \frac{p^2}{\mu^2} + \sdt_{TT} \Big] \label{e:2ptTTren}
\end{align}
where
\begin{align}\label{sdtOO}
\sdt_{\O\O} & = \sl_{\O\O} - 2 (-1)^{\Delta-d/2} \ct_{\O\O}^{(0)}, \\[1ex]
\label{sdtJJ}
\sdt_{JJ} & = \sl_{JJ} - 4 (-1)^{d/2} \ct_{JJ}^{(0)}, \\[1ex]
\sdt_{TT} & = \sl_{TT} - 4 (-1)^{d/2} \ct_{TT}^{(0)}.
\end{align}
From the perspective of the renormalised theory, however, the constants  $\sdt_{\O\O}$, $\sdt_{JJ}$ and $\sdt_{TT}$ represent scheme-dependent terms whose values can be adjusted arbitrarily through a change of the renormalisation scale. 

Unlike our original expressions \eqref{e:2ptOO}\,-\,\eqref{e:2ptTT}, the renormalised correlators \eqref{e:2ptOOren}\,-\,\eqref{e:2ptTTren} are non-analytic functions of the squared momentum, and are hence nonlocal.
Due to the explicit $\mu$-dependence introduced by the counterterms, they acquire a scale-dependence
\begin{align}
\mu \frac{\partial}{\partial \mu} \lla \O^I(\bs{p}) \O^J(-\bs{p}) \rra & = - 2 p^{2\Delta-d} \2_{\O\O}, \\[1ex]
\mu \frac{\partial}{\partial \mu} \lla J^{\mu a}(\bs{p}) J^{\nu b}(-\bs{p}) \rra & = - 2 \delta^{ab} \pi^{\mu\nu}(\bs{p}) p^{d-2} \2_{JJ}, \\[1ex]
\mu \frac{\partial}{\partial \mu} \lla T_{\mu \nu}(\bs{p}) T_{\rho \sigma}(-\bs{p}) \rra & = - 2 \Pi_{\mu\nu\rho\sigma}(\bs{p}) p^{d} \2_{TT}.
\end{align}
The anomalous terms appearing on the right-hand sides of these equations are finite, local, and scheme-independent.

\subsection{Anomalies and beta functions}
\label{sec:anomaliesandbetafns}

The general structure of the anomalies and beta functions can be understood from \eqref{dmuW}.
Let us consider first the case where only anomalies are present, and no beta functions.  This case encompasses all 2-point functions, and 3-point functions for which only type $(---)$ singularities arise.   
The anomaly action is then 
\[\label{anomformulaonlysources}
A = -\lim_{\ep\rightarrow 0} \mu\frac{\p}{\p\mu}S_{\mathrm{ct}},
\]
where the counterterm action depends only on the non-dynamical sources and hence can be extracted from the expectation value in \eqref{Wlim}.
The quadratic anomaly action associated with the 2-point functions above is thus
\begin{align} 
A 
&= -\int \D^{d} \bs{x} \sqrt{g}\,\Big[C_{\O\O}  \phi^I (-\Box)^{\Delta-d/2} \phi^I +\frac{1}{2}C_{JJ} F_{\mu\nu}^a (-\Box)^{(d-4)/2} F^{\mu\nu a}  \nn\\&\qquad\qquad\qquad \quad 
+\frac{1}{2}C_{TT} W_{\mu\nu\rho\sigma}(-\Box)^{(d-4)/2} W^{\mu\nu\rho\sigma}\Big]. \label{e:tranom}
\end{align}

Let us now consider cases involving beta functions.  These arise wherever $(+--)$ singularities of 3-point functions (or permutations thereof) are removed by counterterms involving two sources and an operator.  Where type $(---)$ singularities are also present, then we have both beta functions and anomalies.
For concreteness, let us consider the case of 
a beta function for the gauge field $A^a_\mu$. 
Here, the bare source $A^{a\,\mathrm{(bare)}}_\mu$ for the current $J^{\mu a}$ is renormalised by a cubic counterterm involving $J^{\mu a}$ and some quadratic source combination, 
which we denote schematically by
\[
A^{a\,(\mathrm{ct})}_\mu = \frac{\delta S_{\mathrm{ct}}}{\delta J^{\mu a}}.
\] 
In the full action for counterterms and sources,  the current thus appears as  
\[
S_{\mathrm{source}}+S_{\mathrm{ct}}=
\int\D^d\x \,(A^a_\mu+A^{a\,\mathrm{(ct)}}_\mu)J^{\mu a} +\ldots = \int\D^d\x\, A^{a\,\mathrm{(bare)}}_\mu J^{\mu a} + \ldots
\]
Perturbatively inverting,  the renormalised source $A^a_\mu$ (with respect to which we differentiate to find the renormalised correlators), is now given by  $A^a_\mu = A^{a\,\mathrm{(bare)}}_\mu - A^{a\,\mathrm{(ct)}}_\mu$.  
Since by definition  $A^{a\,\mathrm{(bare)}}_\mu$ is independent of the renormalisation scale $\mu$,  the beta function is now 
\[\label{betagenform}
\beta_{A^a_\mu} 
= - \lim_{\ep\rightarrow 0}\mu\frac{\p}{\p\mu}A^{a\,\mathrm{(ct)}}_\mu. 
\]
From \eqref{dmuW} and \eqref{Wlim}, the corresponding anomaly action is
\begin{align}
A & 
=  \Big( \,\mu\frac{\p}{\p\mu} + \int\D^d\x\, \beta_{A^a_\mu}\frac{\delta}{\delta A^a_\mu}\Big)
\lim_{\ep\rightarrow 0}\Big[\ln\,\<\,e^{-S_{\mathrm{source}}-S_{\mathrm{ct}}}\,\>\Big]
\nn\\[2ex]
&= \lim_{\ep\rightarrow 0}\Big[ -\mu\frac{\p }{\p\mu}S_{\mathrm{ct}}+\int\D^d\x\,\Big(\mu\frac{\p}{\p\mu}A^{a\,\mathrm{(ct)}}_\mu\Big)\Big(J^{\mu a}+\frac{\delta S_{\mathrm{ct}}}{\delta A^a_\mu}\Big) \Big].
\label{anomgenform}
\end{align}
Here, the first term on the second line involves only the counterterm action, since the source action 
has no explicit dependence on $\mu$.
In the second term, the first bracketed factor is equivalent to (minus) the beta function, while the second bracketed factor derives from differentiating the source and counterterm actions with respect to $A^a_\mu$. 
As the beta function is already of quadratic order in the sources, note here we need only  evaluate the linear part of  $\delta S_{\mathrm{ct}}/\delta A^a_\mu$ coming from the 2-point counterterm action \eqref{e:SctOO}. 

The second line of \eqref{anomgenform} now depends only on the sources and not on the current.  Clearly this is true for both the linear part of $\delta S_{\mathrm{ct}}/\delta A^a_\mu$ and for the beta function, while the explicit $J^{\mu a}$ in the second term cancels with the first-term contribution coming from acting with $-\mu(\p/\p\mu)$ on the $\int\D^d\x\, A^{a\,\mathrm{(ct)}}_\mu J^{\mu a}$ piece of $S_{\mathrm{ct}}$.
As the anomaly action thus depends only on the non-dynamical sources, and not on the operators, we were able to remove the expectation value in the second line.  This remains true even in more complicated cases where beta functions for other operators are present, once \eqref{anomgenform} is suitably generalised.

\paragraph{Examples.}
Let us consider the correlator $\<J^{\mu a}\O^{I_2}\O^{I_3}\>$, where $\O^I$ is a marginal scalar.
In $d=3$, as discussed in section \ref{sec:beta3d}, this has only a $(+--)$ singularity and hence a beta function but no anomaly.  In $d=4$, as discussed in section \ref{sec:JOOd4analysis}  and \ref{sec:AnomalousCWIs}, we find both $(+--)$ and $(---)$ singularities, and hence a beta function as well as an anomaly.
Both cases can be understood from the general formulae \eqref{betagenform} and \eqref{anomgenform}.

Working in a scheme with $u=v_1$ and $v_2=v_3$, as dictated by current conservation and permutation symmetry, in three dimensions the relevant counterterm action is 
\[
S_{\mathrm{ct}} = \int\D^{3+2u\ep}\x \, \mathfrak{c}_1 \mu^{2(v_2-u)\ep} ig (T^a_R)^{IJ} \,\phi^I D_\mu\phi^J  J^{\mu a},
\]
where 
the coefficient 
 $\mathfrak{c}_1 =\mathfrak{c}_1^{(-1)}\ep^{-1}+O(\ep^0)$ has a pole removing the  $(+--)$ singularity of the correlator. 
From \eqref{betagenform}, the beta function is then 
\[\label{betafnA}
\beta_{A^a_\mu} = -2(v_2-u) \mathfrak{c}_1^{(-1)}  ig (T^a_R)^{IJ} \,\phi^I D_\mu\phi^J.
\]
From \eqref{anomgenform}, the anomaly vanishes
since $\delta S_{\mathrm{ct}}/\delta A^a_\mu$ has no linear piece in three dimensions, and the remaining terms cancel.

In four dimensions, 
including all terms up to cubic order, we have instead 
\begin{align}
S_{\mathrm{ct}} &= \int\D^{4+2u\ep}\x\,\Big[\mathfrak{c}_{JJ}\mu^{2u\ep}F^{\mu\nu a}F^a_{\mu\nu}+\mathfrak{c}_{\O\O} \mu^{2 v_2\ep}(D^2\phi^I)^2+
\mathfrak{c}_1 \mu^{2(v_2-u)\ep}ig(T^a_R)^{IJ} \phi^{I} D_\mu \phi^{J} J^{\mu a}\nn\\&\qquad\qquad\quad\qquad
+\mathfrak{c}_2 \mu^{2v_2\ep}ig(T^a_R)^{IJ}F^{\mu\nu a} D_\mu\phi^{I}D_\nu \phi^{J}\Big].
\end{align}
As we saw in \eqref{e:SctOO}, the first two counterterms 
are required for the renormalisation of the 2-point functions.
Only the final three terms proportional to $\mathfrak{c}_{\O\O}$, $\mathfrak{c}_1$ and $\mathfrak{c}_2$ contribute to the 3-point function at hand.  
Analysing the divergences, we find  that all counterterm coefficients carry single poles except for $\mathfrak{c}_2$, which has a {\it double} pole 
\[
\mathfrak{c}_2 = \mathfrak{c}_2^{(-2)}\ep^{-2}+\mathfrak{c}_1^{(-1)}\ep^{-1}+O(\ep^0).
\]
The $(+--)$ counterterm proportional to $\mathfrak{c}_1$ leads again to the beta function given in \eqref{betafnA} above.
Using \eqref{anomgenform}, the anomaly action is
\begin{align}
A&=\lim_{\ep\rightarrow 0}\int\D^{4+2u\ep}\x\Big[-2u\ep\mathfrak{c}_{JJ}\mu^{2u\ep}F^{\mu\nu a}F^a_{\mu\nu}-2v_2\ep\mathfrak{c}_{\O\O}\mu^{2v_2\ep}(D^2\phi^I)^2\nn\\[0ex]&\qquad\qquad
+(-2v_2\ep\mathfrak{c}_2+8(v_2-u)\ep\mathfrak{c}_1\mathfrak{c}_{JJ})ig(T^a_R)^{IJ}F^{\mu\nu a}D_{\mu}\phi^ID_{\nu}\phi^J\Big].
\end{align}
For the final term to have a finite limit, the pole of $\ep\mathfrak{c}_2$ must cancel against that of $\ep\mathfrak{c}_1\mathfrak{c}_{JJ}$, the term coming from $\delta S_{\mathrm{ct}}/\delta A^a_\mu$ times the beta function.  As we will see in section \ref{sec:JOOd4analysis}, when we insert the specific counterterm coefficients obtained from our analysis of the divergences, this is indeed precisely what happens.  Making use of \eqref{e:ctOO} and \eqref{e:ctJJ}, the anomaly action is then
\[\label{anomalyactionfor4d}
A = -\int\D^4\x\,\Big[\frac{1}{2}C_{JJ}F^{\mu\nu a}F^a_{\mu\nu}+C_{\O\O}(D^2\phi^I)^2
+ 2a_0 ig(T^a_R)^{IJ}F^{\mu\nu a}D_{\mu}\phi^ID_{\nu}\phi^J\Big],
\]
where 
\[\label{a0firstreln}
a_0 = v_2\mathfrak{c}_2^{(-1)}-4(v_2-u)(\mathfrak{c}_1^{(-1)}\mathfrak{c}_{JJ}^{(0)}+
\mathfrak{c}_1^{(0)}\mathfrak{c}_{JJ}^{(-1)}).
\]
As we will see  in section 
 \ref{sec:JOOd4analysis}, however, this term is scheme-dependent and can be consistently set to zero.  In fact, as we discuss in section \ref{sec:AnomalousCWIs}, the $a_0$ term in \eqref{anomalyactionfor4d} is Weyl exact and hence does not represent a genuine anomaly.

\section{Results for renormalised correlators}

\label{sec:results}

We now present our main results for renormalised 3-point correlators.  In each case, we list the relevant transverse and trace Ward identities; the decomposition of tensor structure into transverse-traceless form factors; the primary and secondary conformal Ward identities; the divergences arising and the counterterms available to us for their disposal.  We generally classify these counterterms according to whether they give rise to beta functions or to conformal anomalies.   For each correlator, we then compute explicit results for the cases $d=3$ and $d=4$, with scalar operators of dimension $\Delta^I=d-2$ and $\Delta^I=d$.

\subsection{\texorpdfstring{$\<J^\mu\O\O\>$}{<JOO>} }
\label{JOOanalysis}

\subsubsection{General analysis}

\paragraph{Decomposition.} Using the transverse Ward identity,
\begin{align} \label{e:pJOO}
& p_{1 \mu_1} \lla J^{\mu_1 a}(\bs{p}_1) \mathcal{O}^{I_2}(\bs{p}_2) \mathcal{O}^{I_3}(\bs{p}_3) \rra \nn\\[1ex]
& \qquad = - g (T_R^a)^{K I_3} \lla \mathcal{O}^K(\bs{p}_2) \mathcal{O}^{I_2}(-\bs{p}_2) \rra - g (T_R^a)^{K I_2} \lla \mathcal{O}^K(\bs{p}_3) \mathcal{O}^{I_3}(-\bs{p}_3) \rra,
\end{align}
we can decompose the 3-point function into  transverse and longitudinal pieces,
\begin{align}\label{JOOdecomp}
& \lla J^{\mu_1 a}(\bs{p}_1) \mathcal{O}^{I_2}(\bs{p}_2) \mathcal{O}^{I_3}(\bs{p}_3) \rra = \lla j^{\mu_1 a}(\bs{p}_1) \mathcal{O}^{I_2}(\bs{p}_2) \mathcal{O}^{I_3}(\bs{p}_3) \rra \nn\\[0.5ex]
& \qquad - \: \frac{p_1^{\mu_1}}{p_1^2} \left[ g (T_R^a)^{K I_3} \lla \mathcal{O}^K(\bs{p}_2) \mathcal{O}^{I_2}(-\bs{p}_2) \rra + g (T_R^a)^{K I_2} \lla \mathcal{O}^K(\bs{p}_3) \mathcal{O}^{I_3}(-\bs{p}_3) \rra \right].
\end{align}

\paragraph{Form factors.} The transverse part can then be written
\begin{equation}\label{JOOformfactors}
\lla j^{\mu_1 a}(\bs{p}_1) \mathcal{O}^{I_2}(\bs{p}_2) \mathcal{O}^{I_3}(\bs{p}_3) \rra = A_1^{a I_2 I_3}\,  \pi^{\mu_1}_{\alpha_1}(\bs{p}_1)  p_2^{\alpha_1},
\end{equation}
where the scalar form factor $A_1$ is a function of the momentum magnitudes and is symmetric under $(p_2, I_2) \leftrightarrow (p_3, I_3)$, \textit{i.e.},
\begin{equation}
A_1^{a I_3 I_2}(p_1, p_3, p_2) = A_1^{a I_2 I_3}(p_1, p_2, p_3).
\end{equation}
Its relation to the complete correlator is
\begin{equation}
A_1^{a I_2 I_3} = \text{coefficient of } p_2^{\mu_1} \text{ in } \lla J^{\mu_1 a}(\bs{p}_1) \mathcal{O}^{I_2}(\bs{p}_2) \mathcal{O}^{I_3}(\bs{p}_3) \rra,
\end{equation}
where before reading off the coefficient we first impose the cyclic rule \eqref{a:momenta}, which in this case amounts to eliminating $p_3^{\mu_1}$ via momentum conservation.

\paragraph{Primary CWIs.}  The primary CWIs are
\begin{equation}\label{primaryCWIJOO}
\K_{ij} A_1^{a I_2 I_3} = 0, \qquad i,j = 1,2,3.
\end{equation}
Their solution in terms of triple-$K$ integrals is
\begin{equation} \label{e:pri1JOO}
A_1^{a I_2 I_3} = \3{1}^{a I_2 I_3} J_{1 \{000\}},
\end{equation}
where $\3{1}^{a I_2 I_3}$ is a primary constant. 

\paragraph{Secondary CWIs.} There is only one independent secondary CWI, which reads
\begin{equation} \label{e:JOOsec}
\Lo_{1} A_1^{a I_2 I_3} = 2(\Delta_1-1) \, p_{1 \mu_1} \lla J^{\mu_1 a}(\bs{p}_1) \mathcal{O}^{I_2}(\bs{p}_2) \mathcal{O}^{I_3}(\bs{p}_3) \rra.
\end{equation}
The right-hand side of this identity can be evaluated using the transverse Ward identity \eqref{e:pJOO} and the scalar 2-point function \eqref{e:2ptOO}.  
The role of this secondary CWI is then to fix the primary constant $\3{1}^{a I_2 I_3}$ in terms of the scalar 2-point normalisation $C_{\O\O}$.  To obtain this relation, it is sufficient to work in the soft limit $p_3\rightarrow 0$.  In this limit, the left-hand side of \eqref{e:JOOsec} can easily be evaluated using \eqref{e:Jid1} and \eqref{e:limJ}.  Divergences, where they arise, can be avoided by working in the regulated theory. 
One finds that the secondary CWI can only be satisfied if 
\begin{equation} \label{e:sec1JOO}
\3{1}^{a I_2 I_3} = \frac{2^{4-d/2} g (T^a_R)^{I_2 I_3}\, \2_{\O\O}(\ep)\,  \sin \left( \pi (\Delta_2 - d/2) \right)}{\pi \Gamma \left( d/2 - 1 \right)}\,\delta_{\Delta_2, \Delta_3}.
\end{equation}
The 3-point function thus vanishes for an uncharged scalar operator, for which $g=0$, or whenever $\Delta_2 \neq \Delta_3$.  For cases where $\Delta_2=d/2+n$, the 3-point function is actually non-vanishing since the zero in the sine function cancels against the pole in the 2-point normalisation $C_{\O\O}(\ep)$, see \eqref{2ptdivcases} and \eqref{2ptnormexp}.

\paragraph{Regularisation.} 

To obtain a nonzero 3-point function, in the following we  restrict to the case of a single scalar operator with $\Delta_2 = \Delta_3$.  This requires a scheme with $v_2 = v_3$.  To preserve current conservation, we also impose $u=v_1$.
Provided $u\neq v_2$, this scheme is then sufficient to regulate all the divergences  arising,  summarised in the following table:

\begin{center}
\begin{tabular}{|c|c|c|c|c|} \hline
Form factor & Integral & $(---)$  & $(+--)$ & $(-++)$  \\ \hline
$A_1$ & $J_{1\{000\}}$ & $\Delta = d/2+1+n $ & $\Delta = d + n$
& $\Delta =d/2-1-n$ \\ \hline
\end{tabular}
\captionof{table}{Singularities arising for $\< J^{\mu_1} \O^{I_2} \O^{I_3} \>$, with $\Delta=\Delta_2 = \Delta_3$.
\label{tab:JOO}}
\end{center}
Note that singularities of type $(++-)$ and $(+++)$ are forbidden as the former requires $d\le 0$ while the latter requires $\Delta_2=\Delta_3\le 0$.
For a unitary theory, the $(-++)$ singularity only occurs for $\Delta_2=\Delta_3=d/2-1$ corresponding to a free theory.

\paragraph{Renormalisation.}  To remove these singularities we need to evaluate the counterterms available to us.  These fall into two classes.  The first class consists of counterterms that are cubic in the sources: these serve to eliminate $(---)$ singularities and give rise to anomalies. 
The second class consists of counterterms involving two sources and one operator.  Counterterms of this type remove $(+--)$ singularities and give rise to beta functions, since adding them to the Lagrangian redefines the source of the operator in question. 

As we will see below, 
counterterms exist for all of the cases in the following table:

\begin{center}
\begin{tabular}{|c|c|}
\hline
Singularity type & Counterterms available when  \\ \hline
$(---)$ & $\Delta_2=\Delta_3 = d/2+1+n$ \\ \hline
$(+--)$ & $\Delta_2=\Delta_3 = d + n$ \\ \hline
\end{tabular}
\captionof{table}{Availability of counterterms for $\< J^{\mu_1} \O^{I_2} \O^{I_3} \>$.\label{tab:physJOO}}
\end{center}
Comparing with table \ref{tab:JOO}, we see that for $(---)$ and $(+--)$ singularities a counterterm is always available.  In the remaining case of a $(-++)$ singularity, the primary constant multiplying the divergence must instead vanish as a suitable power of the regulator.  The resulting finite form factor is then fully nonlocal, see \cite{Bzowski:2015pba}.

\paragraph{Anomalies.} 
To find the possible anomalies, we must therefore classify all counterterms with one $A_{\mu}^a$ and two $\phi^I$.   Since  counterterms must be both Lorentz- and gauge-invariant, the source $A_{\mu}^a$ can only appear through its field strength $F^a_{\mu\nu}$ or  covariant derivatives $D_{\mu}$.  
(We will ignore possible topological terms since we are considering only the parity-even part of correlation functions.)
We therefore have two possible families of counterterms, of which the simplest representatives are
\begin{equation} \label{e:anomalctJOO}
\int \D^{d} \bs{x} D_\mu \phi^I D^\mu \phi^I, \qquad\qquad g (T^a_R)^{IJ} \int \D^{d} \bs{x} F^{\mu \nu a} D_\mu \phi^I D_\nu \phi^J.
\end{equation}
More complicated examples can then be constructed featuring an even number of additional covariant derivatives.  
Counterterms of this type exist and give rise to anomalies whenever
\begin{equation} \label{e:JOO_anomaly_cond}
\Delta_2 = \Delta_3 = \frac{d}{2} + 1 + n, \quad n=0,1,2,\ldots
\end{equation}

\paragraph{Beta functions.} 
There are potentially four types of counterterm involving two sources and one operator:
(i) those containing the current $J^{\mu a}$ and two scalar sources $\phi^I$; (ii) those containing $J^{\mu a}$ plus the sources $\phi^I$ and $A_\mu^a$; (iii) those containing the scalar operator $\O^I$ and two scalar sources $\phi^I$; (iv) those containing $\O^I$ along with $\phi^J$ and $A_\mu^a$.

The simplest counterterm of the form (i) is
\begin{equation} \label{e:formctJOO}
g (T^a_R)^{IJ} \int \D^{d} \bs{x}\, J^{\mu a} \phi^I D_\mu \phi^J.
\end{equation}
Allowing for the possible addition of further derivatives, the existence of this counterterm then leads to the following requirement for a  nontrivial beta function:
\begin{equation} \label{e:condJOO}
\Delta_2 = \Delta_3 = d + n, \quad n=0,1,2,\ldots
\end{equation}
In fact, this counterterm is the only one capable of generating a nontrivial beta function. As we will now verify, all the remaining possibilities are either ruled out or else do not contribute to the 3-point function at hand.

The two simplest counterterms of the form (ii) are
\begin{equation}
r^{aI} \int \D^{d} \bs{x} J^{\mu a} D_{\mu} \phi^I, \qquad\qquad r^{abI} \int \D^{d} \bs{x} F_{\mu}^{\ \nu a} J^{\mu b} D_{\nu} \phi^I,
\end{equation}
where $r^{aI}$ and $r^{abI}$ are some invariant tensors specified by a chosen representation. The first of these counterterms must be rejected as it contributes to the 2-point function $\< J^{\mu a} \O^I \>$, which has to vanish by conformal invariance. 
The second counterterm simply does not contribute,  either to the 2-point functions or to the 3-point function.  (Its contribution to  $\< J^{\mu a} \O^I \O^J \>$ is proportional to $r^{abI} \< J^{\mu b} \O^J \> = 0$.) 

Through similar reasoning, counterterms 
of the form (iii) also make no contribution to the 3-point function. Indeed, the simplest such term is
\begin{equation}
r^{IJK} \int \D^{d} \bs{x}\, \O^I \phi^J \phi^K,
\end{equation}
whose contribution to $\< J^{\mu a} \O^I \O^J \>$ is proportional to $r^{IJK} \< J^{\mu a} \O^K \> = 0$.
Finally, counterterms of the form (iv) are forbidden on dimensional grounds, since the scaling dimension of the combination $A_\mu^{a} \O^I \phi^J$ already exceeds the spacetime dimension.
With these general considerations in place, we now proceed to examine some specific cases of interest.

\subsubsection{\texorpdfstring{$d=3$ and $\Delta_2 = \Delta_3 = 1$}{d=3 and Delta=1}}

The relevant triple-$K$ integral for this case is finite and  \eqref{e:sec1JOO} leads directly to
\begin{equation}
A_1^{a I_2 I_3} = -\frac{2 g (T_R^a)^{I_2 I_3} \2_{\O\O}}{a_{123} b_{23}},
\end{equation}
where the symmetric polynomials appearing in the denominator are defined in \eqref{e:variables}.

\subsubsection{\texorpdfstring{$d=3$ and $\Delta_2 = \Delta_3 = 3$}{d=3 and Delta=3}}
\label{sec:beta3d}

Here, the triple-$K$ integral arising in our solution \eqref{e:pri1JOO} of the primary CWIs is linearly divergent.
Its evaluation in a fully general regularisation scheme yields
\begin{align} \label{e:iJOO}
& \left( \frac{\pi}{2} \right)^{-\frac{3}{2}} I_{\frac{3}{2} + u \epsilon \{ \frac{1}{2} + v_1 \epsilon, \frac{3}{2} + v_2 \epsilon, \frac{3}{2} + v_3 \epsilon \}}  \nn\\[1ex]
& \qquad = -  \frac{p_1}{(u + v_1 - v_2 - v_3) \epsilon} + \frac{p_1}{u + v_1 - v_2 - v_3} \left[ u (-2 + \gamma_E + \ln 2) - v_1 \ln p_1^2 \right]  \nn\\[1ex]
& \qquad\qquad + \: \Big[ p_1 \ln a_{123} + (1 - \ln 2) p_1 - \frac{p_1 (p_2 + p_3) + p_2 p_3 + p_2^2 + p_3^2}{a_{123}} \Big] + O(\epsilon).
\end{align}
This result is obtained by first evaluating the integral in the special scheme $v_j = 0$ for  $j=1,2,3$, in which all Bessel functions reduce to elementary functions.  The scheme can then be changed through the addition of suitable terms using the method described in \cite{Bzowski:2015yxv}.

As discussed above, current conservation and the requirement $\Delta_2=\Delta_3$ impose a regularisation scheme where $u=v_1$ and $v_2=v_3$.
In fact, these conditions are also imposed independently by the secondary Ward identity \eqref{e:JOOsec}.  As it is interesting to see this in operation, we will leave the parameters $u$, $v_j$ generic for the time being.
The primary Ward identities are thus solved by the regulated triple-$K$ integral above, multiplied by an undetermined constant $\3{1}^{aI_2I_3}(\ep)$, which is itself a function of the regulator $\ep$.  
Since the integral is linearly divergent, to keep track of finite terms in the product we need to expand this constant to linear order in the regulator,
\begin{equation}
\3{1}^{aI_2I_3}(\ep) = \3{1}^{(0)aI_2I_3} + \epsilon \: \3{1}^{(1)aI_2I_3} + O(\epsilon^2).
\end{equation}
The left-hand side of the secondary Ward identity \eqref{e:JOOsec} then reads 
\begin{equation} \label{e:secL_JOO}
\Lo_{1} A_1^{aI_2I_3} = \left( \frac{\pi}{2} \right)^{\frac{3}{2}} \3{1}^{(0)aI_2I_3} \Big[ p_2^3 - p_3^3 + \frac{(u - v_1) p_1 (p_2^2 - p_3^2) + (v_2 - v_3) p_1^3}{u + v_1 - v_2 - v_3} + O(\epsilon) \Big].
\end{equation}
As we see, this result does not depend on the subleading term $\3{1}^{(1)aI_2I_3}$. The right-hand side of the Ward identity \eqref{e:JOOsec} is instead
\begin{equation} \label{e:secR_JOO}
-2 g \2_{\O\O} (T_R^a)^{I_2 I_3} (p_2^3 - p_3^3 ) + O(\epsilon),
\end{equation}
after using the transverse Ward identity \eqref{e:pJOO}. 
Comparing these two expressions, we see that the scheme-independent terms match provided
\begin{equation} \label{e:solsecJOO}
\3{1}^{a I_2 I_3} = - 2 \left( \frac{\pi}{2} \right)^{-\frac{3}{2}} g \2_{\O\O} (T_R^a)^{I_2 I_3} + O(\epsilon).
\end{equation}
This result is also consistent with our general formula \eqref{e:sec1JOO}. 
The remaining scheme-dependent terms in \eqref{e:secL_JOO} must then vanish, for arbitrary values of the momenta.  Clearly this is only possible if $u=v_1$ and $v_2=v_3$.  
To solve the secondary Ward identity \eqref{e:JOOsec} thus {\it requires} working in a regularisation scheme respecting current conservation ($u=v_1$) and permutation symmetry of the scalar operators ($v_2=v_3$).

Our remaining task is to renormalise the correlation function. A single counterterm is available, namely
\begin{equation} \label{e:JOO_Sct}
S_{\text{ct}} = \ct_1\,  \int \D^{3 + 2 u \epsilon} \bs{x}\,i g (T^a_R)^{I_2 I_3} J^{\mu a} \phi^{I_2} D_\mu \phi^{I_3} \mu^{2 \epsilon (v_2 - u)}.
\end{equation}
In the regularisation scheme $u=v_1$ and $v_2=v_3$, which we now enforce, the contribution of this counterterm to the 3-point function is
\begin{equation}
\lla J^{\mu a}(\bs{p}_1) \O^{I_2}(\bs{p}_2) \O^{I_3}(\bs{p}_3) \rra_{\text{ct}} = 2 \ct_1 g (T^a_R)^{I_2 I_3} \2_{JJ} \mu^{2 (v_2 - u)\ep} p_1^{1 + 2 u \epsilon}  \pi^\mu_\alpha(\bs{p}_1)p_2^\alpha.
\end{equation}
Choosing the counterterm constant 
\begin{equation}\label{betactcontr1}
\ct_1 = \frac{\2_{\O\O}}{2 \2_{JJ} (v_2-u) \epsilon} + \ct_1^{(0)} + O(\epsilon),
\end{equation}
where $\ct_1^{(0)}$ is arbitrary, we obtain the finite renormalised 3-point function 
\begin{align}
& A_{1}^{a I_2 I_3} = 
2 g (T_R^a)^{I_2 I_3} \2_{\O\O} \Big[ - p_1 \ln \frac{a_{123}}{\mu} + \frac{b_{123} + p_2^2 + p_3^2}{a_{123}} \Big] + \sd{1}^{a I_2 I_3} p_1.
\end{align}
Here, $\sd{1}^{a I_2 I_3}$ is a scheme-dependent constant that can be expressed in terms of subleading quantities in the regulated theory,
\begin{align}
\sd{1}^{a I_2 I_3} & =  \left( \frac{\pi}{2} \right)^{\frac{3}{2}} \frac{\3{1}^{(1)aI_2I_3}}{2(v_2-u)} + 4 \2_{JJ} g (T_R^a)^{I_2I_3} \ct_1^{(0)} \nn\\
& \qquad + \: \2_{\O\O} g (T_R^a)^{I_2I_3} \frac{u (\gamma_E - \ln 2) + 2 v_2 (-1 + \ln 2)}{v_2-u}.
\end{align}
This relationship is not meaningful in the renormalised theory, however, where only the constant $\sd{1}^{a I_2 I_3}$ appears.  The value of this constant can be arbitrarily shifted by a change of the renormalisation scale $\mu$.

\paragraph{Anomalous CWI.}

The $\mu$-dependence introduced by the counterterm ensures the renormalised form factor obeys the anomalous dilatation Ward identity
\begin{equation}\label{JOOanomDWIex1}
\mu \frac{\partial}{\partial \mu} A_{1}^{a I_2 I_3} = 2\2_{\O\O} g (T_R^a)^{I_2 I_3}  p_1.
\end{equation}
The primary CWI are non-anomalous and retain their original homogeneous form \eqref{primaryCWIJOO}.  The secondary Ward identity \eqref{e:JOOsec}, on the other hand, becomes anomalous for the renormalised form factor and reads
\begin{align}\label{JOOsecCWIanom}
\Lo_{1} A_1^{a I_2 I_3} & = 
2 g (T_R^a)^{I_2 I_3} \2_{\O\O} (-p_2^3 + p_3^3) - 2 g (T_R^a)^{I_2 I_3} \2_{\O\O} p_1^3.
\end{align}

As we noted in the introduction, the form of the anomalous dilatation Ward identity \eqref{JOOanomDWIex1}
can easily be understood from \eqref{dmuW}. 
Due to the $(+--)$ counterterm \eqref{e:JOO_Sct}, we have the beta function 
\[
\beta_{A^a_\mu}  =-\frac{C_{\O\O}}{C_{JJ}}ig(T^a_R)^{IJ}\phi^I D_\mu\phi^J,
\]
as can be seen by inserting \eqref{betactcontr1} into   \eqref{betafnA}.
Without $(---)$ counterterms, the anomaly vanishes and from \eqref{dmuW} we find
\begin{align}
\mu\frac{\p}{\p\mu}\<J^{\mu a}(\x_1)\O^{I_2}(\x_2)\O^{I_3}(\x_3)\> = \int\D^3\x\frac{\delta^2\beta_{A^b_\nu}(\x)}{\delta \phi^{I_2}(\x_2)\delta\phi^{I_3}(\x_3)}\Big|_0\<J^{\nu b}(\x)J^{\mu a}(\x_1)\>.
\end{align}
In momentum space, this reads
\[
\mu\frac{\p}{\p\mu}\lla J^{\mu a}(\bs{p}_1)\O^{I_2}(\bs{p}_2)\O^{I_3}(\bs{p}_3)\rra=\frac{C_{\O\O}}{C_{JJ}}2g (T_R^b)^{I_2I_3}p_2^\nu\lla J^{\nu b}(\bs{p}_1)J^{\mu a}(-\bs{p}_1)\rra.
\]
Substituting for the 2-point function and decomposing into form factors using \eqref{JOOdecomp} and \eqref{JOOformfactors}, we  recover precisely \eqref{JOOanomDWIex1}.
The form of the anomalous secondary CWI \eqref{JOOsecCWIanom} can similarly be understood through an analysis analogous to that presented in section \ref{sec:AnomalousCWIs}.

\subsubsection{\texorpdfstring{$d = 4$ and $\Delta_2 = \Delta_3 = 2$}{d=4 and Delta=2}}

Here, the solution to the primary Ward identities reads
\begin{equation}
A_1^{a I_2 I_3} = \3{1}^{a I_2 I_3} I_{2 \{100\}},
\end{equation}
where the triple-$K$ integral on the right-hand side is finite.  Substituting this expression into the secondary Ward identity \eqref{e:JOOsec}, the left-hand side is\footnote{Here, we use (3.12) and (4.19) of \cite{Bzowski:2015yxv} to write $I_{2\{100\}} = -p_1 \partial_1 I_{1\{000\}}$, and the latter is given in \eqref{Ireduce}.}
\begin{equation} \label{e:secL1_JOO}
\Lo_{1} A_1^{a I_2 I_3} = \3{1}^{a I_2 I_3} ( \ln p_3^2 - \ln p_2^2 ).
\end{equation}
Evaluating the right-hand side using the transverse Ward identity \eqref{e:pJOO}, we obtain however a pair of divergent scalar 2-point functions. 
Passing to the regulated theory, we can remove these divergences with the first counterterm in \eqref{e:SctOO}. 
Since the left-hand side of the secondary Ward identity  is finite,  scheme-dependent corrections to \eqref{e:secL1_JOO} can only 
appear at order $\ep$
in the regulated theory.
This means all finite scheme-dependent terms on the right-hand side of the secondary Ward identity, due to the counterterm \eqref{e:SctOO}, must cancel in order to satisfy the Ward identity.
This required cancellation occurs only in the scheme $u = v_1$ and $v_2 = v_3$.  Once again then, we see that the secondary Ward identity forces us to use the appropriate regularisation scheme preserving current conservation and permutation symmetry of the scalar operators.   In this scheme, we then find
\begin{equation}\label{CsolnJOOd4delta2}
\3{1}^{a I_2 I_3} = 4 g (T_R^a)^{I_2 I_3} \2_{\O\O},
\end{equation}
leading to the renormalised form factor 
\begin{align}
A_1^{a I_2 I_3} & = 4 g (T_R^a)^{I_2 I_3} \2_{\O\O} I_{2 \{100\}} 
= - 4 g (T_R^a)^{I_2 I_3} \2_{\O\O} p_1 \frac{\partial}{\partial p_1} I_{1 \{000\}},
\end{align}
where the right-hand side can be evaluated using  \eqref{Ireduce}.
As no singularities arise in the 3-point function itself, the dilatation and primary CWI are non-anomalous.  
The secondary CWI \eqref{e:secL1_JOO} is anomalous only due to the singularities in the scalar 2-point function as we saw above.

\subsubsection{\texorpdfstring{$d = 4$ and $\Delta_2 = \Delta_3 = 4$}{d=4 and Delta=4}}
\label{sec:JOOd4analysis}

In the regulated theory, the primary CWIs are solved by
\begin{equation} \label{e:solJOO4}
A_1^{a I_2 I_3} = \3{1}^{a I_2 I_3} I_{2 \{122\}},
\end{equation}
where the triple-$K$ integral on the right-hand side exhibits a double pole in the regulator.  
Using the method given in \cite{Bzowski:2015yxv}, in the  regularisation scheme $u = v_1$ and $v_2 = v_3$, we find
\begin{align} \label{e:I2122div}
I_{2\{122\}} & = \frac{p_1^2}{2 v_2 (u - v_2) \: \epsilon^2} + \frac{1}{2 v_2 \epsilon} \Big[ \frac{v_2}{u - v_2} p_1^2 \ln p_1^2 
+ \Big(\frac{v_2 + u (\ln 2 - \gamma_E)}{u - v_2}\Big) p_1^2 + p_2^2 + p_3^2 \Big] + O(\epsilon^0).
\end{align}
The scalar 2-point function, on the other hand,  has a single pole as given in \eqref{e:2ptOOreg}.  Evaluating our solution \eqref{e:sec1JOO} of the secondary CWI with $u = v_1$ and $v_2 = v_3$, we find 
\begin{equation} \label{e:solJOO}
\3{1}^{a I_2 I_3} = 4 g (T_R^a)^{I_2 I_3} \left[ \2_{\O\O} + \epsilon \left( \2_{\O\O} u (\gamma_E - \ln 2) + v_2 \sl_{\O\O} \right) + O(\epsilon^2)\right].
\end{equation}
Since the triple-$K$ integral has a double pole, in principle we also need to work out the $\ep^2$ term here.  In practice, however, this term will only generate a  scheme-dependent contribution,  as the double pole it multiplies is ultralocal, so we can avoid evaluating it explicitly.

Up to cubic order in the sources, the relevant counterterms are
\begin{align}\label{JOO_Sct0}
S_{\mathrm{ct}} &= \int\D^{4+2u\ep}\x\,\Big[\mathfrak{c}_{JJ}\mu^{2u\ep}F^{\mu\nu a}F^a_{\mu\nu}+\mathfrak{c}_{\O\O} \mu^{2 v_2\ep}(D^2\phi^I)^2+
\mathfrak{c}_1 \mu^{2(v_2-u)\ep}ig(T^a_R)^{IJ} J^{\mu a}\phi^{I} D_\mu \phi^{J}\nn\\&\qquad\qquad\quad\qquad
+\mathfrak{c}_2 \mu^{2v_2\ep}ig(T^a_R)^{IJ}F^{\mu\nu a}D_\mu\phi^{I}D_\nu \phi^{J}\Big].
\end{align}
Here, the counterterms proportional to $\mathfrak{c}_{JJ}$ and $\mathfrak{c}_{\O\O}$ are responsible for renormalising the 2-point functions. Their coefficients must therefore satisfy  \eqref{e:ctOO}\,-\,\eqref{e:ctJJ}, namely
\[\label{apple}
\mathfrak{c}_{JJ} = \frac{C_{JJ}}{4u\ep}+\mathfrak{c}_{JJ}^{(0)}+O(\ep), \qquad
\mathfrak{c}_{\O\O} =\frac{C_{\O\O}}{2v_2\ep}+\mathfrak{c}_{\O\O}^{(0)}+O(\ep).
\]
Although the $\mathfrak{c}_{JJ}$ counterterm does not contribute to the 3-point function directly, its presence
is necessary to ensure the Weyl covariance of the cubic counterterm action.
As we show in appendix \ref{Weyl_appendix}, this imposes the relation
\[
(v_2-u)\ep \mathfrak{c}_2+2(1+v_2\ep)\mathfrak{c}_{\O\O}-4(v_2-u)\ep\mathfrak{c}_1\mathfrak{c}_{JJ}= 0
\]
which will play an important role in our understanding of the anomalous Ward identities. 

The counterterms proportional to $\mathfrak{c}_{\O\O}$, $\mathfrak{c}_1$ and $\mathfrak{c}_2$ lead to the 3-point contribution
\begin{align} \label{e:JOOct_contrib}
&\lla J^{\mu a}(\bs{p}_1) \O^{I_2}(\bs{p}_2) \O^{I_3}(\bs{p}_3) \rra_{\text{ct}} \nn\\[1ex]& \qquad = 2 g (T^a_R)^{I_2 I_3} \mu^{2 \epsilon v_2} \Big[ - \ct_2  + \ct_1\, \Big( \frac{\2_{JJ}}{u \epsilon} + \sl_{JJ}+O(\epsilon) \Big) \Big( \frac{p_1}{\mu} \Big)^{2 u \epsilon} \Big] p_1^2 p_2^{\alpha} \pi^{\mu}_{\alpha}(\bs{p}_1) \nn\\[1ex]
& \qquad\quad -  2g (T_R^a)^{I_2 I_3} \mu^{2 \epsilon v_2} \Big( \frac{\2_{\O\O}}{2v_2 \epsilon} + \ct_{\O\O}^{(0)} +O(\epsilon) \Big) (p_2^2 + p_3^2) (p_1^\mu + 2 p_2^\mu).
\end{align}
Notice here that the contribution from the $\ct_{\O\O}$ counterterm is not transverse:
\begin{equation}
p_{1\mu} \lla J^{\mu a}(\bs{p}_1) \O^{I_2}(\bs{p}_2) \O^{I_3}(\bs{p}_3) \rra_{\text{ct}} = \Big( \frac{\2_{\O\O}}{v_2 \epsilon} + 2 \ct_{\O\O}^{(0)}  + O(\epsilon)\Big) g (T_R^a)^{I_2 I_3} \mu^{2 \epsilon v_2} (p_3^4 - p_2^4).
\end{equation}
This behaviour can be understood from the  transverse Ward identity \eqref{e:pJOO}, which holds both in the regulated and in the renormalised theory. 
Since the $\ct_{\O\O}$ counterterm contributes to the right-hand side of this identity through the renormalisation of the 2-point function, it must supply an equal contribution to the left-hand side as we see above.

The transverse part of the counterterm contribution \eqref{e:JOOct_contrib} 
must now cancel the divergences of the regulated form factor ({\it i.e.}, the triple-$K$ integral \eqref{e:I2122div} multiplied by the primary constant \eqref{e:solJOO}).  This requires the counterterm coefficients
\begin{align}\label{c0soln00}
\ct_1 & = \frac{\2_{\O\O}}{(v_2 - u) \2_{JJ}\ep}+\mathfrak{c}_1^{(0)}+O(\ep), \\[1ex]
\label{c2soln00}
\ct_2 & = \frac{\2_{\O\O}}{u v_2 \ep^2}
+\frac{1}{\ep}\,\Big[ \frac{\2_{JJ} \ct_1^{(0)}}{u} + \frac{1}{v_2 - u} \Big( \frac{\2_{\O\O} \2_{JJ}^{(0)}}{\2_{JJ}} - \2_{\O\O}^{(0)} - \2_{\O\O} \Big)\Big]+O(\ep^0),
\end{align}
where the leading term in $\mathfrak{c}_1$ is fixed by cancelling the singularities proportional to $p_1^2\ln p_1^2$. 
Inserting these coefficients into the relation \eqref{apple}, Weyl covariance then constrains the finite part of the $\mathfrak{c}_{JJ}$ counterterm so that 
\[\label{Dreln}
\frac{D_{JJ}}{C_{JJ}}=\frac{D_{\O\O}}{C_{\O\O}}.
\]
Here,  $D_{\O\O}$ and $D_{JJ}$ are the scheme-dependent coefficients appearing in the renormalised 2-point functions, as defined in \eqref{sdtOO} and \eqref{sdtJJ}.

Using the reduction scheme for triple-$K$ integrals given in \cite{Bzowski:2015yxv}\footnote{Specifically, $I_{2\{122\}}$ can be related to $I_{0\{111\}}$ as given in Table 1 of \cite{Bzowski:2015yxv}.  Using  (4.2), (4.6), (4.15) and (4.19) of \cite{Bzowski:2015yxv}, we can then re-write the non-local part of this integral as
$I_{0\{111\}}^{\mathrm{(non-local)}}= (1/4)J^2 I_{1\{000\}}$.}, we can now evaluate the renormalised form factor yielding
\begin{align}\label{e:A1resultJOO}
A_1^{a I_2 I_3} & = g (T_R^a)^{I_2 I_3} \2_{\O\O} \Big[  \Big(2 - p_2 \frac{\partial}{\partial p_2} \Big) \Big(2 - p_3 \frac{\partial}{\partial p_3} \Big) \Big(J^2 I_{1 \{000\}} \Big)  \nn\\
& \qquad \qquad\qquad\qquad - \: p_1^2 \Big( \ln \frac{p_1^2}{\mu^2} \ln \frac{p_2^2}{\mu^2} + \ln \frac{p_1^2}{\mu^2} \ln \frac{p_3^2}{\mu^2} - \ln \frac{p_2^2}{\mu^2} \ln \frac{p_3^2}{\mu^2} \Big) \nn\\
& \qquad\qquad\qquad\qquad + \: p_2^2 \ln \frac{p_1^2 p_3^2}{\mu^4} + p_3^2 \ln \frac{p_1^2 p_2^2}{\mu^4} - p_1^2 \ln \frac{p_2^2 p_3^2}{\mu^4}  \Big] \nn\\
& \quad + 2 g (T_R^a)^{I_2 I_3} (a_0+C_{\O\O}-D_{\O\O}) p_1^2 \ln \frac{p_1^2}{\mu^2} \nn\\
& \quad - g (T_R^a)^{I_2 I_3} (\2_{\O\O} - 2 \sdt_{\O\O}) (p_2^2 + p_3^2)  + \sd{1}^{a I_2 I_3} p_1^2.
\end{align}
Besides $\sdt_{\O\O}$, 
this result contains two additional scheme-dependent constants, $\sd{1}^{a I_2 I_3}$ and $a_0$.  
The first   
multiplies a $p_1^2$ term that can always be added with  arbitrary coefficient as it satisfies the homogeneous conformal Ward identities. 
The second 
is related to the data of the regulated theory  by 
\begin{align}
\label{adefn00}
a_0 &= 
C_{JJ}\mathfrak{c}_1^{(0)} + D_{\O\O}+\frac{1}{v_2-u}\Big[\frac{u C_{\O\O}C_{JJ}^{(0)}}{C_{JJ}}-v_2(C_{\O\O}^{(0)}+C_{\O\O})\Big],
\end{align}
and can consistently be set to zero through an appropriate choice of  $\mathfrak{c}_1^{(0)}$. In particular, this choice is preserved under a change of renormalisation scale, since rescaling $\mu^2\rightarrow e^\lambda\mu^2$ 
shifts $D_{\O\O}\rightarrow D_{\O\O}-\lambda C_{\O\O}$ 
and $D_1^{aI_2I_3}\rightarrow D_1^{aI_2I_3}- \big(2\lambda(a_0-D_{\O\O})+\lambda^2 C_{\O\O}\big)g (T^a_R)^{I_2I_3}$, but leaves $a_0$ invariant.

From the perspective of the renormalised theory,  $a_0$ appears as a (scheme-dependent) coefficient in the anomaly action \eqref{anomalyactionfor4d}. 
Indeed,  our earlier expression \eqref{a0firstreln} matches \eqref{adefn00} after plugging in the relevant counterterm coefficients \eqref{e:ctJJ},  \eqref{sdtJJ} and \eqref{c0soln00}\,-\,\eqref{Dreln}.
As its scheme-dependence suggests, however, this term does not represent a genuine anomaly.  
In fact, it is Weyl exact as we will see later in \eqref{Weylexactness}.

%

\paragraph{Anomalous CWI.}

Using the relation \eqref{Dreln}, the anomalous CWI for the renormalised form factor \eqref{e:A1resultJOO} can be written as follows.  First, we have the dilatation Ward identity
\begin{align}\label{DWIA100}
\mu\frac{\p}{\p\mu}A_1^{aI_2I_3} &=
4g(T^a_R)^{I_2I_3}\Big[\frac{C_{\O\O}}{C_{JJ}}\Big(C_{JJ}\ln \frac{p_1^2}{\mu^2}+D_{JJ}\Big)p_1^2
-a_0 p_1^2
-C_{\O\O}(p_2^2+p_3^2)\Big],
\end{align}
then the primary CWIs 
\[\label{pJOOprimaryCWIs00}
K_{23}A_1^{a I_2 I_3}=0,\qquad
K_{12}A_1^{a I_2 I_3}= 8a_0\, g(T_R^a)^{I_2I_3}, 
\]
and the secondary CWI 
\begin{align}\label{pJOOsecondaryCWIsv200}
L_{1}A_1^{a I_2 I_3} 
&= 4g(T^a_R)^{I_2I_3}\Big[
-\frac{C_{\O\O}}{C_{JJ}}\Big(C_{JJ}\ln\frac{p_1^2}{\mu^2}+D_{JJ}\Big)p_1^4 -\Big(C_{\O\O}\ln \frac{p_2^2}{\mu^2}+D_{\O\O}\Big)p_2^4
 \nn\\[1ex]&\qquad
+\Big(C_{\O\O}\ln \frac{p_3^2}{\mu^2}+D_{\O\O}\Big)p_3^4
+2C_{\O\O}p_1^2p_2^2
+a_0\, p_1^2(p_1^2+p_2^2-p_3^2)\Big].
\end{align}
At first sight, the inhomogeneous terms appearing on the right-hand sides of these identities are quite complicated, involving a mix of semilocal terms with momentum dependence matching that of the renormalised 2-point functions, and ultralocal terms related to anomalies.  
As we will show in section \ref{sec:AnomalousCWIs}, however, all these anomalous Ward identities can easily be understood  
from first principles.

In the meantime, the form of the dilatation Ward identity \eqref{DWIA100} can be understood using \eqref{dmuW}.  Since we have both $(+--)$ and $(---)$ counterterms, we have both a beta function and an anomaly as discussed in section \ref{sec:anomaliesandbetafns}. 
From \eqref{betafnA} and \eqref{c0soln00}, the beta function is 
\[
\beta_{A^a_\mu} = -\frac{2C_{\O\O}}{C_{JJ}}\,ig(T^a_R)^{IJ}\phi^I D_\mu\phi^J,
\] 
whereupon \eqref{dmuW} yields 
\begin{align}
&\mu\frac{\p}{\p\mu}\<J^{\mu a}(\x_1)\O^{I_2}(\x_2)\O^{I_3}(\x_3)\> \nn\\
&= \int\D^3\x\left(\frac{\delta^2\beta_{A^b_\nu}(\x)}{\delta \phi^{I_2}(\x_2)\delta\phi^{I_3}(\x_3)}\Big|_0\<J^{\nu b}(\x)J^{\mu a}(\x_1)\>
-\frac{\delta^3\mathcal{A}(\x)}{\delta A_\mu^a(\x_1)\delta\phi^{I_2}(\x_2)\delta\phi^{I_3}(\x_3)}\Big|_0\right).
\end{align}
In momentum space, we then obtain
\begin{align}
\mu\frac{\p}{\p\mu}\lla J^{\mu a}(\bs{p}_1)\O^{I_2}(\bs{p}_2)\O^{I_3}(\bs{p}_3)\rra
&=\frac{4C_{\O\O}}{C_{JJ}}g T^{bI_2I_3}p_{2\nu}\lla J^{\nu b}(\bs{p}_1)J^{\mu a}(-\bs{p}_1)\rra
+\mathcal{A}_{J\O\O}^{a\mu I_2I_3},
\end{align}
where, from the anomaly action \eqref{anomalyactionfor4d},  the anomaly contribution 
\[
(\mathcal{A}_{J\O\O})^{a\mu I_2 I_3}
= -2C_{\O\O} g(T^a_R)^{I_2I_3}(p_2^2+p_3^2)(p_2^\mu-p_3^\mu)-4a_0\, g(T^a_R)^{I_2I_3} p_1^2 \pi^{\mu\nu}(\bs{p}_1)p_{2\nu}.
\]
Inserting the renormalised 2-point function  \eqref{e:2ptJJren} and decomposing into form factors using \eqref{JOOdecomp} and \eqref{JOOformfactors}, we then recover  the anomalous dilatation Ward identity \eqref{DWIA100}.

\subsection{\texorpdfstring{$\<T_{\mu_1\nu_1}\O\O\>$}{<TOO>}}

\subsubsection{General analysis}

\paragraph{Decomposition.} The transverse Ward identity reads
\begin{align} \label{e:pTOO}
&p_1^{\nu_1} \lla T_{\mu_1 \nu_1}(\bs{p}_1) \mathcal{O}^{I_2}(\bs{p}_2) \mathcal{O}^{I_3}(\bs{p}_3) \rra  \nn\\[0.5ex]&\qquad
= p_{3 \mu_1} \lla \mathcal{O}^{I_2}(\bs{p}_3) \mathcal{O}^{I_3}(-\bs{p}_3) \rra + p_{2 \mu_1} \lla \mathcal{O}^{I_2}(\bs{p}_2) \mathcal{O}^{I_3}(-\bs{p}_2) \rra.
\end{align}
Since our renormalisation prescription preserves diffeomorphism invariance, this identity is non-anomalous and takes the same form in both the regulated and the renormalised theory.
Weyl invariance, on the other hand, is not in general preserved in the renormalised theory.  This leads to an anomaly $\mathcal{A}^{I_2 I_3}$ in the trace Ward identity for the renormalised correlator\footnote{Note the trace Ward identity \eqref{e:trTOO} and reconstruction formula \eqref{e:recTOO}  differ from those in \cite{Bzowski:2013sza} since here we define the 3-point function through three functional derivatives, as discussed on p.~15-16 of \cite{Bzowski:2017poo}.}
\begin{align}
 &\lla T(\bs{p}_1) \mathcal{O}^{I_2}(\bs{p}_2) \mathcal{O}^{I_3}(\bs{p}_3) \rra \nn\\[0.5ex]&\qquad = (d- \Delta_3) \lla \mathcal{O}^{I_2}(\bs{p}_2) \mathcal{O}^{I_3}(-\bs{p}_2) \rra +(d-\Delta_2) \lla \mathcal{O}^{I_2}(\bs{p}_3) \mathcal{O}^{I_3}(-\bs{p}_3) \rra   + \mathcal{A}^{I_2 I_3}. \label{e:trTOO}
\end{align}
The specific form of this anomalous contribution  can  be determined on a case-by-case basis.  In fact, only the last of the examples we study here is anomalous and the corresponding anomaly is given in \eqref{AIJ}.

Using these identities, the renormalised 3-point function can be reconstructed from its purely transverse-traceless part according to
\begin{align}\label{e:recTOO}
& \lla T_{\mu_1 \nu_1}(\bs{p}_1) \mathcal{O}^{I_2}(\bs{p}_2) \mathcal{O}^{I_3}(\bs{p}_3) \rra = \lla t_{\mu_1 \nu_1}(\bs{p}_1) \mathcal{O}^{I_2}(\bs{p}_2) \mathcal{O}^{I_3}(\bs{p}_3) \rra \nn\\[1ex]
& \qquad + \Big[\Big( p_2^\alpha \mathscr{T}_{\mu_1 \nu_1\alpha}(\bs{p}_1) + \frac{d-\Delta_3}{d-1} \pi_{\mu_1 \nu_1}(\bs{p}_1) \Big) \lla \mathcal{O}^{I_2}(\bs{p}_2) \mathcal{O}^{I_3}(-\bs{p}_2) \rra + ( 2 \leftrightarrow 3 ) \Big]\nn\\[1ex]
& \qquad + \: \frac{\mathcal{A}^{I_2 I_3}}{d-1} \pi_{\mu_1 \nu_1}(\bs{p}_1),
\end{align}
where
\begin{equation} \label{Iprojdef}
\mathscr{T}_{\mu\nu\alpha} (\bs{p}) = \frac{1}{p^2} \left[ 2 p_{(\mu} \delta_{\nu)\alpha} - \frac{p_\alpha}{d-1} \left( \delta_{\mu\nu} + (d-2) \frac{p_\mu p_\nu}{p^2} \right) \right].
\end{equation}

\paragraph{Form factors.} The tensorial structure of this transverse-traceless part is
\begin{equation}
\lla t_{\mu_1 \nu_1}(\bs{p}_1) \mathcal{O}^{I_2}(\bs{p}_2) \mathcal{O}^{I_3}(\bs{p}_3) \rra =  A_1^{I_2 I_3}\, \Pi_{\mu_1 \nu_1\alpha_1 \beta_1}(\bs{p}_1)  p_2^{\alpha_1} p_2^{\beta_1},
\end{equation}
where $A_1$ is a form factor depending on the momentum magnitudes. This form factor is symmetric under $(p_2, I_2) \leftrightarrow (p_3, I_3)$, \textit{i.e.},
\begin{equation}
A_1^{I_3 I_2}(p_1, p_3, p_2) = A_1^{I_2 I_3}(p_1, p_2, p_3),
\end{equation}
and is related to the full correlator by
\begin{equation}
A_1^{I_2 I_3} = \text{coefficient of } p_{2\mu_1} p_{2\nu_1} \text{ in } \lla T_{\mu_1 \nu_1}(\bs{p}_1) \mathcal{O}^{I_2}(\bs{p}_2) \mathcal{O}^{I_3}(\bs{p}_3) \rra.
\end{equation}
To apply this formula, we must first select the independent momenta according to the cyclic rule \eqref{a:momenta} before extracting the coefficient indicated.

\paragraph{Primary CWIs.} The primary CWIs are
\begin{equation}
\K_{ij} A_1^{I_2 I_3} = 0, \qquad i,j = 1,2,3,
\end{equation}
and their solution in terms of triple-$K$ integrals is
\begin{equation}\label{e:priTOO1}
A_1^{I_2 I_3} = \3{1}^{I_2 I_3} J_{2 \{000\}},
\end{equation}
where $\3{1}^{I_2 I_3}= \3{1}^{I_3 I_2}$ is a constant.  

\paragraph{Secondary CWIs.} The independent secondary Ward identity is
\begin{align}\label{e:secCWITOO}
\Lo_{2} A_1^{I_2 I_3} &= 2 \Delta_1 \cdot \text{coefficient of } p_{2\mu_1} \text{ in } p_{1 }^{\nu_1} \lla T_{\mu_1 \nu_1}(\bs{p}_1) \mathcal{O}^{I_2}(\bs{p}_2) \mathcal{O}^{I_3}(\bs{p}_3) \rra\nn\\[1ex]
&= 2\Delta_1\Big( \lla \mathcal{O}^{I_2}(\bs{p}_2) \mathcal{O}^{I_3}(-\bs{p}_2) \rra- \lla \mathcal{O}^{I_2}(\bs{p}_3) \mathcal{O}^{I_3}(-\bs{p}_3) \rra\Big).
\end{align}
To obtain the second line, we used the  transverse Ward identity \eqref{e:pTOO} with independent momenta as prescribed by  \eqref{a:momenta}.
The constraint imposed by this secondary CWI on the primary constant $C_1^{I_2 I_3}$ can be extracted by analysing the soft limit $p_3\rightarrow 0$, similar to our earlier analysis leading to \eqref{e:sec1JOO}.  Working in the regularised theory where necessary to avoid divergences, we find
\begin{equation} \label{e:sec1TOO}
\3{1}^{I_2 I_3} = \frac{2^{3-d/2} \2_{\O\O}(\ep)  \sin \big( \pi (d/2 -  \Delta_2) \big)}{\pi \Gamma(d/2)}\delta_{\Delta_2, \Delta_3}.
\end{equation}
Once again, 
for non-identical scalars with $\Delta_2 \neq \Delta_3$ the 3-point function vanishes.  
For $\Delta_2=d/2+n$, the 3-point function is  non-vanishing since the zero in the sine function cancels against the pole in the 2-point normalisation $C_{\O\O}(\ep)$, see \eqref{2ptdivcases} and \eqref{2ptnormexp}.

\paragraph{Regularisation.}

From now on, in order to obtain a non-vanishing 3-point function, we will restrict to the case of a single scalar  with $\Delta_2 = \Delta_3$.  This condition requires us to work in a  scheme with $v_2 = v_3$.
In addition, conservation of the stress tensor requires $u = v_1$.  All   singularities are then regulated provided  $u\neq v_2$.  
The cases where singularities occur are summarised in the following table: 
\begin{center}
\begin{tabular}{|c|c|c|c|c|} \hline
Form factor & Integral & $(---)$ & $(+--)$ & $(-++)$ \\ \hline
$A_1$ & $J_{2\{000\}}$ & $\Delta = d/2+1+n $ & $\Delta = d +1+ n$ & $\Delta=d/2-1-n$ \\ \hline
\end{tabular}
\captionof{table}{Singularities arising for $\< T_{\mu_1\nu_2} \O^{I_2} \O^{I_3} \>$, with $\Delta_2=\Delta_3=\Delta$. 
\label{tab:TOO}}
\end{center}
Singularities of types $(++-)$ and $(+++)$ are forbidden as the former requires $d< 0$ while the latter requires $\Delta_2=\Delta_3 < 0$.  For a unitary theory, the $(-++)$ singularity only occurs for $\Delta_2=\Delta_3=d/2-1$ corresponding to a free theory. 

\paragraph{Renormalisation.} 
As discussed previously for $\<J^\mu\O\O\>$, counterterms can be classified by the type of singularity they remove.  Those removing   $(---)$ singularities give rise to anomalies and are cubic in the sources, while those removing $(+--)$ singularities give rise to beta functions and involve two sources and one operator.
As we will see below, counterterms can only be constructed for the cases listed in the following table: 

\begin{center}
\begin{tabular}{|c|c|}
\hline
Singularity type & Counterterms available when  \\ \hline
$(---)$ & $\Delta_2=\Delta_3 = d/2+1+n$ \\ \hline
$(+--)$ & $\Delta_2=\Delta_3 = d + 1+n$ \\ \hline
\end{tabular}
\captionof{table}{Availability of counterterms for $\< T_{\mu_1\nu_1} \O^{I_2} \O^{I_3} \>$.\label{tab:physTOO}}
\end{center}
Comparing with table \ref{tab:TOO}, we see that for $(---)$ and $(+--)$ singularities counterterms are always available.
 For cases with a $(-++)$ singularity, the primary constant must instead vanish as a suitable power of the regulator.  The resulting finite form factor is then fully nonlocal \cite{Bzowski:2015pba}.
 
\paragraph{Anomalies.} Our analysis of anomalies here is similar to that for $\< J^{\mu_1} \O \O \>$.  The counterterms must be Lorentz-invariant, and hence the metric tensor can appear only through the Riemann tensor or covariant derivatives. Moreover, since we functionally differentiate with respect to the source $g^{\mu\nu}$ only once before returning to a flat metric, the Riemann tensor cannot appear more than once. There are therefore two possible types of counterterms, the simplest examples of which are
\begin{equation}
\int \D^{d} \bs{x} \sqrt{g} \nabla_\mu \phi^I \nabla^\mu \phi^I, \qquad\qquad \int \D^{d} \bs{x} \sqrt{g} R^{\mu \nu} \nabla_\mu \phi^I \nabla_\nu \phi^I.
\end{equation}
We can also add an even number of additional covariant derivatives.  The existence of these  counterterms leads to anomalies whenever
\begin{equation} \label{e:TOO_anomaly_cond}
\Delta_2 = \Delta_3 = \frac{d}{2} + 1 + n, \quad n=0,1,2,\ldots
\end{equation}

\paragraph{Beta functions.} 
Beta functions derive from counterterms containing one operator and two sources. 
Here, 
the only such counterterms  
contributing to the 3-point function at hand are those containing the stress tensor and two scalar sources.  
This can be seen through an analysis similar to that discussed earlier for  $\< J^{\mu} \O \O \>$.\footnote{Note that counterterms containing two scalar sources and a scalar operator, while generating a beta function for
$\phi^I$, do not contribute to $\< T_{\mu\nu} \O \O \>$ and so are not relevant to our present discussion.}
The simplest terms of this form are
\begin{equation} \label{e:formctTOO}
\int \D^{d} \bs{x} \sqrt{g}\, T \phi^I \phi^I, \qquad\qquad \int \D^{d} \bs{x} \sqrt{g}\, T^{\mu \nu} \nabla_{\mu} \phi^I \nabla_\nu \phi^I.
\end{equation}
More complicated examples can be constructed by adding an even number of covariant derivatives.
The contribution to the 3-point function  from the first counterterm in \eqref{e:formctTOO} (and its analogues with additional covariant derivatives) is proportional to $\< T T_{\mu\nu} \>$, however, which vanishes by scale invariance.  (The 2-point function has no trace anomaly above two dimensions.) From the second class of counterterms, we expect nontrivial beta functions to appear whenever
\begin{equation} \label{e:condTOO}
\Delta_2 = \Delta_3 = d +1 + n, \quad n=0,1,2,\ldots
\end{equation}

Comparing  \eqref{e:TOO_anomaly_cond} and \eqref{e:condTOO} with the corresponding formulae  \eqref{e:JOO_anomaly_cond} and \eqref{e:condJOO} for $\< J^\mu \O \O \>$, 
we see the conditions for anomalies in the 3-point functions are identical while those to obtain a beta function differ by a unit shift in dimensions. In particular, for a marginal scalar with $\Delta_2 = \Delta_3 = d$, the 3-point function $\< J^\mu \O \O \>$ exhibits a nontrivial beta function
while $\< T_{\mu\nu} \O \O \>$ does not.  For a marginal scalar, both 3-point functions contain anomalies if and only if $d$ is even.

\subsubsection{\texorpdfstring{$d=3$ and $\Delta_2=\Delta_3 = 1$}{d=3 and Delta=1}}

The direct evaluation of \eqref{e:sec1TOO} leads to
\begin{equation}
A_1^{I_2 I_3} = 2 \2_{\O\O} \delta^{I_2 I_3} \frac{(p_1 + a_{123})}{b_{23} a_{123}^2},
\end{equation}
where the symmetric polynomials appearing are defined in \eqref{e:variables}.

\subsubsection{\texorpdfstring{$d=3$ and $\Delta_2=\Delta_3 = 3$}{d=3 and Delta=3}}

In contrast to the case of $\< J^{\mu}\O\O\>$ with $d=3$ and $\Delta_2=\Delta_3 = 3$, here the correlator is represented by a convergent triple-$K$ integral. This is consistent  with \eqref{e:formctTOO}, from which we see that no  counterterms are available.  Evaluating the triple-$K$ integral, we find
\begin{equation} \label{res:TOOd3}
A_1^{I_2 I_3} =  - 2 \2_{\O\O} \delta^{I_2 I_3} \frac{(a_{123}^3 - a_{123} b_{123} - c_{123})}{a_{123}^2}.
\end{equation}

\subsubsection{\texorpdfstring{$d=4$ and $\Delta_2=\Delta_3 = 2$}{d=4 and Delta=2}}

This case is almost identical to the corresponding case for  $\< J^{\mu}\O\O\>$. The triple-$K$ integral is finite and the solution reads
\begin{align}
A_1^{I_2 I_3} & = - 2 \2_{\O\O} \delta^{I_2 I_3} I_{3\{200\}} \nn\\
& = - 2 \2_{\O\O} \delta^{I_2 I_3} p_1 \left( p_1 \frac{\partial}{\partial p_1} - 1 \right) \frac{\partial}{\partial p_1} I_{1 \{000\}},
\end{align}
where $I_{1 \{000\}}$ is given in \eqref{I1000}.
The scalar 2-point function is singular, but the combination appearing on the right-hand side of the secondary Ward identity \eqref{e:secCWITOO} 
is finite and unambiguous.

\subsubsection{\texorpdfstring{$d=4$ and $\Delta_2=\Delta_3 = 4$}{d=4 and Delta=4
}}\label{TOOPaneitzsec}

Expanding the solution \eqref{e:sec1TOO} of the secondary CWI, working in the regulated theory with $C_{\O\O}(\ep)$ as given in \eqref{2ptnormexp}, we find 
\begin{equation} 
\3{1}^{I_2 I_3} = - 2 \2_{\O\O} + 2 \epsilon \left[ u \2_{\O\O} (1 - \gamma_E + \ln 2) - v_2 \sl_{\O\O} \right] + O(\epsilon^2).
\end{equation}
The regulated form factor then reads 
\begin{align}\label{e:solTOO}
A_1^{I_2 I_3} & = \3{1}^{I_2 I_3} I_{3 + u \epsilon \{2 + u \epsilon, 2 + v_2 \epsilon, 2 + v_2 \epsilon\}}  = - \frac{2 \2_{\O\O}}{v_2 \epsilon} (p_1^2 + p_2^2 + p_3^2) + O(\ep^0),
\end{align}
requiring us to choose a scheme with $v_2\neq 0$.

At our disposal we have four linearly independent counterterms built from $g_{\mu\nu}$ and  $\phi^I$ that yield a non-vanishing contribution to $\< T_{\mu \nu} \O \O \>$, namely
\begin{align} \label{e:TOOcts}
& \ct_{\O\O} \int \D^{4+2 u \epsilon} \bs{x} \sqrt{g} ( \Box \phi^I )^2 \mu^{2 v_2 \epsilon}, && \ct_2 \int \D^{4+2 u \epsilon} \bs{x} \sqrt{g} R_{\mu\nu} \nabla^\mu \phi^I \nabla^\nu \phi^I \mu^{2 v_2 \epsilon}, \nn\\
& \ct_3 \int \D^{4+2 u \epsilon} \bs{x} \sqrt{g} R \nabla_\mu \phi^I \nabla^\mu \phi^I \mu^{2 v_2 \epsilon}, && \ct_4 \int \D^{4+2 u \epsilon} \bs{x} \sqrt{g} R \phi^I \Box \phi^I \mu^{2 v_2 \epsilon}.
\end{align}
The coefficient $\ct_{\O\O}$ of the first counterterm is already fixed by the renormalisation of the scalar 2-point function, as given in \eqref{e:ctOO}. The remaining counterterms only contribute to 3- and higher-point functions.  In fact, only the first two counterterms contribute to the transverse-traceless part of the 3-point function. Their contribution to the form factor $A_1^{I_2 I_3}$ is
\begin{equation}
A_{1 \ \text{ct}}^{I_2 I_3} = 4 \ct_{\O\O} (p_2^2 + p_3^2) \mu^{2 v_2 \epsilon} - 2 \ct_2 p_1^2 \mu^{2 v_2 \epsilon}.
\end{equation}
Adding this  contribution to  \eqref{e:solTOO}, the divergence proportional to $p_2^2 + p_3^2$ cancels while to eliminate the divergence proportional to $p_1^2$ requires $\ct_2 = - \2_{\O\O} / (v_2 \epsilon) + \ct_2^{(0)} + O(\epsilon)$.

At this stage the resulting 3-point function is finite and the solution depends on the two undetermined constants $\ct_{\O\O}^{(0)}$ and $\ct_2^{(0)}$.  In addition to cancelling the divergences, however, the counterterm action should also be Weyl covariant as given in \eqref{fullWeylcovcond}. As here there are no beta functions, this condition simplifies to
\[\label{Weylcovonlysources}
\delta_\sigma S_{\mathrm{ct}} = \int\D^d\x\sqrt{g}\sigma\mu\frac{\p}{\p\mu}\mathcal{L}_{\mathrm{ct}}.
\]
Evaluating the Weyl-covariant completion of the first counterterm in \eqref{e:TOOcts}, we find\footnote{See \eqref{Scov} to \eqref{finaltransf} of appendix \ref{Weyl_appendix}. 
Note that when $v_2=0$ the counterterm action $S_{\mathrm{ct}}$ has no dependence on the RG scale $\mu$ and hence it  is Weyl invariant from \eqref{Weylcovonlysources}.  It can then be re-expressed in terms of the Paneitz operator as given in \eqref{ctPanaction}. We emphasise however that the  scheme $v_2=0$ is not admissible since the divergences depend on $1/v_2$. }  
\begin{align}\label{SctcovTOO}
&S_{\mathrm{ct}} 
=\mathfrak{c}_{\O\O}\int\D^{4+2u\ep}\x\sqrt{g}\mu^{2v_2\ep}\Big[(\Box\phi^I)^2-\frac{2(1+v_2\ep)}{(1+u\ep)}R^{\mu\nu}\p_\mu\phi^I \p_\nu\phi^I
+\frac{(v_2-u)\ep}{(3+2u\ep)}R\phi^I \Box\phi^I
\nn\\[1ex] &
+\frac{(2+u\ep)(1+v_2\ep)}{(3+2u\ep)(1+u\ep)}R(\p \phi^I)^2
+\frac{(v_2-u)^2\ep^2}{4(3+2u\ep)^2}R^2(\phi^I)^2-\frac{(v_2-u)\ep(1+v_2\ep)}{4(1+u\ep)(1+2u\ep)}E_4(\phi^I)^2\Big],
\end{align}
where $E_4$ is the four-dimensional Euler density.
 Weyl covariance thus imposes specific relations between the coefficients of the various counterterms listed in \eqref{e:TOOcts}.
In particular, 
\begin{equation}
\ct_2 = \big(-2-2(v_2-u)\ep+O(\ep^2)\big)\ct_{\O\O},
\end{equation}
and hence from \eqref{e:ctOO} and \eqref{sdtOO},
\[
\ct_2 = -\frac{C_{\O\O}}{v_2\ep}+\Big(-\frac{(v_2-u)}{v_2}\2_{\O\O}-\2_{\O\O}^{(0)}+\sdt_{\O\O}\Big)+O(\ep).
\]
The renormalised 3-point function is then given by\footnote{Here, we used  the reduction scheme in Table 1 of \cite{Bzowski:2015yxv} to relate the  triple-$K$ integral $I_{3\{222\}}$  to the integral $I_{0\{111\}}$.  The non-local part of the latter  can then be re-expressed as  $I_{0\{111\}}^{\mathrm{(non-local)}}= (1/4)J^2 I_{1\{000\}}$ using equations (4.2), (4.6), (4.15) and (4.19) of \cite{Bzowski:2015yxv}.}
\begin{align} \label{A1TOO}
A_1^{I_2 I_3} & = - \2_{\O\O} \delta^{I_2 I_3} \Big[ 2 I_{3\{222\}}^{\mathrm{(fin)}} 
+ \: (p_1^2 + p_2^2) \ln \frac{p_3^2}{\mu^2} + (p_2^2 + p_3^2) \ln \frac{p_1^2}{\mu^2} + (p_1^2 + p_3^2) \ln \frac{p_2^2}{\mu^2} \Big] \nn\\[1ex]
& \qquad + 2 (\2_{\O\O}-\sdt_{\O\O}) \delta^{I_2 I_3} (p_1^2 + p_2^2 + p_3^2),
\end{align}
where the finite integral $I_{3\{222\}}^{\mathrm{(fin)}}$ is given in \eqref{I3222fin}.
As we see, the only surviving scheme-dependent constant, $\sdt_{\O\O}$, is that already appearing in the 2-point function. 
This makes sense since the counterterm action is fully fixed by the renormalisation of the 2-point function plus Weyl covariance, so the 3-point function cannot involve any new scheme-dependent constants.
A change of renormalisation scale $\mu^2\rightarrow e^{\lambda}\mu^2$ is equivalent to shifting $D_{\O\O}\rightarrow D_{\O\O}-\lambda C_{\O\O}$, both for the 2- and the 3-point function.

\paragraph{Anomalous CWI.}

The renormalised form factor \eqref{A1TOO} satisfies the following anomalous Ward identities:
\begin{align}\label{anomDWITOO}
\mu \frac{\partial}{\partial \mu} A_1^{I_2 I_3} & = 4 \2_{\O\O} \delta^{I_2 I_3} ( p_1^2 + p_2^2 + p_3^2), \\[1ex]
\K_{ij} A_1^{I_2 I_3} & = 0, \\[1ex]
\Lo_{2} A_1^{I_2 I_3} 
& = 8 \Big[ \2_{\O\O} \Big( p_2^4 \ln \frac{p_2^2}{\mu^2} - p_3^4 \ln \frac{p_3^2}{\mu^2} \Big) + \sdt_{\O\O} (p_2^4 - p_3^4) \Big] 
\nn\\[1ex]& \qquad 
- 4 \2_{\O\O} \delta^{I_2 I_3} p_1^2 (p_1^2 + 2 p_2^2).
\end{align}
In addition, we need to determine the anomaly $\mathcal{A}^{I_2 I_3}$ entering the trace Ward identity \eqref{e:trTOO} and the reconstruction formula \eqref{e:recTOO}.
As only $(---)$ counterterms are present, using 
\eqref{anomformulaonlysources} the anomaly action is
\[\label{Panaction}
A =\int\D^4\x\sqrt{g}\,\mathcal{A}= -C_{\O\O} \int\D^4\x\sqrt{g}\,\phi^I\Delta_4\phi^I,
\]
where the four-dimensional Paneitz operator  
\begin{align}
\Delta_4\phi^I &= \nabla_\mu\Big( (\nabla^\mu\nabla^\nu+2R^{\mu\nu}-\frac{2}{3}Rg^{\mu\nu})\nabla_\nu\phi^I\Big)\nn\\
&=\Box^2\phi^I+\frac{1}{3}\nabla_\mu R\nabla^\mu\phi^I+2R^{\mu\nu}\nabla_\mu\nabla_\nu\phi^I-\frac{2}{3}R\Box \phi^I.
\end{align}
Written in this form, the anomaly action \eqref{Panaction} is manifestly Weyl invariant.
The anomaly $\mathcal{A}^{I_2 I_3}$ then follows by restoring a flat metric and evaluating
\[
\mathcal{A}^{I_2 I_3} =\frac{\delta^2\mathcal{A}(\x_1)}{\delta\phi^{I_2}(\x_2)\delta\phi^{I_3}(\x_3)}\Big|_0.
\]
In momentum space, we thus obtain
\[\label{AIJ}
\mathcal{A}^{I_2 I_3} = -\2_{\O\O}\delta^{I_2I_3}p_2^2p_3^2.
\]
We can also cross-check the form of the anomalous dilatation Ward identity. 
From \eqref{dmuW}, this should be
\[
\mu\frac{\p}{\p\mu}\<T_{\mu_1\nu_1}(\x_1)\O^{I_2}(\x_2)\O^{I_3}(\x_3)\> = -2\frac{\delta^3A}{\delta g^{\mu_1\nu_1}(\x_1)\delta\phi^{I_2}(\x_2)\delta\phi^{I_3}(\x_3)}\Big|_0
\]
Evaluating this, converting to momentum space and decomposing into form factors, we indeed recover \eqref{anomDWITOO}.

\subsection{\texorpdfstring{$\<J^{\mu_1} J^{\mu_2}\O\>$}{<JJO>}}

\subsubsection{General analysis}

\paragraph{Decomposition.} The transverse Ward identity is\footnote{This identity differs from that in \cite{Bzowski:2013sza} since here we define the 3-point function through three functional derivatives, see p.~15-16 of \cite{Bzowski:2017poo}.}
\begin{equation}\label{TWI_JJO}
p_{1 \mu_1} \lla J^{\mu_1 a_1}(\bs{p}_1) J^{\mu_2 a_2}(\bs{p}_2)
\mathcal{O}^I(\bs{p}_3) \rra = 0,
\end{equation}
and so the 3-point function is purely transverse:
\begin{align}
& \lla J^{\mu_1 a_1}(\bs{p}_1) J^{\mu_2 a_2}(\bs{p}_2) \mathcal{O}^I(\bs{p}_3)
\rra = \lla j^{\mu_1 a_1}(\bs{p}_1) j^{\mu_2 a_2}(\bs{p}_2)
\mathcal{O}^I(\bs{p}_3) \rra.
\end{align}

\paragraph{Form factors.} We now have the tensor decomposition
\begin{equation}
\lla j^{\mu_1 a_1}(\bs{p}_1) j^{\mu_2 a_2}(\bs{p}_2) \mathcal{O}^I(\bs{p}_3)
\rra = \pi^{\mu_1}_{\alpha_1}(\bs{p}_1) \pi^{\mu_2}_{\alpha_2}(\bs{p}_2) \left[
A_1^{a_1 a_2 I} p_2^{\alpha_1} p_3^{\alpha_2} + A_2^{a_1 a_2 I} \delta^{\alpha_1
\alpha_2} \right]. 
\end{equation}
The form factors $A_1$ and $A_2$ are functions of the momentum magnitudes. Both form factors are symmetric under $(p_1, a_1) \leftrightarrow (p_2, a_2)$, \textit{i.e.}, they satisfy
\begin{equation}
A_j^{a_2 a_1 I}(p_2, p_1, p_3) = A_j^{a_1 a_2 I}(p_1, p_2, p_3), \qquad j = 1,2.
\end{equation}
These form factors can be extracted from  $\lla J^{\mu_1 a_1}(\bs{p}_1) J^{\mu_2 a_2}(\bs{p}_2) \mathcal{O}^I(\bs{p}_3)
\rra$ using
\begin{align}
A_1^{a_1 a_2 I} & = \text{coefficient of } p_2^{\mu_1} p_3^{\mu_2}, \\
A_2^{a_1 a_2 I} & = \text{coefficient of } \delta^{\mu_1 \mu_2}. \label{e:JJO_A2}
\end{align}
As always, before reading off these coefficients we must first write the 3-point in terms of the independent momenta prescribed by our cyclic rule \eqref{a:momenta}.

\paragraph{Primary CWIs.} The primary CWIs are
\begin{equation} \label{e:K_JJO}
\begin{array}{ll}
\K_{12} A_1^{a_1 a_2 I} = 0, & \qquad\qquad \K_{13} A_1^{a_1 a_2 I} = 0, \\
\K_{12} A_2^{a_1 a_2 I} = 0, & \qquad\qquad \K_{13} A_2^{a_1 a_2 I} = 2 A_1^{a_1a_2 I}. 
\end{array}
\end{equation}
Their solution in terms of triple-$K$ integrals is
\begin{align}
A_1^{a_1 a_2 I} & = \3{1}^{a_1 a_2 I} J_{2 \{000\}}, \label{a:JJO1} \\
A_2^{a_1 a_2 I} & = \3{1}^{a_1 a_2 I} J_{1 \{001\}} + \3{2}^{a_1 a_2 I}
J_{0 \{000\}}, \label{a:JJOlast}
\end{align}
where $\3{j}^{a I_2 I_3}$, $j=1,2$ are constants. In particular $\3{j}^{a_2 a_1 I}
= \3{j}^{a_1 a_2 I}$ for $j=1,2$.

\paragraph{Secondary CWIs.} The independent secondary CWI is
\begin{align}
& \Lo_{1} A_1^{a_1 a_2 I} + 2 \Ro A_2^{a_1 a_2 I}  \nn\\
& \qquad = 2 (\Delta_1 - 1) \cdot \text{coefficient of } p^{\mu_2}_3 \text{ in } p_{1
\mu_1} \lla J^{\mu_1 a_1}(\bs{p}_1) J^{\mu_2 a_2}(\bs{p}_2)
\mathcal{O}^I(\bs{p}_3) \rra, \label{e:secCWI_JJO}
\end{align}
where the right-hand side vanishes from \eqref{TWI_JJO}.
The constraint imposed by this secondary CWI on the primary constants $\3{1}^{a I_2 I_3}$ and $\3{2}^{a I_2 I_3}$ can now be obtained from an analysis of the soft limit $p_3\rightarrow 0$, as used earlier to derive \eqref{e:sec1JOO}.  Working in the regulated theory where necessary to avoid divergences, we find 
\begin{equation}\label{secCWI_JJO_constr}
\3{2}^{a_1 a_2 I} = - \frac{1}{2} \Delta_3 (\Delta_3 - d + 2) \3{1}^{a_1 a_2 I}.
\end{equation}

\paragraph{Regularisation.} 
We work in a scheme with $u = v_1 = v_2$ to maintain current conservation, but set $v_3\neq u$.  This scheme is sufficient to regulate all the singularities we encounter. 
These can be classified according to the choice of $\pm$ signs appearing in the singularity condition \eqref{singcond}. For the 3-point function at hand, the occurrence of $(-++)$ or $(+++)$ singularities violates unitarity by requiring either $\Delta_3 \leq 0$ or $d \leq 0$.  The remaining cases are then given in the following table, using  $''$ to indicate repetition of the line above:
\begin{center}
\begin{tabular}{|c|c|c|c|c|c|} \hline
Factor & Integral & $(---)$ & $(+--)$ & $(--+)$ & $(++-)$ \\ \hline
$A_1$ & $J_{2\{000\}}$ & $\Delta_3 = 4+2n$ & $\Delta_3 = d + 2 + 2 n$ & $\Delta_3 = d - 4 - 2 n$ & $\Delta_3 = 2 d + 2 n$ \\ \hline
$A_2$ & $J_{1\{001\}}$ & $\Delta_3 = 2+2n$ & $\Delta_3 = d + 2 n$ & $\Delta_3 = d - 4 - 2 n$  & $\Delta_3 = 2 d - 2 + 2 n$ \\
& $J_{0\{000\}}$ & $''$ & $''$ & $\Delta_3 = d - 2 - 2 n$ & $''$ \\ \hline
\end{tabular}
\captionof{table}{Singularities arising in triple-$K$ integrals for the form factors of $\< J^{\mu_1} J^{\mu_2} \O^I \>$.   
\label{tab:JJO}}
\end{center}

\paragraph{Renormalisation.}
The counterterms available to remove these singularities are strongly constrained by gauge- and Lorentz-invariance.  In fact, as we will discuss below, counterterms only exist for the cases summarised in the following table: 
\begin{center}
\begin{tabular}{|c|c|} \hline
Singularity type & Counterterms available when  \\ \hline
$(---)$ & $\Delta_3 = 4 + 2n$ \\ \hline
$(--+)$ & $\Delta_3 = d - 4 - 2n$ \\ \hline
$(+--)$ & $\Delta_3 = d + 2 + 2n$ \\ \hline
\end{tabular}
\captionof{table}{Availability of counterterms for $\< J^{\mu_1} J^{\mu_2} \O^I \>$.\label{tab:physJJO}}
\end{center}
 In many examples, we then find there are no counterterms available  to remove a certain type of singularity.  When this occurs, there are two possibilities: either the primary constants multiplying these divergent triple-$K$ integrals must vanish as some suitable power of the regulator, or else the singularities of individual triple-$K$ integrals must cancel when summed together to construct the regulated form factor.
Here, such cancellations can occur between the singularities of $J_{1\{001\}}$ and $J_{0\{000\}}$ when summed to produce the form factor $A_2$. 
In either case, the specific values of the primary constants required to effect these vanishings or cancellations are not chosen by hand; rather, these values are uniquely determined by solving the secondary conformal Ward identities.

\paragraph{Anomalies.} 
Divergences of the type $(---)$ are removed by counterterms that are cubic in the sources, and give rise to anomalies.
Here, we require counterterm involving a single scalar source $\phi^I$ and two gauge fields $A_\mu^a$. 
The two simplest such counterterms take the form
\begin{equation} \label{e:ctJJOmmm}
\delta^{ab} r^{abI} \int \D^d \bs{x}\, D_{\mu} D^{\mu} \phi^I, \qquad\qquad r^{abI} \int \D^d \bs{x}\, F_{\mu\nu}^a F^{\mu\nu b} \phi^I,
\end{equation}
where $r^{abI}$ is some invariant symbol satisfying
\begin{equation} \label{e:r_relation}
-i g (T_R^c)^{IJ} r^{abJ} = f^{acd} r^{dbI} + f^{bcd} r^{daI}.
\end{equation}
As one might expect, however, given its resemblance to a total derivative, the first counterterm in \eqref{e:ctJJOmmm} vanishes: the piece imposed by gauge invariance that is {\it not} a total derivative, namely $-ig\delta^{ab}r^{abI}\int\D^d\bs{x}A_\mu^c T^{cIJ}D^\mu\phi^J$, vanishes since from \eqref{e:r_relation}   $\delta^{ab}r^{abI}$ is proportional to $f^{acd}r^{daI}$, but $r^{daI}$ is symmetric in $a$ and $d$ while $f^{acd}$ is antisymmetric.
All $(---)$ singularities in the form factor $A_2$ with $\Delta_3=2$ must therefore either cancel or else appear with vanishing primary constants, since there are no counterterms able to remove such singularities.  For $(---)$ singularities with $\Delta_3 = 4,6,8, \ldots$, however, we can use the second counterterm in \eqref{e:ctJJOmmm}, along with its analogues containing an even number of additional derivatives.  We therefore obtain anomalies only for $\Delta_3=4+2n$.

\paragraph{Beta functions.} 
Divergences of the type $(--+)$ are removed by  counterterms containing the scalar operator $\O^I$ and two gauge fields, generating a nontrivial beta function for $\O^I$. Their form is similar to the $(---)$ counterterms but with $\phi^I$ replaced by $\O^I$, namely
\begin{equation} \label{e:ctJJOmmp}
\delta^{ab} r^{abI} \int \D^d \bs{x}\, D_{\mu} D^{\mu} \O^I, \qquad r^{abI} \int \D^d \bs{x}\, F_{\mu\nu}^a F^{\mu\nu b} \O^I.
\end{equation}
By the same reasoning as above, however, the first of these counterterms vanishes. 
The second counterterm is only available 
when $\Delta_3 = d - 4 - 2n$, for some non-negative integer $n=0,1,2\ldots$.
Thus, only in these cases can $(--+)$ singularities be removed by a counterterm; in all other cases such singularities must either cancel or else appear with vanishing primary constants.

Type $(+--)$ singularities are removed by counterterms containing the conserved current $J^{\mu a}$ and the sources $A_{\mu}^a$ and $\phi^I$, giving a beta function for $J^{\mu a}$. The simplest such counterterms take the form 
\begin{equation}
r^{aI} \int \D^d \bs{x}\, J^{\mu a} D_{\mu} \phi^I, \qquad\qquad r^{abI} \int \D^d \bs{x}\, J_{\mu}^{a} F^{\mu\nu b} D_{\nu} \phi^I.
\end{equation}
In the first counterterm, we introduced another invariant $r^{aI}$. This counterterm, however, vanishes by current conservation and hence does not contribute to correlation functions.  Only the second counterterm is valid, allowing
$(+--)$ singularities to be removed whenever $\Delta_3 = d + 2 + 2n$ for some non-negative integer $n$.  In all other cases, such singularities must either cancel or appear with vanishing primary constants.

\subsubsection{\texorpdfstring{$d=3$ and $\Delta_3 = 1$}{d=3 and Delta=1
}}

In this example, no counterterms are available to us. 
The triple-$K$ integrals appearing in the form factors are straightforward to calculate:
\begin{align}
J_{2\{000\}} &= I_{\frac{5}{2} \{ \frac{1}{2},\frac{1}{2},{-\frac{1}{2}} \}}  = \left( \frac{\pi}{2} \right)^{3/2} \frac{1}{p_3 a_{123}^2}, \\
J_{1\{001\}} &= I_{\frac{3}{2} \{ \frac{1}{2},\frac{1}{2}, \frac{1}{2} \}}  = \left( \frac{\pi}{2} \right)^{3/2} \frac{1}{a_{123}}, \\
J_{0\{000\}} &= I_{\frac{1}{2} + u \epsilon \{ \frac{1}{2} + u \epsilon, \frac{1}{2}  + u \epsilon, -\frac{1}{2} + v_3 \epsilon \}}  = - \left( \frac{\pi}{2} \right)^{3/2} \frac{1}{(u - v_3) \epsilon}\frac{1}{p_3} + O(\epsilon^0).
\end{align}
Only the integral $J_{0\{000\}}$ has a pole corresponding to a $(--+)$ singularity.  As no other singularities are present,  cancellations cannot occur and so instead the corresponding primary constant $\3{2}^{a_1a_2I}$ must  be of order $\epsilon$.  
Indeed, evaluating the solution  \eqref{secCWI_JJO_constr} of the secondary CWI, we find
\begin{equation}
\3{2}^{a_1a_2I} = \Big[\frac{1}{2}(u-v_3)\ep  +O(\ep^2)\Big]\3{1}^{a_1a_2I}.
\end{equation}
Removing the regulator by sending $\ep\rightarrow 0$, we then find
\begin{align}
A_1^{a_1a_2I} & = \frac{\ren{1}^{a_1a_2I}}{p_3 a_{123}^2}, \\
A_2^{a_1a_2I} & = \ren{1}^{a_1a_2I} \Big( \frac{1}{a_{123}} - \frac{1}{2 p_3} \Big),
\end{align}
where for convenience we have rescaled the primary constant $\ren{1}^{a_1a_2I} \mapsto (\pi/2)^{-3/2} \3{1}^{a_1a_2I}$.

\subsubsection{\texorpdfstring{$d=3$ and $\Delta_3 = 3$}{d=3 and Delta=3}}

In this case, we again lack counterterms. Both the integrals $J_{0\{000\}}$ and $J_{1\{001\}}$ contributing to the form factor $A_2$ satisfy the $(+--)$ condition hence have single poles.
Via \eqref{secCWI_JJO_constr}, the secondary CWI then imposes $C_2^{a_1a_2I}=(-3+O(\ep))C_1^{a_1a_2I}$ and the  singularities cancel.  After removing the regulator and rescaling the primary constant as above, we obtain 
\begin{align}
A_1^{a_1a_2I} & = \ren{1}^{a_1a_2I} \frac{(a_{123} + p_3)}{a_{123}^2}, \\[0.5ex]
A_2^{a_1a_2I} & = - \ren{1}^{a_1a_2I} \frac{(a_{12}^2 + p_3 a_{12} - 2 p_3^2)}{2 a_{123}}.
\end{align}

\subsubsection{\texorpdfstring{$d=3$ and general $\Delta_3$}{d=3 and general Delta}}
\label{JJOchi}

In three dimensions, a particularly useful scheme is $u=v_1=v_2=0$ with $v_3\neq 0$.  Besides regulating all divergences, this scheme ensures that the triple-$K$ integrals appearing in the form factors
\eqref{a:JJO1}\,-\,\eqref{a:JJOlast} all have $\beta_1=\beta_2=1/2$. 
We can then obtain further simplifications using the identities
\begin{align}
I_{\alpha\{\frac{1}{2},\frac{1}{2},\beta_3+1\}}&= -(p_1+p_2)I_{\alpha\{\frac{1}{2},\frac{1}{2},\beta_3\}}+(\alpha+\beta_3-1)I_{\alpha-1\{\frac{1}{2},\frac{1}{2},\beta_3\}}, \label{3dId1} \\
(\beta_3^2-\alpha^2)I_{\alpha\{\frac{1}{2},\frac{1}{2},\beta_3\}} &= -(2\alpha+1)(p_1+p_2)I_{\alpha+1\{\frac{1}{2},\frac{1}{2},\beta_3\}}+2\chi I_{\alpha+2 \{\frac{1}{2},\frac{1}{2},\beta_3\}},\label{3dId2}
\end{align}
where 
\[\label{chidef}
\chi = \frac{1}{2}(p_1+p_2+p_3)(p_1+p_2-p_3).
\]
Here, both identities follows from simple integration by parts: the first after re-expressing all half-integer Bessel functions in terms of elementary functions;  the second after applying the modified Bessel differential operator to the $K_{\beta_3}(p_3 x)$ factor inside $I_{\alpha \{\frac{1}{2},\frac{1}{2},\beta_3\}}$.

Using these identities, our solution \eqref{a:JJO1}\,-\,\eqref{a:JJOlast} plus   \eqref{secCWI_JJO_constr} for the regulated form factors can be re-written as
\begin{align}
A_1^{a_1a_2I} &= C_1^{a_1a_2I} I_{\frac{5}{2}
\{\frac{1}{2},\frac{1}{2}\,\Delta_3-3/2+v_3\ep\}},
\\
A_2^{a_1a_2I} &= -\chi C_1^{a_1a_2I} I_{\frac{5}{2}
\{\frac{1}{2},\frac{1}{2}\,\Delta_3-3/2+v_3\ep\}}.
\end{align}
For {\it general} values of $\Delta_3$, the three-dimensional form factors are thus related by
\begin{align}
0 = \chi A_1^{a_1a_2I}+ A_2^{a_1a_2I},
 \label{chiJJOrel}
\end{align}
and hence the tensor structure takes the universal form
\[
 \lla J^{\mu_1 a_1}(\bs{p}_1) J^{\mu_2 a_2}(\bs{p}_2) \mathcal{O}^I(\bs{p}_3)
\rra = A_1^{a_1a_2I} \pi^{\mu_1}_{\alpha_1}(\bs{p}_1) \pi^{\mu_2}_{\alpha_2}(\bs{p}_2) \Big(
 p_2^{\alpha_1} p_3^{\alpha_2} -\chi \delta^{\alpha_1
\alpha_2} \Big). 
\]
Converting to a helicity basis, as described in section 8.1.1 of \cite{Bzowski:2013sza}, this result reads
\[
 \lla J^{(s_1) a_1}(\bs{p}_1) J^{(s_2) a_2}(\bs{p}_2) \mathcal{O}^I(\bs{p}_3)
\rra = A_1^{a_1a_2I}\chi \delta^{s_1s_2}, 
\]
where the helicities $s_1$ and $s_2$ take values $\pm 1$.  The correlator thus vanishes when the helicities differ.

The relation \eqref{chiJJOrel} is easily checked for the cases $\Delta_3=1$ and $\Delta_3=3$  studied above.  
Where divergences arise, the regulated form factors must satisfy \eqref{chiJJOrel} order by order in the regulator $\ep$.  The same must then be true for all divergent counterterm contributions, meaning that up to finite scheme-dependent terms the relation \eqref{chiJJOrel} extends to the renormalised theory as well.

\subsubsection{\texorpdfstring{$d=4$ and  $\Delta_3 = 2$}{d=4 and Delta=2}}

Once again, in this case we lack counterterms.  The integral $J_{1\{001\}}$ has a single pole associated with a $(---)$ singularity while
$J_{0\{000\}}$ has a double pole corresponding to the presence of both $(---)$ and $(--+)$ singularities:
\begin{align}
J_{1\{001\}}&=I_{2+u\ep \{1+u\ep, 1+u\ep,1+v_3\ep\}}=-\frac{1}{(u+v_3)\ep}+O(\ep^0),\\
J_{0\{000\}}&= I_{1+u\ep\{1+u\ep,1+u\ep,v_3\ep\}} 
\nn\\[0.5ex]&= \frac{1}{(u^2-v_3^2)\ep^2}+\frac{1}{(u-v_3)\ep}\Big[\frac{u(\ln 2-\gamma_E)}{u+v_3}+\ln p_3\Big]+O(\ep^0).
\end{align}
The solution \eqref{secCWI_JJO_constr} of the secondary CWI then imposes $C_2^{a_1a_2 I} = ((u-v_3)\ep +O(\ep^2))C_1^{a_1a_2I}$.  When we assemble the form factor $A_2$, we thus obtain two single poles which cancel.
After removing the regulator, the final result is then
\begin{align}
A_1^{a_1a_2I} & = \ren{1}^{a_1a_2I} p_1 p_2 \frac{\partial^2}{\partial p_1 \partial p_2} I_{1\{000\}}, \\
A_2^{a_1a_2I} & = \ren{1}^{a_1a_2I} \left( I^{\text{(fin)}}_{2\{111\}} + \frac{1}{6} \ln \frac{p_3^4}{p_1^2 p_2^2} + \frac{1}{2} \right),
\end{align}
where the finite integrals $I_{1\{000\}}$ and $I^{\text{(fin)}}_{2\{111\}}$ are given in \eqref{I1000} and \eqref{I2111fin}.

\subsubsection{\texorpdfstring{$d=4$ and  $\Delta_3 = 4$}{d=4 and Delta=4}}

In this case, all the triple-$K$ integrals diverge.  From table \ref{tab:JJO} on page \pageref{tab:JJO},
the integral $J_{2\{000\}}$ for the form factor $A_1$ has a single pole of type $(---)$, while for the form factor $A_2$, both $J_{1\{001\}}$ and $J_{0\{000\}}$ have double poles due to the presence of both $(---)$ and $(+--)$ singularities.
To remove these singularities, we have only a single counterterm of type $(---)$ (see table \ref{tab:physJJO} on page \pageref{tab:physJJO}), and so we can already anticipate that the $(+--)$ type singularities in $A_2$ should cancel.

To verify this, from our solution \eqref{secCWI_JJO_constr} of the secondary CWI, the dependence between the primary constants is
\begin{align}
\3{2}^{(0)a_1a_2I} & = - 4 \3{1}^{(0)a_1a_2I}, \\
\3{2}^{(1)a_1a_2I} & = (u - 3 v_3) \3{1}^{(0)a_1a_2I} - 4 \3{1}^{(1)a_1a_2I}, \\
\3{2}^{(2)a_1a_2I} & = \tfrac{1}{2} (u^2 - v_3^2) \3{1}^{(0)a_1a_2I} + (u - 3 v_3) \3{1}^{(1)a_1a_2I} - 4 \3{1}^{(2)a_1a_2I},
\end{align}
where for convenience we have decomposed
\[
C_j^{a_1a_2I} = C_j^{(0)a_1a_2I} +\ep C_j^{(1)a_1a_2I} +\ep^2 C_j^{(2)a_1a_2I} +O(\ep^3), \qquad j=1,2.
\]
Evaluating the divergences of the triple-$K$ integrals, we then find that the $(+--)$ singularities in $A_2$ do indeed cancel leaving us with only single poles of type $(---)$, namely
\begin{align}
A_1^{a_1a_2I} & = - \frac{2\3{1}^{(0)a_1a_2I}}{(u+v_3) \epsilon} + O(\ep^0), \\[0.5ex]
A_2^{a_1a_2I} & = \frac{\3{1}^{(0)a_1a_2I}}{(u+v_3) \epsilon} (p_1^2 + p_2^2 - p_3^2) + O(\ep^0).
\end{align}
As these form factors still diverge, we use the single available counterterm 
\begin{equation}
S_{\text{ct}} = \ct^{abI} \int \D^{4 + 2 u \epsilon} \bs{x}\, F_{\mu \nu}^a F^{\mu \nu b} \phi^I \mu^{(u+v_3) \epsilon},
\end{equation}
which contributes to the form factors as
\begin{align}
A_1^{a_1a_2I\text{ct}} & = - 4 \ct^{a_1a_2I} \mu^{( u+v_3) \epsilon}, \label{e:JJO_A1ct} \\
A_2^{a_1a_2I\text{ct}} & = 2 \ct^{a_1a_2I} \mu^{(u+v_3) \epsilon} (p_1^2 + p_2^2 - p_3^2).
\end{align}
The divergences can thus be cancelled by setting
\begin{equation}
\ct^{a_1a_2I} = - \frac{\3{1}^{(0)a_1a_2I}}{2(u+v_3) \epsilon} + O(\epsilon^0).
\end{equation}
Removing the regulator and relabelling $C_1^{(0)a_1a_2I}\rightarrow C_1^{a_1a_2I}$, the final renormalised form factors are then
\begin{align}
A_1^{a_1a_2I} & = \ren{1}^{a_1a_2I} \Big(2 - p_3 \frac{\partial}{\partial p_3} \Big) I_{2\{111\}}^{\text{(fin)}} - \frac{1}{3} \ren{1}^{a_1a_2I} \Big( \ln \frac{p_1^2}{\mu^2} + \ln \frac{p_2^2}{\mu^2} + \ln \frac{p_3^2}{\mu^2} \Big) + \sd{1}^{a_1a_2I}, \label{e:JJO_A1ren} \\[0.5ex]
A_2^{a_1a_2I} & = \ren{1}^{a_1a_2I} p_3^2 I_{2\{111\}}^{\text{(fin)}} + \frac{1}{6} \ren{1}^{a_1a_2I} \Big[ (3 p_1^2 - p_3^2) \ln \frac{p_1^2}{\mu^2} + (3 p_2^2 - p_3^2) \ln \frac{p_2^2}{\mu^2} - p_3^2 \ln \frac{p_3^2}{\mu^2} + 3 p_3^2 \Big] \nn\\
& \qquad + \: (p_1^2 + p_2^2 - p_3^2) \Big[ \frac{1}{6} \3{1}^{a_1a_2I} - \frac{1}{2} \sd{1}^{a_1a_2I} \Big], \label{e:JJO_A2ren}
\end{align}
where the finite integral $I_{2\{111\}}^{\text{(fin)}}$ is given in \eqref{I2111fin}.
The scheme-dependent constant $\sd{1}^{a_1a_2I}$ is related to the data of the regulated theory  by
\begin{equation}
\sd{1}^{a_1a_2I} = -\Big( \frac{2}{3} - \gamma_E + \ln 2 \Big)\3{1}^{(0)a_1a_2I}  -  \frac{\3{1}^{(1)a_1a_2I}}{u} - 4 \ct^{(0)a_1a_2I} .
\end{equation}
In the renormalised theory, $D_1^{a_1a_2I}$ can be shifted arbitrarily by a change of renormalisation scale.  (Scaling $\mu^2\rightarrow e^\lambda\mu^2$ is equivalent to shifting $D_1^{a_1a_2I}\rightarrow D_1^{a_1a_2I}+\lambda C_1^{a_1a_2I}$.)

\paragraph{Anomalous CWI.}
The renormalised form factors satisfy the anomalous dilatation Ward identities
\begin{align}
\mu \frac{\partial}{\partial \mu} A_1^{a_1a_2I} & = 2 \3{1}^{a_1a_2I}, \label{e:JJO_dmu1} \\
\mu \frac{\partial}{\partial \mu} A_2^{a_1a_2I} & = - \3{1}^{a_1a_2I} (p_1^2 + p_2^2 - p_3^2), \label{e:JJO_dmu2}
\end{align}
and the anomalous primary CWIs
\begin{align}
\K_{12} A_1^{a_1 a_2 I} & = 0, && \K_{13} A_1^{a_1 a_2 I} = 0, \nn\\
\K_{12} A_2^{a_1 a_2 I} & = 0, && \K_{13} A_2^{a_1 a_2 I} = 2 A_1^{a_1 a_2 I} + 4 \3{1}^{a_1 a_2 I}.
\end{align}
The secondary CWI is however non-anomalous, and retains its original homogeneous form
\begin{equation}
\Lo_{1} A_1^{a_1 a_2 I} + 2 \Ro A_2^{a_1 a_2 I} = 0.
\end{equation}
As only a $(---)$ counterterm is present, from \eqref{anomformulaonlysources} the anomaly action is
\[
A = \int\D^4\x\, \frac{1}{2}C_1^{a b I}\phi^I F^a_{\mu\nu}F^{\mu\nu b}.
\]
On a conformal manifold, for cases where $C_1^{abI}=C_1^I\delta^{ab}$ this can be interpreted as a moduli-dependent shift  of the 2-point normalisation $C_{JJ}\rightarrow C_{JJ}- C_1^{I}\phi^I$ in the quadratic anomaly action \eqref{e:tranom}.

The anomalous dilatation Ward identities \eqref{e:JJO_dmu1}\,-\,\eqref{e:JJO_dmu2} now correspond to
\[
\mu\frac{\p}{\p\mu}\<J^{\mu_1 a_1}(\x_1)J^{\mu_2 a_2}(\x_2)\O^I(\x_3)\> = -\frac{\delta^3 A}{\delta A^{a_1}_{\mu_1}(\x_1)\delta A^{a_2}_{\mu_2}(\x_2)\delta\phi^I(\x_3)}\Big|_0.
\]

\subsection{\texorpdfstring{$\<T_{\mu_1\nu_1}T_{\mu_2\nu_2}\O\>$}{<TTO>}}

\subsubsection{General analysis}

\paragraph{Decomposition.} The renormalised 3-point function satisfies the following transverse and trace Ward identities\footnote{Note these identities differ from those in \cite{Bzowski:2013sza} since here we define the 3-point function through three functional derivatives, see p.~15-16 of \cite{Bzowski:2017poo}.}
\begin{align}
& p_1^{\nu_1} \lla T_{\mu_1 \nu_1}(\bs{p}_1) T_{\mu_2 \nu_2}(\bs{p}_2)
\mathcal{O}^I(\bs{p}_3) \rra =0, \label{pTTO} \\[1ex]
& \lla T(\bs{p}_1) T_{\mu_2 \nu_2}(\bs{p}_2) \mathcal{O}^I(\bs{p}_3) \rra = \mathcal{A}_{\mu_2\nu_2}^I.\label{pTTOTWI}
\end{align}
In general, the trace Ward identity features an anomaly $\mathcal{A}_{\mu_2\nu_2}^I$.  Taking the trace of the transverse Ward identity we see this anomaly must be transverse, $p_2^{\mu_2}\mathcal{A}_{\mu_2\nu_2}^I=0$, and clearly its trace $\delta^{\alpha\beta}\mathcal{A}^I_{\alpha\beta}$ must be symmetric under $p_1\leftrightarrow p_2$. 
We will evaluate this anomaly on a case-by-case basis; for most of the cases we study here, no counterterms are present and hence this anomaly is absent.  In the final case we study ($d=\Delta_3=4$) there is a nontrivial counterterm but, as it turns out, $A_{\mu_2\nu_2}^I$ also vanishes.  Nevertheless, this anomaly will be present in general.
The renormalised 3-point function can then be reconstructed from its transverse-traceless part using the formula
\begin{align}
& \lla T_{\mu_1 \nu_1}(\bs{p}_1) T_{\mu_2 \nu_2}(\bs{p}_2)
\mathcal{O}^I(\bs{p}_3) \rra  = \lla t_{\mu_1 \nu_1}(\bs{p}_1) t_{\mu_2
\nu_2}(\bs{p}_2) \mathcal{O}^I(\bs{p}_3) \rra  +\frac{1}{d-1}\pi_{\mu_1\nu_1}(\bs{p}_1)\mathcal{A}_{\mu_2\nu_2}^I \nn\\& \qquad\qquad +\frac{1}{d-1}\pi_{\mu_2\nu_2}(\bs{p}_2)\mathcal{A}_{\mu_1\nu_1}^I(p_1\leftrightarrow p_2)
-\frac{1}{(d-1)^2}\pi_{\mu_1\nu_1}(\bs{p}_1)\pi_{\mu_2\nu_2}(\bs{p}_2)\delta^{\alpha\beta}\mathcal{A}_{\alpha\beta}^I.
\end{align}
When the anomaly is absent, the 3-point function is purely transverse-traceless.

\paragraph{Form factors.} The transverse-traceless part of the correlator can be decomposed as
\begin{align}\label{TTOformfactordecomp}
 &\lla t_{\mu_1 \nu_1}(\bs{p}_1) t_{\mu_2 \nu_2}(\bs{p}_2) \mathcal{O}^I(p_3)
\rra\\[1ex] &\quad  = \Pi_{\mu_1 \nu_1\alpha_1 \beta_1}(\bs{p}_1) \Pi_{\mu_2
\nu_2\alpha_2 \beta_2}(\bs{p}_2) \Big[
A_1^I p_2^{\alpha_1} p_2^{\beta_1} p_3^{\alpha_2} p_3^{\beta_2}  + \: A_2^I \delta^{\alpha_1 \alpha_2} p_2^{\beta_1}
p_3^{\beta_2} + A_3^I \delta^{\alpha_1 \alpha_2} \delta^{\beta_1 \beta_2}
\Big],\nn
\end{align}
where the form factors $A_j$, $j=1,2,3$ are functions of the momentum magnitudes. All
form factors are symmetric under $p_1 \leftrightarrow p_2$, \textit{i.e.}, they
satisfy
\begin{equation}
A_j^I(p_2, p_1, p_3) = A_j^I(p_1, p_2, p_3), \qquad j=1,2,3.
\end{equation}
Given the full correlator $\lla T_{\mu_1 \nu_1}(\bs{p}_1) T_{\mu_2 \nu_2}(\bs{p}_2)
\mathcal{O}^I(\bs{p}_3) \rra$, with the independent momenta chosen according to our cyclic rule \eqref{a:momenta}, the form factors can be extracted as follows:
\begin{align}
A_1^I & = \text{coefficient of } p_{2\mu_1} p_{2\nu_1} p_{3\mu_2}
p_{3\nu_2}, \\
A_2^I & = 4 \cdot \text{coefficient of } \delta_{\mu_1 \mu_2} p_{2\nu_1}
p_{3\nu_2}, \\
A_3^I & = 2 \cdot \text{coefficient of } \delta_{\mu_1 \mu_2} \delta_{\nu_1
\nu_2}. \label{e:TTO_A3}
\end{align}

\paragraph{Degeneracy.}\label{degexpl}
In three dimensions, the form factor basis above is degenerate as discussed in appendix A.4 of \cite{Bzowski:2017poo}.
Specifically, the following combination vanishes since in three dimensions one of the indices in the 4-form must necessarily be repeated:
\begin{align}
&\Pi_{\mu_1\nu_1}{}^{\alpha_1}{}_{\beta_1}(\bs{p}_1)\Pi_{\mu_2\nu_2}{}^{\alpha_2}{}_{\beta_2}(\bs{p}_2)\,4!\,\delta_{[\alpha_1}^{\beta_1}\delta_{\alpha_2}^{\beta_2}p_{1\alpha_3}p_{2\alpha_4]} p_1^{\alpha_3}p_2^{\alpha_4} \nn\\[0.5ex]&
 =\Pi_{\mu_1\nu_1\alpha_1\beta_1}(\bs{p}_1)\Pi_{\mu_2\nu_2\alpha_2\beta_2}(\bs{p}_2)\,\Big[
  p_2^{\alpha_1} p_2^{\beta_1} p_3^{\alpha_2} p_3^{\beta_2} \nn\\&\qquad\qquad\qquad\qquad\qquad\qquad \quad
   - (p_1^2+p_2^2-p_3^2) \delta^{\beta_1 \beta_2} p_2^{\alpha_1} p_3^{\alpha_2}
  -\frac{1}{4}J^2 \delta^{\alpha_1 \alpha_2} \delta^{\beta_1 \beta_2} \Big]. \label{3ddeg}
\end{align}
Multiplying through by an arbitrary function  $F^I(p_1,p_2,p_3)=F^I(p_2,p_1,p_3)$, we obtain a degenerate set of form factors which yield zero contribution to the 3-point function:
\begin{align}\label{Fdeg1}
A_1^I = F^I, \qquad 
A_2^I = -(p_1^2+p_2^2-p_3^2)F^I, \qquad
A_3^I = -\frac{1}{4}J^2 F^I .
\end{align}
In principle, one can use this degeneracy to eliminate one of the form factors, however we will not do so here since the resulting conformal Ward identities 
then take a more complicated form.
(The degeneracy does not apply in the dimensionally regulated theory, so using it to eliminate a form factor in the renormalised theory means the renormalised CWIs are no longer simply those in the regulated theory plus potential anomalous terms.)

The existence of this degeneracy also raises the interesting possibility of novel three-dimensional type A anomalies, where the degenerate form factor combination appears with a linearly divergent coefficient.  Such a $0/0$ structure is directly responsible for the four-dimensional Euler anomaly, as discussed in \cite{Deser:1993yx, Bzowski:2017poo}.
Here, a specific example where a $0/0$ structure of this type occurs is the case of $\Delta_3=4$, as  we will discuss  in section \ref{typeAex1}.

\paragraph{Primary CWIs.} The primary CWIs
are
\begin{equation}
\begin{array}{ll}
\K_{12} A_1^I = 0, &\qquad\qquad \K_{13} A_1^I = 0, \\
\K_{12} A_2^I = 0, &\qquad\qquad \K_{13} A_2^I = 8 A_1^I, \\
\K_{12} A_3^I = 0, &\qquad\qquad \K_{13} A_3^I = 2 A_2^I.\label{KeqnsTTO}
\end{array}
\end{equation}
Their solution in terms of triple-$K$ integrals is
\begin{align}
A_1^I & = \3{1}^I J_{4 \{000\}}, \label{a:TTO1} \\
A_2^I & = 4 \3{1}^I J_{3 \{001\}} + \3{2}^I J_{2 \{000\}}, \label{a:TTO2}\\
A_3^I & = 2 \3{1}^I J_{2 \{002\}} + \3{2}^I J_{1 \{001\}} + \3{3}^I J_{0 \{000\}},
\label{a:TTO3}
\end{align}
where $\3{j}^I$, $j=1,2,3$ are constants.

\paragraph{Secondary CWIs.} The independent
secondary CWIs are
\begin{align}
& \Lo_{2} A_1^I + \Ro A_2^I  \nn\\
& \qquad = 2 \Delta_1 \cdot \text{coefficient of } p_{2\mu_1} p_{3\mu_2} p_{3\nu_2}
\text{ in } p_1^{\nu_1} \lla T_{\mu_1 \nu_1}(\bs{p}_1) T_{\mu_2 \nu_2}(\bs{p}_2)
\mathcal{O}^I(\bs{p}_3) \rra, \label{secTTO1} \\[2ex]
& \Lo_{2} A_2^I + 4 \Ro A_3^I  \nn\\
& \qquad = 8 \Delta_1 \cdot \text{coefficient of } \delta_{\mu_1 \mu_2} p_{3\nu_2}
\text{ in } p_1^{\nu_1} \lla T_{\mu_1 \nu_1}(\bs{p}_1) T_{\mu_2 \nu_2}(\bs{p}_2)
\mathcal{O}^I(\bs{p}_3) \rra, \label{secTTO2}
\end{align}
where both right-hand sides vanish by \eqref{pTTO}.
As in our earlier analysis leading to \eqref{e:sec1JOO}, these secondary Ward identities can conveniently be solved in the soft limit $p_3\rightarrow 0$.  Working in the regulated theory where necessary to avoid divergences, we obtain the constraints
\begin{align}
\3{2}^{I} & = (\Delta_3 + 2)(d - \Delta_3 - 2) \3{1}^{I}, \label{secsolTOO1} \\
\3{3}^{I} & = \frac{1}{4} \Delta_3 (\Delta_3 + 2)(d - \Delta_3)(d - \Delta_3 - 2) \3{1}^{I}. \label{secsolTOO2}
\end{align}
The regulated 3-point function thus depends on a single theory-specific constant $\3{1}^I$.

\paragraph{Regularisation.} 
Conservation of the stress tensor requires $u = v_1 = v_2$.  To regulate the various singularities arising in triple-$K$ integrals, however, we must retain $v_3\neq u$.
These singularities are summarised in the table below. 
Divergences of the type $(+++)$ and $(-++)$  are excluded by unitarity since they require either $\Delta_3 \leq 0$ or $d \leq 0$. 
\begin{center}
\begin{tabular}{|c|c|c|c|c|c|} \hline
Factor &Integral& $(---)$ & $(+--)$ & $(--+)$ & $(++-)$ \\ \hline
$A_1$ & $J_{4\{000\}}$ & $\Delta_3 = 4+2n$ & $\Delta_3 = d + 4 + 2 n$ & $\Delta_3 = d - 4 - 2 n$ & $\Delta_3 = 2d + 4 + 2 n$ \\ \hline
$A_2$ & $J_{3\{001\}}$ & $\Delta_3 = 2+2n$ & $\Delta_3 = d + 2 + 2 n$ & $\Delta_3 = d - 4 - 2 n$ & $\Delta_3 = 2d + 2 + 2 n$ \\
& $J_{2\{000\}}$ & $''$ & $''$ & $\Delta_3 = d - 2 - 2 n$ & $''$ \\ \hline
$A_3$ & $J_{2\{002\}}$ & $\Delta_3 = 2n$ & $\Delta_3 = d + 2 n$ & $\Delta_3 = d - 4 - 2 n$  & $\Delta_3 = 2d + 2 n$ \\
& $J_{1\{001\}}$ & $''$ & $''$ & $\Delta_3 = d - 2 - 2 n$ & $''$ \\
& $J_{0\{000\}}$ & $''$ & $''$ & $\Delta_3 = d - 2 n$ & $''$ \\ \hline
\end{tabular}
\captionof{table}{Singularities arising in triple-$K$ integrals for the form factors of $\< T_{\mu_1 \nu_1} T_{\mu_2 \nu_2} \O^I \>$. 
\label{tab:TTO}}
\end{center}

\paragraph{Renormalisation.} 
As we discuss below, counterterms are only available for the cases summarised in  table \ref{tab:physTTO}.  In certain instances, there are then no counterterms able to remove a particular singularity.  When this occurs, there are two possibilities: either the primary constants multiplying these divergent triple-$K$ integrals must vanish as some suitable power of the regulator, or else the singularities of individual triple-$K$ integrals cancel against one another when summed to construct the regulated form factor.  As we will see in specific examples later, these two possibilities are not mutually exclusive.
In fact, the precise combination of cancellations and/or vanishing that occurs is dictated by the secondary conformal Ward identities, which impose specific relations between the  primary constants.  
The elimination of singularities in this manner thus introduces no arbitrariness.

\begin{center}
\begin{tabular}{|c|c|} \hline
Singularity type & Counterterms available when  \\ \hline
$(---)$ & $\Delta_3 = 4+2n  $ \\ \hline
$(+--)$ & $\Delta_3 = d +2+ 2n$ \\ \hline
$(--+)$ & $\Delta_3 = d -4- 2n$ \\ \hline
\end{tabular}
\captionof{table}{Availability of counterterms for $\< T_{\mu_1\mu_2} T_{\mu_2\nu_2} \O^I \>$.\label{tab:physTTO}}
\end{center}

\paragraph{Anomalies.}
\label{TTOctdiscussion}
Singularities of type $(---)$ are removed by counterterms that are cubic in the sources, giving rise to anomalies.  Here, the relevant sources are a single scalar source $\phi^I$ and two metric perturbations. The three simplest such counterterms are
\begin{equation}
\int \D^d \bs{x} \sqrt{g}\, \phi^I, \qquad \int \D^d \bs{x} \sqrt{g}\, R \phi^I, \qquad \int \D^d \bs{x} \sqrt{g}\, W^2 \phi^I,
\end{equation}
and more complicated examples can be constructed by adding an even number of covariant derivatives.
The first two of these counterterms are forbidden, however, since they contribute respectively to the 1-point function $\< \O^I \>$ and to the 2-point function $\< T \O^I \>$, both of which must vanish to preserve conformal invariance.   Only the third counterterm (and its analogues with additional covariant derivatives) is therefore permitted.  All such counterterms contain at least two Riemann tensors, which requires $\Delta_3 = 4+2n$, for $n=0,1,2\ldots$.   Anomalies can then only arise in these cases.

\paragraph{Beta functions.}
Singularities of type $(+--)$ are removed by a counterterm containing the stress tensor $T_{\mu\nu}$ and two sources, the metric and $\phi^I$. The three simplest such counterterms are
\begin{equation}
\int \D^d \bs{x} \sqrt{g}\, T \phi^I, \qquad \int \D^d \bs{x} \sqrt{g}\, \nabla_{\mu} \nabla_{\nu} T^{\mu\nu} \phi^I, \qquad \int \D^d \bs{x} \sqrt{g} \,T_{\mu\nu} R^{\mu\nu} \phi^I,
\end{equation}
while more complicated examples follow by adding an even number of additional covariant derivatives.
The first two of these counterterms introduce however a mixing between $\O$ and either $T$ or $\nabla_{\mu} \nabla_{\nu} T^{\mu\nu}$. While these latter operators are local, they are not conformal primaries and so these counterterms cannot be added while keeping $O^I$ primary. Only the third counterterm (and its analogues with additional covariant derivatives) is therefore acceptable; such counterterms only exist for $\Delta_3 = d + 2 + 2n$. 

Finally, $(--+)$ singularities can be removed by counterterm involving the scalar operator $\O^I$ and two metric perturbations. The three simplest such counterterms are
\begin{equation}
\int \D^d \bs{x} \sqrt{g}\, \O^I, \qquad \int \D^d \bs{x} \sqrt{g}\, \O^I R, \qquad \int \D^d \bs{x} \sqrt{g}\, W^2 \O^I,
\end{equation}
and again more complicated examples can be constructed by adding additional covariant derivatives.
The first counterterm simply represents a constant deformation of the original CFT by a marginal operator, and so does not need to be considered.
The second counterterm is the so-called improvement term and contributes to the mixed 2-point function $\< T_{\mu\nu} \O^I \>$.  As this must vanish by conformal invariance, this counterterm can also be excluded.
Only the third counterterm and its analogues with additional covariant derivatives are therefore permitted.  All such counterterms involve at least two Riemann tensors, and hence only exist for $\Delta_3 = d - 4 - 2n$.

\subsubsection{\texorpdfstring{$d = 3$ and $\Delta_3 = 1$}{d=3 and Delta=1}}

From table \ref{tab:TTO}, all triple-$K$ integrals multiplying the primary constant $C_1^I$ are finite, while those multiplying  $C_2^I$ and $C_3^I$ all have $(--+)$ singularities and hence $\ep^{-1}$ poles.   From our solution \eqref{secsolTOO1}\,-\,\eqref{secsolTOO2} of the secondary Ward identities, however, we see that the primary constants $C_2^I$ and $C_3^I$ are both suppressed by a factor of $\ep$ relative to $C_1^I$.   
The leading term in $C_1^I$ is thus of order $\ep^0$ while the expansions of $C_2^I$ and $C_3^I$ begin at order $\ep$.
The regulated form factors are then finite as $\ep\rightarrow 0$, as indeed must be the case given the absence of counterterms.
After redefining $C_1^I$ to absorb an overall numerical factor,  the result is
\begin{align}
A_1^I &= \frac{C_1^I}{a_{123}^4 p_3} \mathcal{E}_1,
\\[2ex]
A_2^I &=\frac{C_1^I}{a_{123}^3 p_3} \big( - \mathcal{E}_1 (a_{12} - p_3) + 2 \mathcal{E}_2 b_{12} \big),
\\[2ex]
A_3^I &=\frac{C_1^I(a_{12}-p_3)}{4a_{123}^2 p_3} \big( \mathcal{E}_1 (a_{12} - p_3) - 4 \mathcal{E}_2 b_{12}\big),
\end{align}
where the polynomials
\begin{align}
\mathcal{E}_1 & = 3 a_{12}^2 + 6 b_{12} + 4 a_{12} p_3 + p_3^2, \qquad
\mathcal{E}_2 =3 a_{12}+p_3.
\end{align}
As $d=3$, we can also add to these form factors the degenerate combination \eqref{Fdeg1}.

\subsubsection{\texorpdfstring{$d = 3$ and $\Delta_3 = 3$}{d=3 and Delta=3}}

In this case, all the triple-$K$ integrals appearing in the form factors $A_1^I$ and $A_2^I$ are finite, while those appearing in $A_3^I$ all have $\ep^{-1}$ poles due to the presence of $(+--)$ singularities. (The integral $J_{0\{000\}}$ has in addition a $(--+)$ singularity, however since this differs only by a permutation, it does not lead to a double pole; see \cite{Bzowski:2015pba}).
Explicitly, we find
\begin{align}
J_{2\{002\}} &= I_{\frac{5}{2}+u\ep\{\frac{3}{2}+u\ep,\frac{3}{2}+u\ep,\frac{7}{2}+v_3\ep\}}= \Big(\frac{\pi}{2}\Big)^{3/2}\frac{5}{(u-v_3)\ep}\,(p_1^3+p_2^3)+O(\ep^0),\\[1ex]
J_{1\{001\}} &= I_{\frac{3}{2}+u\ep\{\frac{3}{2}+u\ep,\frac{3}{2}+u\ep,\frac{5}{2}+v_3\ep\}} =\Big(\frac{\pi}{2}\Big)^{3/2}\frac{1}{(u-v_3)\ep}\,(p_1^3+p_2^3)+O(\ep^0),\\[1ex] 
J_{0\{000\}} &= I_{\frac{1}{2}+u\ep\{\frac{3}{2}+u\ep,\frac{3}{2}+u\ep,\frac{3}{2}+v_3\ep\}} = \Big(\frac{\pi}{2}\Big)^{3/2}\frac{1}{3(u-v_3)\ep}\,(p_1^3+p_2^3-p_3^3)+O(\ep^0).
\end{align}
As again there are no counterterms available, the regulated form factors must be finite as $\ep\rightarrow 0$.  
Here, the secondary Ward identities \eqref{secsolTOO1}\,-\,\eqref{secsolTOO1}  tell us that to leading order
\[
C_2^I=\big(-10+O(\ep)\big)C_1^I, \qquad
C_3^I = \big(-\frac{15}{2}(u-v_3)\ep+O(\ep^2)\big)C_1^I.
\]
With these primary constants, the poles in $J_{2\{002\}}$ and $J_{1\{001\}}$ now cancel when summed to construct the form factor $A_3^I$, while the expansion of $C_3^I$ begins at order $\ep$ eliminating the pole contribution from  $J_{0\{000\}}$.
Sending $\ep\rightarrow 0$ and reabsorbing an overall numerical factor into $C_1^I$, the final result is
\begin{align}
A_1^I &= \frac{C_1^I}{a_{123}^4} \mathcal{E}_1,
\\[2ex]
A_2^I &=\frac{C_1^I}{a_{123}^3} \big( - \mathcal{E}_1 (a_{12} - p_3) + 2 \mathcal{E}_2 b_{12} \big),
\\[2ex]
A_3^I &=\frac{C_1^I(a_{12}-p_3)}{4a_{123}^2} \big( \mathcal{E}_1 (a_{12} - p_3) - 4 \mathcal{E}_2 b_{12} \big),
\end{align}
where the polynomials
\begin{align}
\mathcal{E}_1 & = a_{123}^3 + a_{123} b_{123} + 3 c_{123}, \qquad
\mathcal{E}_2  = a_{12}^2+3a_{12}p_3+p_3^2.
\end{align}
Once again, as $d=3$, we are free to add to these form factors the degenerate combination \eqref{Fdeg1}.

\subsubsection{\texorpdfstring{$d=3$ and $\Delta_3=4$}{d=3 and Delta=4}}
\label{typeAex1}

We examine this additional case   
since it 
raises the novel possibility of a  
three-dimensional type A anomaly.
In dimensional regularisation, type A anomalies originate from a $0/0$ limit in which an $\ep^{-1}$ pole multiplies an evanescent tensorial structure that vanishes as $\ep\rightarrow 0$. 
Here, we find exactly this: the regulated form factors are
\begin{align}\label{anomdiv1}
A_1^I &= \frac{c^I}{u\ep}+O(\ep^0),\\[1ex]
A_2^I &= -\frac{c^I}{u\ep}(p_1^2+p_2^2-p_3^2) + O(\ep^0),\\[1ex]\label{anomdiv3}
A_3^I &= -\frac{c^I}{4 u\ep} J^2 +O(\ep^0),
\end{align}
where 
\[\label{divcIdef}
c^I = -\Big(\frac{\pi}{2}\Big)^{3/2}\frac{3  u}{(u+v_3)}C^{(0)I}_1.
\]
The divergences thus correspond to a pole multiplying the degenerate form factor combination \eqref{Fdeg1}.
This combination derives from the contraction of a 4-form (namely \eqref{3ddeg}), and so vanishes in the limit of three spacetime dimensions. 

As discussed in section 4.1 of \cite{Bzowski:2017poo}, the divergent form factors \eqref{anomdiv1}\,-\,\eqref{anomdiv3} are equivalent to a finite anomalous contribution to the 3-point function of the form   
\begin{align}\label{TTOanomcontr}
&\lla T_{\mu_1\nu_1}(\bs{p}_1)T_{\mu_2\nu_2}(\bs{p}_2)\O^I(\bs{p}_3)\rra_{\mathrm{anom.}}  \nn\\[1ex] & = c^I p_2^2\, \pi_{\mu_1\nu_1}(\bs{p}_1)\Pi_{\mu_2\nu_2\alpha_2\beta_2}(\bs{p}_2)p_3^{\alpha_2}p_3^{\beta_2} 
+c^I p_1^2\, \pi_{\mu_2\nu_2}(\bs{p}_2)\Pi_{\mu_1\nu_1\alpha_1\beta_1}(\bs{p}_1)p_2^{\alpha_1}p_2^{\beta_1} \nn\\[1ex]&\quad
-\frac{c^I}{8}J^2 \pi_{\mu_1\nu_1}(\bs{p}_1)\pi_{\mu_2\nu_2}(\bs{p}_2).
\end{align}
One way to see this is to note that, in dimensional regularisation, the {\it external} Lorentz indices $\mu_1, \nu_1, \mu_2,\nu_2$ run only over the physical values $1, 2,3$, while all {\it internal} Lorentz indices ({\it i.e.,} those that are contracted) run over the full $d$ dimensions.  
Proceeding now to evaluate the left-hand side of \eqref{3ddeg}, we note first that the 4-form vanishes when contracted with any momentum.  From the definition \eqref{TTprojdef} of the transverse-traceless projectors, when contracted with the 4-form the only nonzero contributions then come from the terms 
\[
\Pi_{\mu\nu\alpha\beta}(\bs{p}) = \delta_{\mu(\alpha}\delta_{\beta)\nu} - \frac{1}{d-1}\pi_{\mu\nu}(\bs{p})\delta_{\alpha\beta} + \ldots
\]
Here, in the first $\delta_{\mu(\alpha}\delta_{\beta)\nu}$ term, the $\alpha$ and $\beta$ indices take physical values since the $\mu$ and $\nu$ are external indices.  In the second term, the $\alpha$ and $\beta$ indices are instead internal and hence run over $d$ dimensions.  Consequently, when we evaluate the left-hand side 
of \eqref{3ddeg}, the $\delta_{\mu_1(\alpha_1}\delta_{\beta_1)\nu_1}\delta_{\mu_2(\alpha_2}\delta_{\beta_2)\nu_2}$ piece from the product of the two transverse-traceless projectors makes no contribution, since all indices in the 4-form are forced to take physical values.  (The  $\alpha_3$ and $\alpha_4$ indices are attached to external momenta hence also take physical values.)  The remaining terms all contain at least one $d$-dimensional trace over the 4-form, generating a factor of $(d-3)=2u\ep$.  This zero cancels the overall pole multiplying the form factors, yielding the finite result given in \eqref{TTOanomcontr}.

An alternative way to obtain this same result is to 
consider adding a counterterm
\[\label{Sdeg}
S_{\mathrm{ct}} = \mathfrak{a}^I\int\D^{3+2u\ep}\x\sqrt{g} \mu^{(u+v_3)\ep}\phi^I E_4,
\]
whose contribution to the 3-point function is
\begin{align}\label{TTOEulerctcontr}
&\lla T_{\mu_1\nu_1}(\bs{p}_1)T_{\mu_2\nu_2}(\bs{p}_2)\O^I(\bs{p}_3)\rra_{\mathrm{ct}} 
\nn\\[1ex]&\qquad\qquad\quad 
= 192\mu^{(u+v_3)\ep}\mathfrak{a}^I \delta_{(\mu_1}^{\alpha_1}\delta_{\nu_1)\beta_1} 
\delta_{(\mu_2}^{\alpha_2}\delta_{\nu_2)\beta_2} 
\delta_{[\alpha_1}^{\beta_1}\delta_{\alpha_2}^{\beta_2}p_{1\alpha_3}p_{2\alpha_4]} p_1^{\alpha_3}p_2^{\alpha_4}.
\end{align}
We then re-write the right-hand side using the identity
\[\label{Idecomp}
\delta_{\mu(\alpha}\delta_{\beta)\nu}=
\Pi_{\mu\nu\alpha\beta}(\bs{p})+\mathscr{T}_{\mu\nu (\alpha}(\bs{p})\,p_{\beta)}+\frac{1}{d-1}\pi_{\mu\nu}(\bs{p})\delta_{\alpha\beta},
\]
where $\mathscr{T}_{\mu\nu\alpha}(\bs{p})$ is defined in 
\eqref{Iprojdef}.
The contribution from $\mathscr{T}_{\mu\nu (\alpha}(\bs{p})p_{\beta)}$ vanishes, however, since the 4-form on the right-hand side of \eqref{TTOEulerctcontr} is transverse. 
The terms containing two transverse-traceless projectors are proportional to \eqref{3ddeg}, and are equivalent to the form factors listed in \eqref{TTOctcontr1}\,-\,\eqref{TTOctcontr3}.  Setting 
\[\label{baboon}
\mathfrak{a}^I = -\frac{c^I}{8u\ep} +O(\ep^0),
\]
we can cancel the divergences in the regulated form factors \eqref{anomdiv1}\,-\,\eqref{anomdiv3}.
The remaining contribution from the counterterm then corresponds to the anomalous terms given in 
\eqref{TTOanomcontr}.  In this second approach, therefore, rather than evaluating the 4-form we simply cancel it with an appropriate counterterm; on the other hand, from the first approach it is clear that adding counterterms is not {\it required} in order to remove the divergences.  This is generally the case for type A anomalies, as originally emphasised in \cite{Deser:1993yx}.

Resuming our analysis, tracing over \eqref{TTOanomcontr} the anomalous contribution 
to the trace Ward identity \eqref{pTTOTWI} can be written
\begin{align}
\mathcal{A}^I_{\mu_2\nu_2} 
&= -2c^I\,\delta_{(\mu_1}^{\alpha_1}\delta_{\nu_2)\beta_1} \big(\ep_{\alpha_1\alpha_2\alpha_3}p_2^{\alpha_2}p_3^{\alpha_3}\big)\big(\ep_{\beta_1\beta_2\beta_3}p_2^{\beta_2}p_3^{\beta_3}\big).
\end{align}
Both this contribution and \eqref{TTOanomcontr} follow from  the full trace anomaly 
\[
\<T\>_s 
= -3 c^I \nabla_{[\alpha_1}\nabla^{[\alpha_1}\Big( R_{\alpha_2\alpha_3]}^{\alpha_2\alpha_3]}\phi^I\Big)
= 2 c^I G^{\mu\nu}\nabla_{\mu}\nabla_\nu\phi^I, 
\]
where $G_{\mu\nu}$ is the Einstein tensor,
as can easily be verified by writing $g_{\mu\nu}=\delta_{\mu\nu}+\delta g_{\mu\nu}$ and $h^\alpha_\beta = \delta^{\alpha\sigma}\delta g_{\beta\sigma}$, whereupon
\[
\<T\>_s = 6 c^I \,\p_{[\alpha_1}\p^{[\alpha_1}h_{\alpha_2}^{\alpha_2}\p_{\alpha_3]}\p^{\alpha_3]}\phi^I 
+O(h^2\phi).
\]

Unfortunately, as this anomaly is exact, it must however be of the trivial variety that can be removed through the addition of counterterms. 
These counterterms should be scale-invariant and give rise to a virial current $\mathcal{J}^\mu$ such that $\<T\>_{\mathrm{ct}} = \nabla_\mu\mathcal{J}^\mu$ cancels the anomaly.
Independently, we could have anticipated the triviality of this anomaly from the scheme-dependence of the coefficient $c^I$ in \eqref{divcIdef}.

To find the counterterms cancelling the anomaly,
notice that the square of the Weyl tensor in $d=3+2u\ep$ is
\begin{align}
W_d^2 
&= E_4 +
\frac{8u\ep}{1+2u\ep}R_{\mu\nu}R^{\mu\nu}-\frac{u\ep(3+2u\ep)}{(1+u\ep)(1+2u\ep)}R^2,
\end{align}
and the Weyl variation of the corresponding counterterm is
\[\label{Weylctvar}
\delta_\sigma\Big(\mathfrak{a}^I\int\D^{d}\x\sqrt{g}\mu^{(u+v_3)\ep} \phi^I W_d^2\Big) = (u+v_3)\ep\,\mathfrak{a}^I\int\D^{d}\x\sqrt{g}\mu^{(u+v_3)\ep} \phi^I W_d^2\sigma. 
\]
In the limit $\ep\rightarrow 0$, this variation then vanishes since the prefactor on the right-hand side is finite from \eqref{baboon}, while the Weyl tensor vanishes in three dimensions.
We recognise the first part of this counterterm as 
\eqref{Sdeg}, so the finite remainder must therefore supply the counterterms required to remove the anomaly,
\[\label{typeActs}
S_{\mathrm{ct}}  = -c^I\int\D^{d}\x\sqrt{g}\mu^{(u+v_3)\ep} \phi^I\Big(R_{\mu\nu}R^{\mu\nu}-\frac{3}{8}R^2\Big).
\]
Indeed, in three dimensions, 
these have exactly the Weyl variation we seek:
\[
\delta_\sigma S_{\mathrm{ct}} = 2c^I\int\D^3\x \sqrt{g}\,\sigma G^{\mu\nu}\nabla_\mu\nabla_\nu\phi^I.
\]

Finally, for completeness, let us  evaluate the renormalised correlator.  From our discussion above, we can effectively remove both the divergences in the original regulated form factors \eqref{anomdiv1}\,-\,\eqref{anomdiv3} and the anomaly through the combined counterterm
\[\label{weylcttypeA}
S_{\mathrm{ct}} = \mathfrak{c}^I\int\D^{d}\x\sqrt{g}\mu^{(u+v_3)\ep} \phi^I W_d^2, \qquad \mathfrak{c}^I = -\frac{c^I}{8u\ep}+c^I_{(0)}+O(\ep).
\]
This combined counterterm is also Weyl covariant (in the sense of \eqref{Weylcovonlysources}), 
as follows from \eqref{Weylctvar}.
Relabelling  $C_1^{(0)I} \rightarrow (\pi/2)^{3/2}(C_1^I/4)$ and removing the regulator, we then obtain the renormalised form factors 
\begin{align}\label{typeAA1}
A_1^I &= \frac{C_1^I}{a_{123}^4} \mathcal{E}_1 - 6 C_1^I\ln\frac{a_{123}^2}{\mu^2}+D_1^I,
\\[2ex]
A_2^I &=\frac{C_1^I}{a_{123}^3} \big( - \mathcal{E}_1 (a_{12} - p_3) + 2 \mathcal{E}_2 b_{12} \big)
+(p_1^2+p_2^2-p_3^2)\Big(6C_1^I\ln\frac{a_{123}^2}{\mu^2}-D_1^I\Big),
\\[2ex]
A_3^I &=\frac{C_1^I(a_{12}-p_3)}{4a_{123}^2} \big( \mathcal{E}_1 (a_{12} - p_3) - 4 \mathcal{E}_2 b_{12} \big) + \frac{1}{4}J^2 \Big(6C_1^I\ln\frac{a_{123}^2}{\mu^2}-D_1^I\Big),\label{typeAA3}
\end{align}
where the polynomials
\begin{align}
\mathcal{E}_1 & = a_{12}^2(a_{12}^2+12b_{12})+16a_{12}(a_{12}^2+3b_{12})p_3+6(7a_{12}^2+10b_{12})p_3^2+32a_{12}p_3^3+5p_3^4, \\[1ex]
\mathcal{E}_2 & = a_{12}^3  + 15 a_{12}^2 p_3 + 27 a_{12} p_3^2 + 5 p_3^3.
\end{align}
%

The scheme-dependent constant $D_1^I$ can be adjusted arbitrarily by rescaling $\mu$, and in fact these terms, as well as all logarithmic terms, are of the degenerate form \eqref{Fdeg1}.  When we reconstruct the renormalised correlator, all dependence on $D_1^I$ and $\mu$ therefore drops out and the result depends on $C_1^I$ only. 

This is also evident from the dilatation Ward identities, in which the right-hand sides are again of the degenerate form meaning the reconstructed correlator is scale-invariant:
\begin{align}\label{an_TTO1}
\mu\frac{\p}{\p\mu}A_1^I &= 12 C_1^I \\
\mu\frac{\p}{\p\mu}A_2^I &= -12 C_1^I (p_1^2+p_2^2-p_3^2)\\
\mu\frac{\p}{\p\mu}A_3^I &= -3C_1^I J^2. \label{an_TTO3}
\end{align}
The primary CWIs read
\begin{equation}\label{typeAprims}
\begin{array}{ll}
\K_{12} A_1^I = 0, &\qquad\qquad \K_{13} A_1^I = 0, \\
\K_{12} A_2^I = 0, &\qquad\qquad \K_{13} A_2^I = 8 A_1^I+48 C_1^I, \\
\K_{12} A_3^I = 0, &\qquad\qquad \K_{13} A_3^I = 2 A_2^I-24 C_1^I(p_1^2+p_2^2-p_3^2),
\end{array}
\end{equation}
while the secondary CWIs retain their original homogeneous form,
\begin{align}
\Lo_{2} A_1^I + \Ro A_2^I  = 0,
\qquad
\Lo_{2} A_2^I + 4 \Ro A_3^I =0.
\end{align}
The trace Ward identity \eqref{pTTOTWI} is satisfied with  $\mathcal{A}_{\mu\nu}^I=0$.

While two of the primary CWI in \eqref{typeAprims}  are apparently anomalous,  we know that in reality the anomaly has been removed by the counterterm \eqref{weylcttypeA}.   The `anomalous' terms appearing in these identities are in fact an artefact of the degeneracy: as discussed in \cite{Bzowski:2013sza}, special conformal transformations are represented, not by the $K_{ij}$ per se, but rather by these operators acting in combination with dilatations.
For the two `anomalous' primary CWI in \eqref{typeAprims}, the corresponding identities associated purely with special conformal transformations are
\begin{align}
0 = \K_{13} A_2^I + \frac{2}{p_3} \frac{\partial}{\partial p_3} \mathcal{D}_2 A_2^I - 8 A_1^I, \\
0 = \K_{13} A_3^I + \frac{2}{p_3} \frac{\partial}{\partial p_3} \mathcal{D}_4 A_3^I - 2 A_2^I,
\end{align}
where $\mathcal{D}_\alpha$ is the dilatation operator
\begin{equation}
\mathcal{D}_\alpha = - \alpha + \sum_{j=1}^3 p_j \frac{\partial}{\partial p_j} = - \mu \frac{\partial}{\partial \mu}.
\end{equation}
When evaluated on the solution \eqref{typeAA1}\,-\,\eqref{typeAA3}, these equations are satisfied  without any anomalous terms.  
Using \eqref{an_TTO1}\,-\,\eqref{an_TTO3}, we can then check that the two terms involving the dilatation operator $\mathcal{D}_\alpha$ are responsible for producing the `anomalous' terms in \eqref{typeAprims}.
These terms are thus simply the result of cross-contamination from the corresponding `anomalous' terms in the dilatation Ward identities \eqref{an_TTO1}\,-\,\eqref{an_TTO3}, which are manifestly of the degenerate form.
When the renormalised correlator is reconstructed from its constituent form factors, it therefore obeys the homogeneous conformal Ward identities in their full tensorial form.

\subsubsection{\texorpdfstring{$d=3$ and general $\Delta_3$}{d=3 and general Delta}}
\label{TTOchi}

As we saw in section \ref{JJOchi}, additional simplifications in $d=3$ can be obtained through use of the scheme $u=v_1=v_2=0$ with $v_3\neq 0$.  In this scheme all triple-$K$ integrals appearing in our solution 
\eqref{a:TTO1}\,-\,\eqref{a:TTO3} for the regulated form factors have $\beta_1=\beta_2=3/2$.
These can then converted to integrals with $\beta_1=\beta_2=1/2$ using the identity
\begin{align}
I_{\alpha\{\frac{3}{2},  \frac{3}{2}, \beta_3\}} 
&= I_{\alpha-2\{\frac{1}{2},\frac{1}{2},\beta_3\}}+(p_1+p_2)I_{\alpha-1\{\frac{1}{2},\frac{1}{2},\beta_3\}}+p_1p_2I_{\alpha\{\frac{1}{2},\frac{1}{2},\beta_3\}},
\end{align}
which follows from writing out all half-integer Bessel functions as elementary functions.  Making repeated use of our earlier identities \eqref{3dId1}\,-\,\eqref{3dId2}, one can then show that our solution for the regulated form factors,  \eqref{a:TTO1}\,-\,\eqref{a:TTO2} plus \eqref{secsolTOO1}, is equivalent to
\begin{align}\label{TTOF1}
A_1^I &= p_1p_2 \F_1^I + \F_2^I,\\
A_2^I &=  -2\chi p_1p_2 \F_1^I+2(p_1p_2-\chi) \F_2^I,\\
A_3^I &= \chi^2 p_1p_2 \F_1^I+\chi(\chi-2p_1p_2) \F_2^I,\label{TTOF3}
\end{align}
with $\chi$ as given in \eqref{chidef} and 
\begin{align}
\F_1^I &=C_1^I I_{\frac{9}{2}\{\frac{1}{2},\frac{1}{2},\Delta_3-\frac{3}{2}+v_3\ep\}},\\
\F_2 &= C_1^I I_{\frac{5}{2}\{\frac{1}{2},\frac{1}{2},\Delta_3-\frac{3}{2}+v_3\ep\}}+(p_1+p_2)C_1^I I_{\frac{7}{2}\{\frac{1}{2},\frac{1}{2},\Delta_3-\frac{3}{2}+v_3\ep\}}.\label{F2exp}
\end{align}
Since 
\[
2(p_1p_2-\chi)=-(p_1^2+p_2^2-p_3^2), \qquad \chi(\chi-2p_1p_2)=-J^2/4,
\]  
all terms proportional to $\F_2^I$ are thus of the degenerate form \eqref{Fdeg1}.  When the correlator is reconstructed from the form factors, the result is thus given by $\F_1^I$ times a single universal tensor structure that is independent of $\Delta_3$. Due to the degeneracy, this structure can be written in a number of equivalent ways.  Perhaps the most efficient is to use the degeneracy to set $A_1^I$ to zero, whereupon we obtain
\begin{align}\label{TTOuniversal}
& \lla t_{\mu_1 \nu_1}(\bs{p}_1) t_{\mu_2 \nu_2}(\bs{p}_2) \mathcal{O}^I(p_3)
\rra = \nn\\[0.5ex]&\qquad\qquad 
-2p_1^2 p_2^2 \F_1^I \,\Pi_{\mu_1 \nu_1\alpha\, \beta_1}(\bs{p}_1) \Pi_{\mu_2
\nu_2}{}^{\alpha}{}_{\beta_2}(\bs{p}_2) \Big(
  p_2^{\beta_1}
p_3^{\beta_2} -\chi  \delta^{\beta_1 \beta_2}
\Big).
\end{align}
Projecting into a helicity basis as described in section 8.1.1 of \cite{Bzowski:2013sza}, we find
\[
\lla T^{(s_1)}(\bs{p}_1)T^{(s_2)}(\bs{p}_2)\O^I(\bs{p}_3)\rra = \frac{1}{2}p_1p_2\chi^2 \F_1^I \delta^{s_1 s_2}
\]
where the helicities $s_1$ and $s_2$ take values $\pm 1$.  The correlator thus vanishes for opposite helicities. 
More generally, even when we retain the degenerate $\F_2^I$ terms, the form factors satisfy 
\[\label{TTOchirel}
0 = \chi^2 A_1^I+\chi A_2^I+A_3^I.
\]
This remarkable relation is valid for {\it general} values of the scalar dimension $\Delta_3$.  A quick check shows it holds for all the specific cases $\Delta_3=1,3,4$ studied above.

Finally, \eqref{TTOF1}\,-\,\eqref{TTOF3} also tells us  the functional dependence of the form factors on the symmetric polynomials $a_{12}=p_1+p_2$  and $b_{12}=p_1p_2$.
Since 
\[
I_{\alpha\{\frac{1}{2},\frac{1}{2},\beta_3\}}
=\frac{\pi}{2}\int_0^\infty\D x\, x^{\alpha} e^{-(p_1+p_2)x}p_3^{\beta_3}K_{\beta_3}(p_3 x),
\]
both $\F_1^I$ and $\F_2^I$ are functions of $a_{12}$ and $p_3$ only, and indeed the same is true of $\chi$.   
The dependence of the form factors on $b_{12}$ is thus limited to that appearing explicitly in \eqref{TTOF1}\,-\,\eqref{TTOF3}.  

Our discussion thus far has been in the regulated theory.  Where divergences arise, the relation \eqref{TTOchirel} must hold order by order in the regulator $\ep$.  The counterterm contributions removing these divergences must then satisfy \eqref{TTOchirel} also, at least up to finite scheme-dependent terms.  The relation \eqref{TTOchirel} then extends to the renormalised form factors, as indeed we saw for the case of $\Delta_3=4$ in section \ref{typeAex1} above.

The occurrence of a 0/0 limit for this case is also clear, since $\F_2^I$ has a pole from the $(---)$ singularity of the first triple-$K$ integral in \eqref{F2exp}, but multiplies the degenerate combination of form factors. 
Similar 0/0 limits will clearly arise for all other  $\Delta_3$ such that $\F_2^I$ is singular, however the residual scheme-dependence of the result suggests that no genuine type A anomalies can arise.

\subsubsection{\texorpdfstring{$d = 4$ and $\Delta_3 = 2$}{d=4 and Delta=2}}

Here, all triple-$K$ integrals multiplying the primary constants $C_2^I$ and $C_3^I$ have $\ep^{-2}$ poles, due to the presence of both $(---)$ and $(--+)$ singularities.  
The secondary Ward identities \eqref{secsolTOO1}\,-\,\eqref{secsolTOO2}, however, tell us that these primary constants are suppressed by a factor of $\ep$ relative to $C_1^I$.  To leading order,
\[\label{C2C3TTO}
C_2^I = \big(4(u-v_3)\ep+O(\ep^2)\big)C_1^I,\qquad C_3^I =\big(4(u-v_3)\ep+O(\ep^2)\big)C_1^I.
\]
The corresponding contributions to the form factors $A_2^I$ and $A_3^I$ are therefore only linearly divergent.
Meanwhile, of the triple-$K$ integrals multiplying the primary constant $C_1^I$, those appearing in the form factors $A_2^I$ and $A_3^I$ have single poles (since only $(---)$ singularities are present), while the triple-$K$ integral for the form factor $A_1^I$ is finite.

As no counterterms are available (see table \ref{tab:physTTO}), we now anticipate a grand cancellation of singularities.    
First, for the form factor $A_2^I$, we have
\begin{align}
J_{3\{001\}} &= I_{4+u\ep\{2+u\ep,2+u\ep,1+v_3\ep\}}=
-\frac{4}{(u+v_3)\ep}+O(\ep^0),\\[1ex]
J_{2\{000\}} &=I_{3+u\ep\{2+u\ep,2+u\ep,v_3\ep\}}
=\frac{4}{(u^2-v_3^2)\ep^2}+O(\ep^{-1}).
\end{align}
From \eqref{C2C3TTO} and \eqref{a:TTO2}, the two pole contributions to $A_2^I$  then cancel as required.
Next, for the form factor $A_3^I$, the leading singularities are
\begin{align}
J_{2\{002\}} &= I_{3+u\ep\{2+u\ep,2+u\ep,2+v_3\ep\}} = \frac{2}{(u+v_3)\ep}(p_1^2+p_2^2+p_3^2)+O(\ep^0),
\\[1ex]
J_{1\{001\}} &=I_{2+u\ep\{2+u\ep,2+u\ep,1+v_3\ep\}}= -\frac{2}{(u^2-v_3^2)\ep^2}p_3^2+O(\ep^{-1}),\\[1ex]
J_{0\{000\}} &= I_{1+u\ep\{2+u\ep,2+u\ep,v_3\ep\}} = -\frac{1}{(u^2-v_3^2)\ep^2}(p_1^2+p_2^2-p_3^2)+O(\ep^{-1}).
\end{align}
From \eqref{C2C3TTO} and \eqref{a:TTO3}, we then obtain three cancelling pole contributions to $A_3^I$.
Thus, all the regulated form factors are indeed finite as $\ep\rightarrow 0$.
 
Evaluating the subleading contributions to the triple-$K$ integrals, after removing the regulator and reabsorbing an overall numerical factor into $C_1^I$, we obtain the final result
\begin{align}
A_1^I & = \ren{1}^I \Big(2 - p_1 \frac{\partial}{\partial p_1} \Big) \Big(2 - p_2 \frac{\partial}{\partial p_2} \Big) p_1 p_2 \frac{\partial^2}{\partial p_1 \partial p_2} I_{1\{000\}}, \\[2ex]
A_2^I & = 4 \ren{1}^I \Big[ \Big(2 - p_1 \frac{\partial}{\partial p_1} \Big) \Big(2 - p_2 \frac{\partial}{\partial p_2} \Big) I_{2 \{111\}}^{\text{(fin)}} - \frac{2}{3} \ln \frac{p_1^2 p_2^2}{p_3^4} + \frac{7}{3} \Big], \\[2ex]
A_3^I & = \frac{96 \ren{1}^I p_1^4 p_2^4 p_3^4 }{J^4}I_{1\{000\}} \\[1ex]
& \quad - \frac{2 \ren{1}^I}{J^4} \Big[ p_1^4 (p_1^2 - p_2^2 - p_3^2)(J^2+6p_2^2p_3^2) 
\ln \frac{p_1^2}{p_3^2} + (p_1 \leftrightarrow p_2) \Big] \nn\\[1ex]
& \quad + \frac{\ren{1}^I}{J^2} \Big[ 3 a_{12}^6 - a_{12}^4 (18 b_{12} + 7 p_3^2) + a_{12}^2 (24 b_{12}^2 + 28 b_{12} p_3^2 + 5 p_3^4) - 28 b_{12}^2 p_3^2 - 10 b_{12} p_3^4 - p_3^6 \Big].\nn
\end{align}
The integrals $I_{2\{111\}}^{\text{(fin)}}$ and  $I_{1\{000\}}$ are given in \eqref{I2111fin0} and \eqref{I1000}.

\subsubsection{\texorpdfstring{$d = 4$ and $\Delta_3 = 4$}{d=4 and Delta=4}}

From table \ref{tab:TTO}, the triple-$K$ integrals appearing in the form factors $A_1$ and $A_2$  diverge as $\ep^{-1}$ due to the presence of $(---)$ singularities.  Those entering the form factor $A_3$  diverge as $\ep^{-2}$ due to the presence of $(---)$ and $(+--)$ singularities. ($J_{0\{000\}}$ in addition has a $(--+)$ singularity.)
Evaluating this latter form factor  explicitly, we find
\begin{equation}
A_3^I = \frac{1}{2 (u^2 - v_3^2) \: \epsilon^2} \left[ -  \big(48 \3{1}^{(0)I} + 4 \3{2}^{(0)I} + \3{3}^{(0)I}\big)(p_1^4 + p_2^4)+\3{3}^{(0)I} p_3^4  \right] + O(\epsilon^{-1}),
\end{equation}
where we have expanded all primary constants as
\[
C_j^I = C_j^{(0)I}+\ep C_j^{(1)I}+\ep^2 C_j^{(2)I}+O(\ep^3)
\]
for $j=1,2,3$.
From \eqref{secsolTOO1}\,-\,\eqref{secsolTOO2},
on the other hand, the secondary Ward identities
tell us that
\begin{align}
C_2^{(0)I}&=-12C_1^{(0)I}, \\
C_2^{(1)I} &= -12 C_1^{(1)I}+4(u-2v_3)C_1^{(0)I},\\
C_2^{(2)I} &= -12 C_1^{(2)I}+4(u-2v_3)C_1^{(1)I} + (u^2 - v_3^2) C_1^{(0)I},\\
C_3^{(0)I}&=0,\\
C_3^{(1)I} &= -12(u-v_3)C_1^{(0)I}, \\
C_3^{(2)I} &= -12(u-v_3)C_1^{(1)I} + (u-v_3)(u - 11 v_3) C_1^{(0)I}
\end{align}
and so we see immediately that the leading $\ep^{-2}$ pole in $A_3^I$ in fact vanishes.
The remaining divergences are now all of order $\ep^{-1}$, and read
\begin{align}\label{p:A1div}
A_1^I &= -\frac{8C_1^{(0)I}}{(u+v_3)\ep}+O(\ep^0),\\[1ex]  
\label{p:A2div}
A_2^I &= \frac{8 C_1^{(0)I}}{(u+v_3)\ep}(p_1^2+p_2^2 -p_3^2)+O(\ep^0),\\[1ex]\label{p:A3div}  
A_3^I &= \frac{2C_1^{(0)I}}{(u+v_3)\ep}(
J^2-2p_1^2p_2^2)+O(\ep^{0}).
\end{align}
To eliminate these remaining divergences, we have at our disposal the $(---)$ counterterm
\[\label{Weylctpear}
S_{\mathrm{ct}}=
\mathfrak{c}^I \int\D^{4+2u\ep}\x\sqrt{g}\,\phi^I W_d^2 \mu^{(u+v_3)\ep}.
\]
Here, we use the Weyl tensor in $d=4+2u\ep$ dimensions, rather than four, to ensure the counterterm action is Weyl covariant ({\it i.e.,} of the form \eqref{Weylcovonlysources}).  
Using
\[
W_d^2 = W_4^2 +u\ep (W_4^2-E_4+\frac{1}{9}R^2)+O(\ep^2),
\]
and the results of 
appendix \ref{sec:evalctcontr}, we then obtain the counterterm contributions
\begin{align}\label{TTOctcontr}
A^{\mathrm{ct}\,I}_1 &= 8\mathfrak{c}^I \mu^{(u+v_3) \epsilon},\\[1ex]
A^{\mathrm{ct}\,I}_2 &= -8\mathfrak{c}^I(p_1^2+p_2^2-p_3^2) \mu^{(u+v_3) \epsilon},\\[1ex]
A^{\mathrm{ct}\,I}_3 &= 2\mathfrak{c}^I \Big(2(1+u\ep) p_1^2p_2^2- J^2
\Big) \mu^{(u+v_3) \epsilon}.
\end{align}
We can now eliminate the divergences by choosing
\begin{equation}\label{acchoice}
\mathfrak{c}^I = \frac{\3{1}^{(0)I}}{ (u+v_3) \epsilon} + \mathfrak{c}^{(0)I} + O(\epsilon).
\end{equation}
Relabelling $C_1^{(0)I}\rightarrow C_1^I$ and
evaluating the renormalised form factors, we find
\begin{align}\label{p:A1renTTO}
A_1^I & = \ren{1}^{I} \Big(2 - p_1 \frac{\partial}{\partial p_1} \Big) \Big(2 - p_2 \frac{\partial}{\partial p_2} \Big) \Big(2 - p_3 \frac{\partial}{\partial p_3} \Big) I_{2\{111\}}^{\text{(fin)}} \nn\\[0.5ex]
& \quad - \: \frac{4}{3} \ren{1}^{I} \Big[ \ln \frac{p_1^2}{\mu^2} + \ln \frac{p_2^2}{\mu^2} + \ln \frac{p_3^2}{\mu^2} \Big] - 8 \ren{1}^{I} - 4 \sd{1}^I, \\[2ex]
A_2^I & = 4 \ren{1}^{I} \Big(1 - p_3 \frac{\partial}{\partial p_3} \Big) I_{3\{222\}}^{\text{(fin)}} \nn\\
& \quad + \: 2 \ren{1}^{I} \Big[ (p_1^2 + p_2^2) \ln \frac{p_3^2}{\mu^2} + (p_1^2 - p_3^2) \ln \frac{p_2^2}{\mu^2} + (p_2^2 - p_3^2) \ln \frac{p_1^2}{\mu^2} \Big] \nn\\
& \quad + \: 4 \ren{1}^{I} p_3^2 + \: 4 (p_1^2 + p_2^2 - p_3^2) \sd{1}^I, \\[2ex]
A_3^I & = 2 \ren{1}^{I} p_3^2 I_{3\{222\}}^{\text{(fin)}} + \ren{1}^{I} \Big[ ( p_2^2 p_3^2 - p_1^4 + p_3^4 ) \ln \frac{p_1^2}{\mu^2} + ( p_1^2 p_3^2 - p_2^4 + p_3^4 ) \ln \frac{p_2^2}{\mu^2} \nn\\
& \qquad  + \: p_3^2( p_1^2 + p_2^2  - 3 p_3^2 ) \ln \frac{p_3^2}{\mu^2} - 2 (p_1^4 + p_2^4 -p_1^2p_2^2+ 2 p_3^4) \Big]\nn\\&\quad 
 + ( J^2 - 2 p_1^2 p_2^2) \sd{1}^I, \label{p:A3renTTO}
\end{align}
where the finite integrals $I_{2\{111\}}^{\text{(fin)}}$ and $I_{3\{222\}}^{\text{(fin)}}$ are given in \eqref{I2111fin} and \eqref{I3222fin}.
The scheme-dependent constant $\sd{1}^I$ 
is a linear combination of the regulated theory data, 
\begin{equation}
\sd{1}^I = -2 \mathfrak{c}^{(0)I} + \frac{2\3{1}^{(1)I}}{(u+v_3)} + \frac{u(3 - 2 \gamma_E + 2 \ln  2) - v_3}{u + v_3} \3{1}^{(0)I}.
\end{equation}
In the renormalised theory, $D_1^I$ can be shifted arbitrarily by rescaling $\mu$.

\paragraph{Anomalous CWI.} 
Due to the counterterms, the renormalised form factors satisfy the anomalous dilatation Ward identities 
\begin{align}\label{anomDWITTO1}
\mu \frac{\partial}{\partial \mu} A_1^I & = 8 \ren{1}^I, \\[1ex]
\mu \frac{\partial}{\partial \mu} A_2^I & = -8 \ren{1}^I (p_1^2 + p_2^2-p_3^2), \\[1ex]
\mu \frac{\partial}{\partial \mu} A_3^I & = 2 \ren{1}^I (2 p_1^2 p_2^2 - J^2).\label{anomDWITTO3}
\end{align}
The conformal Ward identities are similarly anomalous: the primary CWI read
\begin{align}
& \K_{12} A_1^I = 0, && \K_{13} A_1^I = 0, \nn\\[1ex]
& \K_{12} A_2^I = 0, && \K_{13} A_2^I = 8 A_1^I + 32 \ren{1}^I, \\[1ex]
& \K_{12} A_3^I = 0, && \K_{13} A_3^I = 2 A_2^I - 16 \ren{1}^I (  p_1^2 + p_2^2 -  p_3^2), \nn
\end{align}
while the secondary CWI retain their original homogeneous form
\begin{align}
\Lo_{2} A_1^I + \Ro A_2^I = 0, \qquad
\Lo_{2} A_2^I + 4 \Ro A_3^I = 0.
\end{align}

From \eqref{anomformulaonlysources}, 
the anomaly action is 
\[\label{TTOanomact}
A = -C_1^I \int\D^4\x\sqrt{g}\,\phi^IW_4^2,
\]
and from \eqref{dmuW}, the anomalous dilatation Ward identities \eqref{anomDWITTO1}\,-\,\eqref{anomDWITTO3} are  equivalent to 
\begin{align}
\mu\frac{\p}{\p\mu}\<T_{\mu_1\nu_1}(\x_1)T_{\mu_2\nu_2}(\x_2)\O^I(\x_3)\> = -4\frac{\delta^3A}{\delta g^{\mu_1\nu_1}(\x_1)\delta g^{\mu_2\nu_2}(\x_2)\delta\O^I(\x_3)}\Big|_0.
\end{align}

On a conformal manifold, \eqref{TTOanomact} represents a moduli-dependent shift $c\rightarrow c-C_1^I\phi^I$ in the $c$-coefficient of the Weyl anomaly, as discussed recently in  \cite{Nakayama:2017oye}.  Since $c=-C_{TT}/2$, this is equivalent to a shift 
$C_{TT}\rightarrow C_{TT}+2C_1^I\phi^I$ in the 2-point normalisation entering the quadratic anomaly action \eqref{e:tranom}.
The same result also follows from conformal perturbation theory, where one finds the renormalised 2-point function takes the expected form \eqref{e:2ptTTren}, but with $C_{TT}$ shifted as above.\footnote{For this analysis, one evaluates
$\lla T_{\mu_1\nu_1}(\bs{p})T_{\mu_2\nu_2}(-\bs{p})\O(\bs{0})\rra = A_3(p,p,0)\Pi_{\mu_1\nu_1\mu_2\nu_2}(\bs{p})$ using \eqref{p:A3renTTO}.}

Finally, the anomalous contribution $\mathcal{A}_{\mu_2\nu_2}^I$ appearing in the trace Ward identity \eqref{pTTOTWI} can be obtained by functionally differentiating the anomaly action  \eqref{TTOanomact} with respect to the metric and the scalar source $\phi^I$.   
As this action is quadratic in the metric perturbations, however, $\mathcal{A}_{\mu_2\nu_2}^I$ vanishes and hence
\[
\lla T(\bs{p}_1) T_{\mu_2 \nu_2}(\bs{p}_2) \mathcal{O}^I(\bs{p}_3) \rra =  0.
\]

\subsection{\texorpdfstring{$\< T^{\mu_1 \nu_1} J^{\mu_2} \mathcal{O} \>$}{<TJO>}}

\paragraph{Decomposition.} The transverse and trace Ward
identities are
\begin{align}
 p_1^{\nu_1} \lla T_{\mu_1 \nu_1}(\bs{p}_1) J^{\mu_2 a}(\bs{p}_2)
\mathcal{O}^I(\bs{p}_3) \rra &= 0, \label{p1TJO} \\[1ex]
 p_{2 \mu_2} \lla T_{\mu_1 \nu_1}(\bs{p}_1) J^{\mu_2 a}(\bs{p}_2)
\mathcal{O}^I(\bs{p}_3) \rra &= 0, \label{p2TJO} \\[1ex]
 \lla T(\bs{p}_1) J^{\mu_2 a}(\bs{p}_2) \mathcal{O}^I(\bs{p}_3) \rra &= 0, \label{p3TJO}
\end{align}
so the 3-point function is purely transverse,
\begin{equation}
\lla T^{\mu_1 \nu_1}(\bs{p}_1) J^{\mu_2 a}(\bs{p}_2) \mathcal{O}^I(\bs{p}_3) \rra = \lla t^{\mu_1 \nu_1}(\bs{p}_1) j^{\mu_2 a}(\bs{p}_2) \mathcal{O}^I(\bs{p}_3) \rra.
\end{equation}

\paragraph{Form factors.} The tensor structure can be decomposed as
\begin{equation}
\lla t^{\mu_1 \nu_1}(\bs{p}_1) j^{\mu_2 a}(\bs{p}_2) \mathcal{O}^I(\bs{p}_3)
\rra = \Pi^{\mu_1 \nu_1}_{\alpha_1 \beta_1}(\bs{p}_1)
\pi^{\mu_2}_{\alpha_2}(\bs{p}_2) \left[ A_1^{aI} p_2^{\alpha_1} p_2^{\beta_1}
p_3^{\alpha_2} + A_2^{aI} \delta^{\alpha_1 \alpha_2} p_2^{\beta_1} \right].
\end{equation}
The form factors $A_1$ and $A_2$ are functions of  the momentum magnitudes, with no symmetries under permutations of these momenta.
Given $\lla T^{\mu_1 \nu_1}(\bs{p}_1) J^{\mu_2 a}(\bs{p}_2) \mathcal{O}^I(\bs{p}_3)
\rra$, with momenta chosen according to our cyclic rule \eqref{a:momenta}, the form factors can be extracted using
\begin{align}
A_1^{aI} & = \text{coefficient of } p_2^{\mu_1} p_2^{\nu_1} p_3^{\mu_2}, \\
A_2^{aI} & = 2 \cdot \text{coefficient of } \delta^{\mu_1 \mu_2} p_2^{\nu_1}.
\end{align}

\paragraph{Primary CWIs.} The primary CWIs are
\begin{equation}
\begin{array}{ll}
\K_{12} A_1^{aI} = 0, & \qquad\qquad \K_{13} A_1^{aI} = 0, \\
\K_{12} A_2^{aI} = 0, & \qquad\qquad \K_{13} A_2^{aI} = 4 A_1^{aI}.
\end{array}
\end{equation}
The solution in terms of triple-$K$ integrals is
\begin{align}
A_1^{aI} & = \3{1}^{aI} J_{3 \{000\}}, \label{A1TJO} \\
A_2^{aI} & = 2 \3{1}^{aI} J_{2 \{001\}} + \3{2}^{aI} J_{1 \{000\}}, \label{A2TJO}
\end{align}
where $\3{j}^{aI}$, $j=1,2$ are constants.

\paragraph{Secondary CWIs.} The independent secondary CWIs are
\begin{align} \label{TJO_fullSec1}
& \Lo_{2} A_1^{aI} + \Ro A_2^{aI} = 2 d \cdot \text{coefficient of }
p_2^{\mu_1} p_3^{\mu_2} \text{ in } p_{1 \nu_1} \lla T^{\mu_1 \nu_1}(\bs{p}_1)
J^{\mu_2 a}(\bs{p}_2) \mathcal{O}^I(\bs{p}_3) \rra, \\[1ex]
& \Lo'_{1} A_1^{aI} + 2 \Ro' A_2^{aI}  \nn\\
& \qquad = - 2 (d - 2) \cdot \text{coefficient of } p_2^{\mu_1} p_2^{\nu_1}
\text{ in } p_{2 \mu_2} \lla T^{\mu_1 \nu_1}(\bs{p}_1) J^{\mu_2 a}(\bs{p}_2)
\mathcal{O}^I(\bs{p}_3) \rra, \\[1ex]
& \Lo_{2} A_2^{aI} = 4 d \cdot \text{coefficient of } \delta^{\mu_1 \mu_2}
\text{ in } p_{1 \nu_1} \lla T^{\mu_1 \nu_1}(\bs{p}_1) J^{\mu_2 a}(\bs{p}_2)
\mathcal{O}^I(\bs{p}_3) \rra, \label{TJO_fullSec3}
\end{align}
where all right-hand sides vanish using \eqref{p1TJO} and \eqref{p2TJO}. This leads to an over-determined system of equations. Evaluating  the soft limit $p_3 \rightarrow 0$ using \eqref{e:Jid1} and \eqref{e:limJ}, keeping generic values of dimensions, we find
\begin{align}
0 & = \3{2}^{aI} + \Delta_3 (\Delta_3 + 2 - d) \3{1}^{aI}, \label{TJOsec1} \\
0 & = \3{2}^{aI} + \frac{1}{2} (\Delta_3 + 2)(\Delta_3 + 2 - d) \3{1}^{aI}, \\
0 & = \3{2}^{aI} + 2 (2 \Delta_3 - d) \3{1}^{aI}, \label{TJOsec3}
\end{align}
which requires\footnote{While $\Delta_2=2$ with  $v_3=-u$ appears to yield a solution, for this specific case the derivation leading to \eqref{TJOsec1}\,-\,\eqref{TJOsec3} is invalidated by the presence of singularities in \eqref{e:l}.  A careful analysis of the  soft limit for this specific case shows instead that the secondary CWIs are {\it not} satisfied.}  
\begin{equation} \label{TJOzerosol}
\3{1}^{aI} = \3{2}^{aI} = 0.
\end{equation}
The correlation function therefore vanishes:
\begin{equation}
\lla T_{\mu_1 \nu_1}(\bs{p}_1) J^{\mu_2 a}(\bs{p}_2) \mathcal{O}^I(\bs{p}_3) \rra = 0.
\end{equation}
This result is consistent with our original analysis in \cite{Bzowski:2013sza} after taking into account the different definition of the 3-point function employed here.\footnote{In \cite{Giombi:2011}, 
a nontrivial result for this correlator was proposed based on a position space analysis (at non-coincident points) for the case $d=3$ and $\Delta_3=1$.  
Our results indicate however that this is only possible  by relaxing {\it both} of the transverse Ward identities \eqref{p1TJO} and \eqref{p2TJO} to allow additional semilocal terms.}  

\section{Anomalous terms in the conformal Ward identities}  
\label{sec:AnomalousCWIs}

In this section, we return to our goal of understanding  
the {\it form} of the anomalous conformal Ward identities obeyed by the renormalised correlators.   
In our analysis above, we found the inhomogeneous terms entering these identities simply by inserting the renormalised form factors back into the original homogeneous 
Ward identities.
Here, our aim is to understand 
how these anomalous terms arise from the underlying beta functions and anomalies. 

For concreteness, we  focus on the most interesting  of the cases encountered above, namely $\<J^\mu\O\O\>$ for $d=\Delta_2=\Delta_3=4$, which features 
both an anomaly and a beta function.  
The general principles of our discussion can then be applied to other correlators. 
To set the scene, recall from  
equations \eqref{DWIA100}\,-\,\eqref{pJOOsecondaryCWIsv200} in section
\ref{sec:JOOd4analysis} that the renormalised form factor for this correlator obeys the anomalous dilatation Ward identity
\begin{align}\label{DWIA100_v2}
\mu\frac{\p}{\p\mu}A_1^{aI_2I_3} &=
4g(T^a_R)^{I_2I_3}\Big[
\frac{C_{\O\O}}{C_{JJ}}\Big(C_{JJ}\ln\frac{p_1^2}{\mu^2}+D_{JJ}\Big)p_1^2
-a_0 p_1^2
-C_{\O\O}(p_2^2+p_3^2)\Big],
\end{align}
the anomalous primary CWIs
\[\label{pJOOprimaryCWIs00_v2}
K_{23}A_1^{a I_2 I_3}=0,\qquad
K_{12}A_1^{a I_2 I_3}= 8a_0\, g(T_R^a)^{I_2I_3}, 
\]
and the anomalous secondary CWI 
\begin{align}\label{pJOOsecondaryCWIsv200_v2}
L_{1}A_1^{a I_2 I_3} 
&= 4g(T^a_R)^{I_2I_3}\Big[
-\frac{C_{\O\O}}{C_{JJ}}\Big(C_{JJ}\ln\frac{p_1^2}{\mu^2}+D_{JJ}\Big)p_1^4 -\Big(C_{\O\O}\ln \frac{p_2^2}{\mu^2}+D_{\O\O}\Big)p_2^4
 \nn\\[1ex]&\qquad
+\Big(C_{\O\O}\ln \frac{p_3^2}{\mu^2}+D_{\O\O}\Big)p_3^4
+2C_{\O\O}p_1^2p_2^2
+a_0\, p_1^2(p_1^2+p_2^2-p_3^2)\Big].
\end{align}
In contrast, in the dimensionally regulated theory, all these right-hand sides are absent.
Our task is thus to understand the origin 
of these nontrivial right-hand sides.

\subsection{Generating relation}

We can view a conformal transformation as a diffeomorphism  followed by a Weyl rescaling.  The diffeomorphism is such as to produce an initial rescaling of the flat metric, $\delta g_{\mu\nu} = 2\partial_{(\mu}\xi_{\nu)}=(2/d)(\p\cdot\xi)\delta_{\mu\nu}$, which is then eliminated by a Weyl transformation of the opposing sign, $\delta g_{\mu\nu}=2\sigma\delta_{\mu\nu}$ with $\sigma=-(1/d)(\p\cdot\xi)$.   The net transformation of the sources, following our discussion of the trace Ward identity in section \ref{sec:Definition}, is then
\begin{align}
\delta g_{\mu\nu}&=0, \\  \delta A^a_\mu &= \xi^\nu \p_\nu A_\mu^a+A^a_\nu\p_\mu \xi^\nu-\frac{1}{d}(\p\cdot\xi)\beta_{A^a_\mu}, \\
 \delta\phi^I &= \xi^\mu\p_\mu \phi^I-\frac{1}{d}(\p\cdot\xi)\Big[-(d-\Delta^I)\phi^I+\beta_{\phi^I}\Big].
\end{align}
In particular, the beta function contributions $\beta_{A^a_\mu}$ and $\beta_{\phi^I}$ have expansions beginning at quadratic order in the sources.  (We will obtain explicit expressions for these shortly.)
The renormalised generating functional, meanwhile, transforms anomalously as
\[
\delta W = -\frac{1}{d}\int \D^d \x \,(\p\cdot\xi)\mathcal{A}.
\]
Combining these equations and integrating by parts, we obtain the generating relation
\begin{align}
0 &= \int \D^d\x \,\Big[\frac{1}{d}(\p\cdot\xi)\mathcal{A}+\Big(\frac{1}{d}(\p\cdot\xi)(\Delta^I\phi^I+\beta_{\phi^I})+\phi^I\xi^\mu\p_\mu\Big)\<O^I\>_s\nn\\&\qquad\qquad\quad+\Big(\frac{1}{d}(\p\cdot\xi)(d A^a_\mu+\beta_{A^a_\mu})+A^a_\mu\xi^\nu\p_\nu-A^a_\nu\, (\partial_\mu\xi^\nu)\Big)\<J^{\mu a}\>_s\Big],
\end{align}
which holds about a flat background but with arbitrary scalar and gauge field sources.\footnote{
For correlators with stress tensor insertions, an analogous generating relation can be found by first functionally differentiating with respect to the metric to create the required insertions.}

The conformal Ward identities for the renormalised 3-point function now follow by functional differentiation.  Noting that once the sources are switched off, (i) all 1-point functions, and 2-point functions of different operators vanish; 
(ii) the beta functions and their first derivatives vanish (as they are of quadratic or higher order), we find
\begin{align}\label{JOO_CWI_general}
&\mathcal{L}^\mu_\nu\,\<J^{\nu a}(\x_1)\O^{I_2}(\x_2)\O^{I_3}(\x_3)\> 
\nn\\[1ex]&=
\int\D^d\x\,\frac{1}{d}(\p\cdot\xi)_{\x}\left(
\left(\frac{\delta^2\beta_{\phi^J}(\x)}{\delta\phi^{I_2}(\x_2)\delta A^a_\mu(\x_1)}\Big|_0 \<\O^{J}(\x)\O^{I_3}(\x_3)\>+(2\leftrightarrow 3)\right)
\right. \nn\\[1ex]&\qquad\qquad\left.
+\frac{\delta^2\beta_{A^b_\nu}(\x)}{\delta\phi^{I_2}(\x_2)\delta\phi^{I_3}(\x_3)}\Big|_0 \<J^{\nu b}(\x)J^{\mu a}(\x_1)\>  -\frac{\delta^3\mathcal{A}(\x)}{\delta A^a_\mu(\x_1)\delta \phi^{I_2}(\x_2)\delta \phi^{I_3}(\x_3)}\Big|_0
\right).
\end{align}
The subscript zero here denotes switching off the sources, and the differential operator 
\[
\mathcal{L}^\mu_\nu = \Big[\xi^\alpha(\x_1)\frac{\p}{\p x_1^\alpha}+(\p\cdot\xi)_{\x_1} + \sum_{i=2}^3\Big( \xi^\alpha(\x_i)\frac{\p}{\p x_i^\alpha}+\frac{1}{d}(\p\cdot\xi)_{\x_i}\Delta^{I_i}\Big)\Big]\delta^\mu_\nu
-\frac{\p\xi^\mu(\x_1)}{\p x_1^\nu}.
\]
For dilatations $\xi^\mu = \lambda x^\mu$, this operator reduces to 
$\mathcal{L}^\mu_\nu=-\lambda\delta^\mu_\nu \mathcal{D}$ where
\[
\mathcal{D} = -\Delta_t-\sum_{i=1}^3 x_i^\alpha\frac{\p}{\p x_i^\alpha}
\]
is the usual dilatation operator, with $\Delta_t= d-1+ \Delta^{I_2}+\Delta^{I_3}$ the total dimension of $\<J^\mu\O\O\>$.  
On dimensional grounds, we also have the relation $\mathcal{D}\<J^\mu\O\O\>= -\mu(\p/\p\mu)\<J^\mu\O\O\>$.

For special conformal transformations $\xi^\mu = x^2 b^\mu-2(b\cdot x)x^\mu$, we recover the expected special conformal operator for $\<J^\mu\O\O\>$.  This can be written as 
\[\label{LSct}
\mathcal{L}^\mu_\nu = b_\alpha \big[2x_1^\mu\delta^\alpha_\nu-2x_{1\nu}\delta^{\alpha\mu}+(2dx_1^\alpha
 - \mathcal{K}^\alpha)\delta^\mu_\nu
\big],
\]
where 
\[
\mathcal{K}^\alpha = \sum_{i=1}^3\Big[2\Delta_i x_i^\alpha+ \Big(2x_i^\alpha x_i^\kappa-x_i^2\delta^{\alpha\kappa}\Big)\frac{\p}{\p x_i^\kappa}\Big]
\]
is the special conformal operator that one would find when acting on correlators of three scalars.  (We take $\Delta_1=d-1$ and $\Delta_i=\Delta^{I_i}$ for $i=2,3$.)  The additional terms in \eqref{LSct} are a result of the vectorial nature of $\<J^\mu\O\O\>$.  

In summary then, the left-hand side of \eqref{JOO_CWI_general} is equivalent to the homogeneous conformal Ward identities.  The right-hand side consists of inhomogeneous terms that encode the breaking of conformal symmetry due to the beta functions and the conformal anomaly.  The precise form of these inhomogeneous terms can be determined from the nature of the counterterm action, as we discuss next. 

\subsection{Finding the beta function}
\label{sec:betasandanomaly}

We now focus specifically on the case at hand, $d=\Delta^{I_2}=\Delta^{I_3}=4$.
To preserve gauge invariance, as well as the symmetry of interchanging $\O^{I_2}$ and $\O^{I_3}$, we adopt a regularisation scheme where $u=v_1$ and $v_2=v_3$.
As we saw  in \eqref{JOO_Sct0},
including all terms up to cubic order in the sources, the counterterm action is then  
\begin{align}\label{JOO_Sct}
S_{\mathrm{ct}} &= \int\D^{4+2u\ep}\x\,\Big[\mathfrak{c}_{JJ}\mu^{2u\ep}F^{\mu\nu a}F^a_{\mu\nu}
+\mathfrak{c}_{\O\O} \mu^{2 v_2\ep}(D^2\phi^I)^2
\nn\\&\qquad
+\mathfrak{c}_1 \mu^{2(v_2-u)\ep}ig(T^a_R)^{IJ} J^{\mu a}\phi^{I} D_\mu \phi^{J}
+\mathfrak{c}_2 \mu^{2v_2\ep}ig(T^a_R)^{IJ}F^{\mu\nu a}D_\mu\phi^{I}D_\nu \phi^{J}\Big].
\end{align}
The coefficients $\mathfrak{c}_{JJ}$ and $\mathfrak{c}_{\O\O}$ are fixed from the renormalisation of the 2-point function, as given in \eqref{apple}.
In addition, we have of course the usual source terms
\[\label{JOO_Sources}
S_{\mathrm{source}} = \int\D^{4+2u\ep}\x\,\Big[\O^I\phi^I+J^{\mu a}A^a_{\mu}\Big].
\]

Recalling our earlier discussions in sections \ref{sec:anomaliesandbetafns} and \ref{sec:JOOd4analysis}, 
only those terms proportional to $\mathfrak{c}_{\O\O}$,  $\mathfrak{c}_1$ and $\mathfrak{c}_2$ contribute to $\<J^\mu\O\O\>_{\mathrm{ct}}$. 
The  $\mathfrak{c}_1$ counterterm  acts to renormalise the gauge field $A^a_\mu$ generating a nonzero beta function $\beta_{A^a_\mu}$.
As there are no cubic counterterms involving $\O^I$, however, the scalar source $\phi^I$ is not renormalised and $\beta_{\phi^I}$ vanishes.

Since differentiating the generating functional $W$ with respect to $A^a_\mu$ generates  renormalised current insertions,  $A^a_\mu$ is by definition the renormalised source.
The {\it bare} source corresponds instead to the overall coefficient multiplying $J^{\mu a}$ in the subtracted action,
\[
(A^\mathrm{bare})^a_\mu = A^a_\mu+\mathfrak{c}_1\mu^{2(v_2-u)\ep}ig(T^a_R)^{IJ}\phi^{I}D_\mu \phi^{J}.
\]
Since $(A^\mathrm{bare})^a_\mu$ and $\phi^I$ are independent of the renormalisation scale $\mu$, to quadratic order the beta function is then
\[\label{betaA}
\beta_{A^a_\mu} = \lim_{\ep\rightarrow 0}\, \mu\frac{\D}{\D\mu}A_\mu^a = -2(v_2-u)\mathfrak{c}_1^{(-1)}ig(T^a_R)^{IJ}\phi^{I}\p_\mu \phi^{J},
\]
where we expanded $\mathfrak{c}_1 = \mathfrak{c}_1^{(-1)}\ep^{-1}+\mathfrak{c}_1^{(0)}+O(\ep)$.
Alternatively, under a Weyl variation 
\[\label{Weylofphi}
\delta_\sigma \phi^I = (v_2-u)\ep \phi^I\sigma, \qquad \delta_\sigma (A^{\mathrm{bare}})^a_\mu = 0,
\]
hence 
\[\label{WeylofA}
\delta_\sigma A^a_\mu = -2(v_2-u)\ep\mu^{2(v_2-u)\ep} \mathfrak{c}_1 ig(T^a_R)^{IJ}\phi^I D_\mu\phi^J \sigma,
\]
where the $\partial_\mu\sigma$ term vanishes due to the antisymmetry\footnote{Starting from the gauge transformation $\delta\phi^I = -ig\alpha^a(T^a_R)^{IJ}\phi^J$, for $\phi^I$ and $\alpha^a$ to be real we require the combination $i(T^a_R)^{IJ}$ to be real.  Since $(T^a_R)^{IJ}$ is also Hermitian, we  have $i(T^a_R)^{IJ}=-i({T^{a}_R}^*)^{IJ}=-i(T^a_R)^{JI}$.}  of $(T^a_R)^{IJ}$.  The vanishing of this term is important since it ensures that 
\[\label{musigmaA}
\delta_\sigma A^a_\mu = \sigma \mu\frac{\D}{\D\mu}A^a_\mu.
\]
As we remove the regulator by sending $\ep\rightarrow 0$, we now recover $\delta_\sigma A^a_\mu = \sigma \beta_{A^a_\mu}$ which was our starting point in deriving the anomalous conformal Ward identities above.

\subsection{Finding the anomaly}
\label{sec:findingtheanomaly}

Having found the beta function, we now need to find the anomaly.
For this, we need to know the Weyl variation of the renormalised generating functional.
In appendix \ref{Weyl_appendix}, we show that imposing the relation
\[\label{specialreln}
(v_2-u)\ep \mathfrak{c}_2+2(1+v_2\ep)\mathfrak{c}_{\O\O}-4(v_2-u)\ep\mathfrak{c}_1\mathfrak{c}_{JJ}= 0
\]
ensures the Weyl covariance of the action for counterterms and sources:
\[\label{Weylcov}
\delta_\sigma (S_{\mathrm{ct}}+S_{\mathrm{source}}) = \int\D^{d}\x \sqrt{g}\,\sigma \mu\frac{\D}{\D\mu}(\mathcal{L}_{\mathrm{ct}}+\mathcal{L}_{\mathrm{source}}).
\]
Evaluating this expression to cubic order in the sources about a flat background then gives
\begin{align}\label{flatWeylvar}
\delta_\sigma (S_{\mathrm{ct}}+S_{\mathrm{source}}) &= 
\int\D^{4+2u\ep}\x\,\sigma\Big[2u\ep\mathfrak{c}_{JJ}\mu^{2u\ep}F^{\mu\nu a}F^a_{\mu\nu}
+2v_2\ep\mathfrak{c}_{\O\O}\mu^{2 v_2\ep}(D^2\phi^I)^2
\nn\\&\qquad
 +2\big(v_2\ep\mathfrak{c}_2 -4(v_2-u)\ep\mathfrak{c}_1\mathfrak{c}_{JJ}\big) \mu^{2v_2\ep}ig(T^a_R)^{IJ}F^{\mu\nu a}D_\mu\phi^{I}D_\nu \phi^{J}\Big],
\end{align}
where the final term proportional to $\mathfrak{c}_1\mathfrak{c}_{JJ}$ derives from the implicit $\mu$-dependence of the renormalised gauge field $A^a_\mu$ in the $F^{\mu\nu a}F^a_{\mu\nu}$ counterterm,
\begin{align}\label{Fsqvar}
\mu\frac{\D}{\D\mu} (F^{\mu\nu a}F^a_{\mu\nu})
=-8(v_2-u)\ep \mu^{2(v_2-u)\ep}\mathfrak{c}_1 ig(T^a_R)^{IJ}F^{\mu\nu a}D_\mu\phi^I D_\nu\phi^J.
\end{align}
Notice also that all dependence on the operators has dropped out, since
\[\label{scaledepofc0}
\mu\frac{\D}{\D\mu}\int\D^{4+2u\ep}\x\,\Big[\mathfrak{c}_1 \mu^{2(v_2-u)\ep}ig(T^a_R)^{IJ} \phi^{I} D_\mu \phi^{J}+A^a_\mu\Big]J^{\mu a}=\mu\frac{\D}{\D\mu}\int\D^{4+2u\ep}\x\,(A^{\mathrm{bare}})^a_\mu J^{\mu a} = 0.
\]
As the Weyl variation \eqref{flatWeylvar} thus depends only on the sources, we can pull it outside the path integral in the generating functional to obtain
\begin{align}
\delta_\sigma W &= \lim_{\ep\rightarrow 0}\Big[\delta_\sigma \ln\, \<e^{-S_{\mathrm{ct}}-S_{\mathrm{source}}}\>\Big] 
=\lim_{\ep\rightarrow 0}\Big[ -\delta_\sigma (S_{\mathrm{ct}}+S_{\mathrm{source}})\Big].
\end{align}
Since $W$ is the renormalised generating function, this Weyl variation must be finite as $\ep\rightarrow 0$.  From \eqref{flatWeylvar}, we see that $\mathfrak{c}_{JJ}$ and $\mathfrak{c}_{\O\O}$ can have at most single poles in $\ep$, consistent with the renormalisation of the 2-point functions.  In addition, the scheme-dependent constant 
\begin{align}
a_0 & = \lim_{\ep\rightarrow 0}\Big[(v_2\ep\mathfrak{c}_2 -4(v_2-u)\ep\mathfrak{c}_1\mathfrak{c}_{JJ}\big) \mu^{2v_2\ep}\Big]=\lim_{\ep\rightarrow 0}\Big[(u\ep\mathfrak{c}_2-2(1+v_2\ep)\mathfrak{c}_{\O\O})\mu^{2v_2\ep}\Big]
\end{align}
must be finite, where the second equation follows from  \eqref{specialreln}.
As $\mathfrak{c}_{\O\O}$ has a single pole, this requires that $\mathfrak{c}_2$ has a {\it double pole} with coefficient
\[\label{c2m2reln}
\mathfrak{c}_2^{(-2)} = \frac{2}{u}\mathfrak{c}_{\O\O}^{(-1)}=\frac{C_{\O\O}}{u v_2}.
\]
Solving \eqref{specialreln} at orders $\ep^{-1}$ and $\ep^0$ with the aid of  \eqref{sdtOO} and \eqref{sdtJJ}, we then find that $\mathfrak{c}_1$ has only a single pole, and obtain the relations
\begin{align}\label{c0m1reln}
\mathfrak{c}_1^{(-1)}&=\frac{C_{\O\O}}{(v_2-u)C_{JJ}},\\[2ex] \label{c2m1reln}
\mathfrak{c}_2^{(-1)}
&=\frac{C_{JJ}}{u}\mathfrak{c}_1^{(0)}+\frac{1}{(v_2-u)}\Big[\frac{C_{\O\O}}{C_{JJ}}(C_{JJ}^{(0)}-D_{JJ})+D_{\O\O}-C_{\O\O}^{(0)}-C_{\O\O}\Big], 
\end{align}
and hence
\begin{align}
\label{adefn}
a_0 
&=
C_{JJ}\mathfrak{c}_1^{(0)} + \frac{1}{(v_2-u)}\Big[\frac{u C_{\O\O}}{C_{JJ}}(C_{JJ}^{(0)}-D_{JJ})+v_2(D_{\O\O}-C_{\O\O}^{(0)}-C_{\O\O})\Big].
\end{align}
These relations are equivalent to those we obtained earlier through a direct analysis of the 3-point function divergences, namely \eqref{c0soln00}, \eqref{c2soln00} and \eqref{adefn00}, after using \eqref{Dreln} to eliminate terms proportional to $D_{JJ}$.

The anomaly action can now be read off using
\begin{align}\label{anomaly_formula}
A =\int\D^4\x\,\mathcal{A} 
=\lim_{\ep\rightarrow 0}\Big[-\delta_\sigma(S_{\mathrm{ct}}+S_{\mathrm{source}})\Big]_{\sigma=1}
=
\lim_{\ep\rightarrow 0}\Big[-\mu\frac{\D}{\D\mu}(S_{\mathrm{ct}}+S_{\mathrm{source}})\Big].
\end{align}
As we saw above, for the Weyl covariance \eqref{Weylcov} used in the last step here,
it is crucial to take into account the implicit $\mu$-dependence of the renormalised sources, both in $S_{\mathrm{ct}}$ and $S_{\mathrm{source}}$.
We thus find the anomaly action
\begin{align}\label{anomaly_action_JOO}
A&= -\int\D^{4}\x\,\Big[\frac{C_{JJ}}{2}F^{\mu\nu a}F^a_{\mu\nu}
+C_{\O\O}(D^2\phi^I)^2 
+2a_0\, ig(T^a_R)^{IJ}F^{\mu\nu a}D_\mu\phi^{I}D_\nu \phi^{J}\Big].
\end{align}
As we noted earlier in section \ref{sec:anomaliesandbetafns}, due to its scheme dependence $a_0$ does not parametrise a genuine anomaly.  Instead, this term is Weyl exact and can be obtained from the variation
\begin{align}\label{Weylexactness}
&\delta_\sigma\int \D^4\x \,\frac{a_0}{2}\Big(\frac{C_{JJ}}{C_{\O\O}}F^{\mu\nu a}F_{\mu\nu}^a+(D^2\phi^I)^2\Big) 
=-\int\D^4\x\, 2a_0\,i g (T^a_R)^{IJ}F^{\mu\nu a}D_\mu\phi^I D_\nu\phi^J \sigma,
\end{align} 
as can be seen from \eqref{cJJnoncov} and \eqref{c1noncov} in appendix \ref{Weyl_appendix}.

\subsection{Anomalous dilatation Ward identity}
\label{DWIsec}

Knowing the anomaly action and the beta functions, we now have all the ingredients we need to reconstruct the right-hand sides of the anomalous conformal Ward identities.
From our results \eqref{betaA} and \eqref{c0m1reln} above, the beta function is
\[\label{newbeta}
\beta_{A^a_\mu} =  -\frac{2C_{\O\O}}{C_{JJ}}\,ig(T^a_R)^{IJ}\phi^{I}\p_\mu \phi^{J}.
\]
Inserting this into the dilatation Ward identity following from \eqref{JOO_CWI_general}, we obtain
\begin{align}\label{JOO_DWI_a}
&\mathcal{D}\<J^{\mu a}(\x_1)\O^{I_2}(\x_2)\O^{I_3}(\x_3)\> \nn\\& =
\Big[\frac{2C_{\O\O}}{C_{JJ}}ig(T^b_R)^{I_2I_3}\<J^{\mu a}(\x_1)J^{\nu b}(\x_2)\>\frac{\p}{\p x_2^\nu}\delta(\x_2-\x_3) + (2\leftrightarrow 3)\Big] - (\mathcal{A}_{J\O\O})^{a \mu  I_2I_3 },
\end{align}
where the anomalous contribution 
\[\label{anomcontrdefn}
(\mathcal{A}_{J\O\O})^{a\mu I_2I_3}=-\frac{\delta^3A}{\delta A^a_\mu(\x_1)\delta \phi^{I_2}(\x_2)\delta \phi^{I_3}(\x_3)}\Big|_0. 
\]
Examining \eqref{anomaly_action_JOO}, we see in fact only the last two terms contribute.
Transforming to momentum space,  this Ward identity reads
\begin{align}\label{JOO_DWI_tensorial}
&\Big( 2d-\Delta_t +\sum_{i=2}^3 p_i^\nu\frac{\p}{\p p_i^\nu}\Big)
\lla J^{\mu a}(\bs{p}_1)\O^{I_2}(\bs{p}_2)\O^{I_3}(\bs{p}_3)\rra \nn\\[1ex] & \qquad \qquad\qquad =
-\frac{4C_{\O\O}}{C_{JJ}}g(T^b_R)^{I_2I_3}p_{2\nu}\lla J^{\nu b}(\bs{p}_1)J^{\mu a}(-\bs{p}_1)\rra-(\mathcal{A}_{J\O\O})^{a\mu I_2 I_3},
\end{align}
where the renormalised current 2-point function is given in \eqref{e:2ptJJren} and the anomalous contribution\footnote{Here, we have chosen $\bs{p}_2$ and $\bs{p}_3$ as independent momenta to make the $2\leftrightarrow 3$ symmetry manifest.
Despite appearances, \eqref{JOO_DWI_tensorial} is actually symmetric under $2\leftrightarrow 3$ since 
$(T_R^b)^{I_3I_2}=-(T_R^b)^{I_2 I_3}$ 
 and the current 2-point function is transverse.}
\begin{align}
\label{AJOOresult}
(\mathcal{A}_{J\O\O})^{a\mu I_2 I_3}
&= -2C_{\O\O} g(T^a_R)^{I_2I_3}(p_2^2+p_3^2)(p_2^\mu-p_3^\mu)-4a_0\, g(T^a_R)^{I_2I_3} p_1^2 \pi^{\mu\nu}(\bs{p}_1)p_{2\nu}.
\end{align}

To recover \eqref{DWIA100_v2}, 
we need to recast this dilatation Ward identity in terms of the form factor $A_1^{aI_2I_3}$.
From \eqref{JOOdecomp} and \eqref{JOOformfactors}, we recall the 3-point function takes the form
\begin{align}
&\lla J^{\mu a}(\bs{p}_1)\O^{I_2}(\bs{p}_2)\O^{I_3}(\bs{p}_3)\rra   = A_1^{aI_2I_3}\pi^{\mu\nu}(\bs{p}_1)p_{2\nu}
\nn\\[1ex]&\qquad\qquad
-\frac{p_1^\mu}{p_1^2}\Big[g(T^a_R)^{JI_3}\lla \O^{J}(\bs{p}_2)\O^{I_2}(-\bs{p}_2)\rra+g(T^a_R)^{JI_2}\lla \O^{J}(\bs{p}_3)\O^{I_3}(-\bs{p}_3)\rra\Big].
\end{align}
Inserting this decomposition into \eqref{JOO_DWI_tensorial}, we find the longitudinal part is automatically satisfied while the transverse part yields
\begin{align}\label{DWIA1}
\mu\frac{\p}{\p\mu}A_1^{aI_2I_3} &=-\Big[2d-\Delta_t+1 +\sum_{i=1}^3 p_i\frac{\p}{\p p_i} \Big] A_1^{aI_2I_3} \nn\\[1ex]&= 4g(T^a_R)^{I_2I_3}\Big[
\frac{C_{\O\O}}{C_{JJ}}\Big(C_{JJ}\ln\frac{p_1^2}{\mu^2}+D_{JJ}\Big)p_1^2
-a_0 p_1^2
-C_{\O\O}(p_2^2+p_3^2)\Big].
\end{align}
This is indeed the expected anomalous dilatation Ward identity \eqref{DWIA100_v2}.
The logarithm of the renormalisation scale $\mu$ appearing on the right-hand side means that the  form factor $A_1^{aI_2I_3}$ contains a product of such logarithms.  From \eqref{JOO_DWI_tensorial}, we see this behaviour is a result of the beta function  introducing a dependence on the renormalised current 2-point function.

\subsection{Anomalous special conformal Ward identities}

The special conformal Ward identity following from \eqref{JOO_DWI_tensorial} reads
\begin{align}\label{fullSCWIJOO}
&\big[2x_1^\mu\delta^\alpha_\nu-2x_{1\nu}\delta^{\alpha\mu}+(2dx_1^\alpha
 - \mathcal{K}^\alpha)\delta^\mu_\nu
\big]\<J^{\nu a}(\x_1)\O^{I_2}(\x_2)\O^{I_3}(\x_3)\>
 \nn\\[1ex]& \qquad=
\Big[\frac{4C_{\O\O}}{C_{JJ}}ig(T^b_R)^{I_2I_3}\<J^{\mu a}(\x_1)J^{\nu b}(\x_2)\>x_2^\alpha\frac{\p}{\p x_2^\nu}\delta(\x_2-\x_3) + (2\leftrightarrow 3)\Big] \nn\\[1ex]&\qquad\quad -2\frac{\delta^3}{\delta A^a_\mu(\x_1)\delta \phi^{I_2}(\x_2)\delta \phi^{I_3}(\x_3)}\Big|_0\int\D^d\x\,x^{\alpha}\mathcal{A}(\x)
\end{align}
where we used the beta function \eqref{newbeta}.
The anomalous contribution can be evaluated from the action \eqref{anomaly_action_JOO}, giving
\begin{align}
&\frac{\delta^3}{\delta A^a_\mu(\x_1)\delta \phi^{I_2}(\x_2)\delta \phi^{I_3}(\x_3)}\Big|_0 \int\D^d\x\,x^\alpha \mathcal{A}(\x)\nn\\[2ex]&= 
-2C_{\O\O}ig(T_R^a)^{I_2I_3}\big[2 x_1^\alpha \p_1^2\delta(\x_2-\x_1)\p_1^\mu \delta(\x_3-\x_1)+x_3^\alpha \p_3^\mu\delta(\x_1-\x_3)\p_3^2\delta(\x_2-\x_3) - (2\leftrightarrow 3)\big]\nn\\[2ex]&\quad
-4a_0\,ig(T_R^a)^{I_2I_3}\big[\p_1^\alpha\delta(\x_2-\x_1)\p_1^\mu\delta(\x_3-\x_1)+x_1^\alpha \p_{1\nu}(\p_1^\nu\delta(\x_2-\x_1)\p_1^\mu\delta(\x_3-\x_1)) - (2\leftrightarrow 3)\big].
\end{align}
The conversion of \eqref{fullSCWIJOO}
to momentum space is simplified by first translating $\x_1\rightarrow 0$, reducing the tensor structure on the left-hand side. To keep the $2\leftrightarrow 3$ symmetry manifest, we choose $\bs{p}_2$ and $\bs{p}_3$ as the independent momenta. The result is
\begin{align}\label{tensorSCWIJOO}
&\sum_{i=2}^3 \Big[p_i^\alpha\frac{\p}{\p p_i^\nu}\frac{\p}{\p p_{i\nu}}-2p_i^\nu\frac{\p}{\p p_i^\nu}\frac{\p}{\p p_i^\alpha}\Big]\lla J^{\mu a}(\bs{p}_1)\O^{I_2}(\bs{p}_2)\O^{I_3}(\bs{p}_3)\rra\nn\\[1ex]
&=\frac{4C_{\O\O}}{C_{JJ}}g(T^b_R)^{I_2I_3}\Big(p_2^\nu\frac{\p}{\p p_2^\alpha}-p_3^\nu\frac{\p}{\p p_3^\alpha}\Big)\lla J^{\mu a}(\bs{p}_1)J^{\nu b}(-\bs{p}_1)\rra\nn\\[1ex]&\quad 
-4C_{\O\O}g(T_R^a)^{I_2I_3}(p_2^2-p_3^2)\delta^{\alpha\mu}
+8 a_0\,g(T_R^a)^{I_2I_3}(p_2^\alpha p_3^\mu-p_3^\alpha p_2^\mu).
\end{align}
Inserting once again our form factor decomposition \eqref{JOOdecomp}, and using the renormalised current 2-point function \eqref{e:2ptJJren}, with some algebra one can show 
this Ward identity is equivalent to the primary CWIs
\[\label{pJOOprimaryCWIs}
K_{23}A_1^{a I_2 I_3}=0,\qquad
K_{12}A_1^{a I_2 I_3}= 8a_0\, g(T_R^a)^{I_2I_3}, 
\]
and the secondary CWI
\begin{align}
&L_{1}A_1^{a I_2 I_3} - 4 p_{1\mu}\lla J^{a\mu}(\bs{p}_1)\O^{I_2}(\bs{p}_2)\O^{I_3}(\bs{p}_3)\rra \nn\\[1ex]&\quad
=4g(T_R^a)^{I_2I_3}p_1^2\Big[
-\frac{C_{\O\O}}{C_{JJ}}\Big(C_{JJ}\ln \frac{p_1^2}{\mu^2}+D_{JJ}\Big)p_1^2 
+2C_{\O\O}p_2^2
+a_0(p_1^2+p_2^2-p_3^2)\Big].
\end{align}
Using the transverse Ward identity \eqref{e:pJOO}, 
we can then rewrite this as
\begin{align}\label{pJOOsecondaryCWIsv2}
L_{1}A_1^{a I_2 I_3} 
&= 4g(T^a_R)^{I_2I_3}\Big[
-\frac{C_{\O\O}}{C_{JJ}}\Big(C_{JJ}\ln\frac{p_1^2}{\mu^2}+D_{JJ}\Big)p_1^4 -\Big(C_{\O\O}\ln \frac{p_2^2}{\mu^2}+D_{\O\O}\Big)p_2^4
 \nn\\[1ex]&\qquad
+\Big(C_{\O\O}\ln \frac{p_3^2}{\mu^2}+D_{\O\O}\Big)p_3^4
+2C_{\O\O}p_1^2p_2^2
+a_0\, p_1^2(p_1^2+p_2^2-p_3^2)\Big].
\end{align}
Comparing \eqref{pJOOprimaryCWIs} and \eqref{pJOOsecondaryCWIsv2} with  \eqref{pJOOprimaryCWIs00_v2} and \eqref{pJOOsecondaryCWIsv200_v2}, we see these are precisely the anomalous CWI we wished to derive.
We can equivalently express these identities, along with the anomalous dilatation Ward identity \eqref{DWIA1}, in terms of the renormalised 2-point functions:
\begin{align}
\mu\frac{\p}{\p\mu}A_1^{aI_2I_3} &=
4g(T^b_R)^{I_2I_3}\Big[\frac{C_{\O\O}}{3C_{JJ}}\lla J^{\nu b}(\bs{p}_1)J^{a}_{\nu}(-\bs{p}_1)\rra -\delta^{ab}\big(a_0p_1^2+C_{\O\O}(p_2^2+p_3^2)\big)\Big], \\[2ex]
 L_{1}A_1^{a I_2 I_3} &
= - 4g (T_R^a)^{K I_3} \lla \mathcal{O}^K(\bs{p}_2) \mathcal{O}^{I_2}(-\bs{p}_2) \rra + 4g (T_R^a)^{I_2 K} \lla \mathcal{O}^K(\bs{p}_3) \mathcal{O}^{I_3}(-\bs{p}_3) \rra
\nn\\[1ex]&\quad
+4\big(C_{\O\O}+a_0\big)g(T_R^a)^{I_2I_3} p_1^2(p_2^2-p_3^2) -p_1^2\,\mu\frac{\p}{\p\mu}A_1^{aI_2I_3},
\end{align}
As discussed earlier, with an appropriate choice of scheme, we can moreover set $a_0$ to zero.
In conclusion then, our generating relation \eqref{JOO_CWI_general} correctly accounts for the inhomogeneous terms appearing on the right-hand sides of all the anomalous conformal Ward identities, tracing their form back to the underlying beta functions and conformal anomalies of the renormalised theory.

\section{Discussion}
\label{sec:Discussion}

Our first steps in understanding momentum-space CFT are now complete.
We know the form of 2- and 3-point correlators,  both for general values of the spacetime and operator dimensions \cite{Bzowski:2013sza}, and for the special cases requiring renormalisation \cite{Bzowski:2015pba, Bzowski:2017poo}.  
With the results of this paper, we can  now construct renormalised 3-point functions involving a mix of scalars, stress tensors and conserved currents.  Besides obtaining compact and explicit expressions for all major cases of interest,  we have identified the relevant anomalous Ward identities, conformal anomalies and beta functions. 

We hope these results will find many interesting applications.  
Promising candidates include 
 the analysis of inflationary correlators in holographic cosmology (see {\it e.g.}, \cite{
 Antoniadis:2011ib, Maldacena:2011nz, Schalm:2012pi,  Bzowski:2012ih, Mata:2012bx, McFadden:2013ria, Ghosh:2014kba, Anninos:2014lwa, Kundu:2014gxa,  
Arkani-Hamed:2015bza,  Baumann:2015xxa, Isono:2016yyj})
and 
extending studies of quantum critical transport \cite{Chowdhury:2012km, Huh:2013vga, Jacobs:2015fiv, Myers:2016wsu, Lucas:2016fju, Lucas:2017dqa} 
to cases where divergences arise.
It may also be interesting to re-interpret recent  bounds, such as those derived from conformal collider physics and the average null energy condition (see {\it e.g.,} \cite{Chowdhury:2017vel, Cordova:2017zej, Meltzer:2017rtf}), from our present momentum-space perspective.
Our results should further be relevant for  
the analysis of conformal manifolds \cite{Gerchkovitz:2014gta, Gomis:2015yaa, Baggio:2017mas, Schwimmer:2018hdl}, particularly for cases where the dimension of some  scalar operator varies continuously as a function of the moduli.

Indeed, where such a  manifold exists, 
all the specific results we have obtained in the present series of papers \cite{Bzowski:2013sza, Bzowski:2015pba, Bzowski:2015yxv, Bzowski:2017poo} can be reinterpreted as providing nontrivial information about the analytic structure of the CFT data as a function of the moduli.  
Moving about on the conformal manifold, the conformal dimensions in general will vary and  the constants that determine 2- and 3-points functions will contain poles wherever the scalar dimensions are such that the corresponding correlators diverge.
Our results reveal the location of these poles, and whether they are of single or higher order \cite{Bzowski:2015pba}.  It would be interesting to connect this local analytic structure  with  global information  
from integrability or supersymmetry.

Another interesting question raised by our work is whether type A anomalies exist in three spacetime dimensions.  For 
 $\<T_{\mu_1\nu_1}T_{\mu_2\nu_2}\O\>$ with $\Delta_3=4$, we found in section \ref{typeAex1} an evanescent tensor appearing with a divergent coefficient.  While such a 0/0 mechanism is characteristic of type A anomalies \cite{Deser:1993yx, Bzowski:2017poo}, here the result we obtained was exact and thus could be removed with a counterterm.
Similar $0/0$ limits exist for other values of $\Delta_3$, however in all cases the result remains scheme-dependent and should likewise be removable.
This suggests no true type A anomalies can be found, but leaves room for deeper investigation.

Looking to the bigger picture, beyond extending our 3-point results to parity-odd and higher-spin examples, the next  
challenge 
in the development of momentum-space CFT is 
clearly the 4-point function \cite{Isono:2018rrb, Gillioz:2018mto}.
What are the momentum-space analogues of conformal cross-ratios?  Can momentum-space methods be of service to the bootstrap programme?

\section*{Acknowledgements}

We thank James Drummond and Eric Perlmutter for  discussion.
The work of 
AB is supported in part 
by the National Science Foundation
of Belgium (FWO) grant G.001.12 Odysseus,  
the European Research
Council grant ERC-2013-CoG 616732 HoloQosmos, 
the COST Action MP1210 ``The String Theory Universe'' and the CEA Enhanced Eurotalents Fellowship.
PM is supported by the STFC through an Ernest Rutherford Fellowship, and the Consolidated Grant, ``M-theory, Cosmology and Quantum Field Theory'', 
ST/L00044X/1.  
KS is supported in part by the STFC Consolidated Grant
``New Frontiers in Particle Physics and Cosmology", ST/P000711/1.
This project has received funding from the European Union's Horizon 2020 research and innovation programme under the Marie Sk\l{}odowska-Curie grant agreement No 690575.
AB and PM thank the University of Southampton for hospitality.

\appendix

\section{Appendices}

\subsection{Weyl covariance of the action for counterterms and sources}\label{Weyl_appendix}

In this appendix, we discuss how to impose
the Weyl covariance relation
\[\label{Weylcov_A}
\delta_\sigma (S_{\mathrm{ct}}+S_{\mathrm{source}}) = \int\D^d\x \sqrt{g}\,\sigma \mu\frac{\D}{\D\mu}(\mathcal{L}_{\mathrm{ct}}+\mathcal{L}_{\mathrm{source}})
\]
on the counterterm action for  $\<J^{\mu a}\O^{I_2}\O^{I_3}\>$ in the case of $d=\Delta^{I_2}=\Delta^{I_3}=4$.
From section \ref{sec:JOOd4analysis}, 
the relevant counterterm and source actions are
\begin{align}\label{JOO_Sct_A}
S_{\mathrm{ct}} &= \int\D^{4+2u\ep}\x\sqrt{g}\,\Big[\mathfrak{c}_{JJ}\mu^{2u\ep}F^{\mu\nu a}F^a_{\mu\nu}+\mathfrak{c}_{\O\O}\mu^{2 v_2\ep}(D^2\phi^I)^2
\nn\\&\qquad
+
\mathfrak{c}_1 \mu^{2(v_2-u)\ep}ig(T^a_R)^{IJ} J^{\mu a}\phi^{I} D_\mu \phi^{J}
+\mathfrak{c}_2 \mu^{2v_2\ep}ig(T^a_R)^{IJ}F^{\mu\nu a}D_\mu\phi^{I}D_\nu \phi^{J}\Big],\\[1ex]
\label{JOO_Sources_A}
S_{\mathrm{source}} &= \int\D^{4+2u\ep}\x\sqrt{g}\,\Big[\O^I\phi^I+J^{\mu a}A^a_{\mu}\Big],
\end{align}
where gauge invariance and permutation symmetry of the scalars impose a scheme where $u=v_1$ and $v_2=v_3$. 
As we will now show, Weyl covariance requires the counterterm coefficients satisfy the additional relation
\[\label{covreln}
(v_2-u)\ep \mathfrak{c}_2+2(1+v_2\ep)\mathfrak{c}_{\O\O}-4(v_2-u)\ep\mathfrak{c}_1\mathfrak{c}_{JJ}= 0,
\]
which we have made use of in sections \ref{sec:JOOd4analysis} and  \ref{sec:findingtheanomaly} of the main text. 
(Further discussion of this correlator also appears in section \ref{sec:anomaliesandbetafns}.)

We begin first with the counterterm
\begin{align}
S_{\mathrm{ct}}^{(JJ)} &= \mathfrak{c}_{JJ}\int\D^{4+2u\ep}\x\,\sqrt{g}\,\mu^{2u\ep}F^{\mu\nu a}F^a_{\mu\nu}.
\end{align}
Under a Weyl transformation, $\delta_\sigma \sqrt{g}=(4+2u\ep)\sigma\sqrt{g}$ while to quadratic order
\begin{align}
\delta_\sigma F^a_{\mu\nu}&=2\partial_{[\mu}\delta_\sigma A_{\nu]}^a+2g f^{abc}A_{[\mu}^b \delta_\sigma A_{\nu]}^c\nn\\[1ex]
&=-4(v_2-u)\ep\mu^{2(v_2-u)\ep}\mathfrak{c}_1 ig(T^a_R)^{IJ}\Big[\p_{\mu}\phi^I\p_{\nu}\phi^J\sigma
+\phi^I\p_{[\nu}\phi^J\p_{\mu]}\sigma\Big],
\end{align}
where the Weyl variation of $A^a_\mu$ (derived from the corresponding beta function) is given in \eqref{WeylofA}. 
To cubic order, using the Weyl variation of $\phi^I$ given in \eqref{Weylofphi}, we then find
\begin{align}\label{cJJnoncov}
\delta_\sigma S_{\mathrm{ct}}^{(JJ)}&= \mathfrak{c}_{JJ}\int\D^{4+2u\ep}\x\,\sqrt{g}\,\Big[ 2u\ep \mu^{2u\ep} F^{\mu\nu a}F^a_{\mu\nu}\sigma
\nn\\&\qquad\qquad
-8(v_2-u)\ep\mu^{2v_2\ep}\mathfrak{c}_1 ig(T^a_R)^{IJ}F^{\mu\nu a}\big(\p_{\mu}\phi^I\p_\nu\phi^J\sigma+\phi^I\p_\nu\phi^J\p_\mu\sigma\big)\Big].
\end{align}
The implicit $\mu$-dependence of the renormalised source $A^a_\mu$ is encoded in the beta function \eqref{betaA}.  The corresponding $\mu$-dependence of the squared field strength is then  \eqref{Fsqvar}.
Under change of the renormalisation scale, 
we thus find
\begin{align}\label{cJJmudep}
&\int\D^{4+2u\ep}\x\sqrt{g}\,\sigma
 \mu\frac{\D}{\D\mu}\mathcal{L}_{\mathrm{ct}}^{(JJ)}
 \nn\\[1ex]&\qquad  =\mathfrak{c}_{JJ}
 \int\D^{4+2u\ep}\x\sqrt{g}\,\sigma\Big[2u\ep\mu^{2u\ep}F^{\mu\nu a}F^a_{\mu\nu}
 -8(v_2-u)\ep\mu^{2v_2\ep}\mathfrak{c}_1 ig(T^a_R)^{IJ}F^{\mu\nu a}\p_\mu\phi^I\p_\nu\phi^J\Big]\nn\\[1ex]&
 \qquad =\delta_\sigma S_{\mathrm{ct}}^{(JJ)}+ 8(v_2-u)\ep\mathfrak{c}_1\mathfrak{c}_{JJ}\int\D^{4+2u\ep}\x\,\sqrt{g}\,\mu^{2v_2\ep}F^{\mu\nu a}ig(T^a_R)^{IJ}\phi^I\p_\nu\phi^J\p_\mu\sigma.
\end{align}

Next, as we noted in \eqref{scaledepofc0}, 
$\mu (\D/\D\mu)$ of the $\mathfrak{c}_1$ counterterm  is zero when combined with the current source term, and its Weyl variation similarly vanishes:
\begin{align}
0&=\delta_\sigma \int\D^{4+2u\ep}\x\sqrt{g}\,
\Big[\mathfrak{c}_1\mu^{2(v_2-u)\ep}ig(T^a_R)^{IJ}\phi^ID_\mu\phi^J +A^a_\mu\Big]J^{\mu a} 
\nn\\[0.5ex]&
= \delta_\sigma\int\D^{4+2u\ep}\x\sqrt{g}\,J^{\mu a}(A^{bare})_\mu^a.
\end{align}
The  $\phi^I\O^I$ source term is likewise invariant under both Weyl transformations and $\mu(\D/\D\mu)$.

We now deal with the counterterm 
\[
S_{\mathrm{ct}}^{(2)}=\mathfrak{c}_2\int\D^{4+2u\ep}\x\,\sqrt{g}\, \mu^{2v_2\ep}ig(T^a_R)^{IJ}F^{\mu\nu a}D_\mu\phi^I D_\nu\phi^J.
\]
To cubic order in the sources, the Weyl variation of this term is 
\begin{align}\label{WeylSct2}
\delta_\sigma S_{\mathrm{ct}}^{(2)} =
\mathfrak{c}_2 \int\D^{4+2u\ep}\x\,\sqrt{g}\mu^{2v_2 \ep}ig(T^a_R)^{IJ}&\Big[
2v_2\ep F^{\mu\nu a}\p_\mu\phi^I\p_\nu\phi^J\sigma
\nn\\&
\quad
+2(v_2-u)\ep F^{\mu\nu a}\phi^I\p_\nu\phi^J\p_\mu\sigma\Big],
\end{align}
while under a change of renormalisation scale,
\begin{align}\label{Sct2mudep}
&\int\D^{4+2u\ep}\x\sqrt{g}\sigma
 \mu\frac{\D}{\D\mu}\mathcal{L}_{\mathrm{ct}}^{(2)}
=\mathfrak{c}_2 \int\D^{4+2u\ep}\x\,\sqrt{g}\,\mu^{2v_2 \ep}ig(T^a_R)^{IJ} 2v_2\ep  F^{\mu\nu a}\p_\mu\phi^I \p_\nu\phi^J \sigma\nn\\[1ex]
&\qquad \qquad =\delta_\sigma S_{\mathrm{ct}}^{(2)}-2(v_2-u)\ep\mathfrak{c}_2 \int\D^{4+2u\ep}\x\,\sqrt{g}\mu^{2v_2 \ep}ig(T^a_R)^{IJ} F^{\mu\nu a}\phi^I\p_\nu\phi^J\p_\mu\sigma.
\end{align}
Neither the Weyl variation nor the $\mu$-dependence of $F^a_{\mu\nu}$ contribute here as both begin at quadratic order in the sources.

Finally, to deal with the counterterm proportional to $\mathfrak{c}_{\O\O}$, a little more work is required.  
First, we can construct a Weyl covariant analogue of $(\Box\phi)^2$ by introducing the following additional couplings to spacetime curvature, 
\begin{align}\label{Scov}
S_{\mathrm{cov}}&=\mathfrak{c}_{\O\O}\int\D^d\x\sqrt{g}\mu^{d-4+2s}\Big[(\Box\phi)^2-\frac{2(d-2+2s)}{(d-2)}R^{\mu\nu}\p_\mu\phi\p_\nu\phi+\frac{d(d-2+2s)}{2(d-1)(d-2)}R(\p\phi)^2
\nn\\[1ex]&\qquad\qquad\quad
+\frac{s}{(d-1)}R\phi\Box\phi
+\frac{s^2}{4(d-1)^2}R^2\phi^2-\frac{s(d-2+2s)}{4(d-2)(d-3)}E_4\phi^2\Big],
\end{align}
where $E_4$ is the four-dimensional Euler density.
Using the transformations
\begin{align}
\delta_\sigma\phi&=s\sigma\phi,\\
\delta_\sigma (\Box\phi) &= (s-2)\sigma\Box\phi+(d-2+2s)\p_\mu\phi\p^\mu\sigma+s\phi\Box\sigma,\\
\delta_\sigma R&=-2\sigma R-2(d-1)\Box\sigma,\\
\delta_\sigma R_{\mu\nu}&=-(d-2)\nabla_\mu\p_\nu\sigma-g_{\mu\nu}\Box\sigma,\\
\delta_\sigma E_4 &=-4\sigma E_4+8(d-3)R^{\mu\nu}\nabla_\mu\p_\nu\sigma-4(d-3)R\Box\sigma,
\end{align}
one can then verify that
\[\label{finaltransf}
\delta_\sigma S_{\mathrm{cov}}=(d-4+2s)\int\D^d\x\sqrt{g}\,\sigma \mathcal{L}_{\mathrm{cov}}=\int\D^d\x\sqrt{g}\,\sigma \mu\frac{\p}{\p\mu}\mathcal{L}_{\mathrm{cov}}.
\]

In the regulated theory,  $s=(v_2-u)\ep$ and the counterterm \eqref{Scov} is not Weyl invariant (as one would anticipate). 
The corresponding anomaly action is however Weyl invariant, and can be obtained (for general $d$) by setting  $s=-(d-4)/2$.  
Up to a constant, \eqref{Scov} can then be rewritten 
\[\label{ctPanaction}
\int \D^d\x\sqrt{g}\Big[\frac{(d-4)}{4(d-3)(d-3)}W_d^2\phi^2+\phi \Delta_4 \phi\Big],
\]
where $W_d^2$ is the square of the $d$-dimensional Weyl tensor
and $\Delta_4$ is the $d$-dimensional Paneitz operator\footnote{Our conventions are those of Wald \cite{Wald}, hence relative to \cite{Paneitz} we have sent $R_{\mu\nu}\rightarrow -R_{\mu\nu}$.}
\begin{align}
\Delta_4 \phi = \Box^2\phi-\nabla_\mu\Big[\Big((d-2)J g^{\mu\nu}-4P^{\mu\nu}\Big)\nabla_\nu\phi\Big]+(d-4) Q\phi,
\end{align}
with the Schouten tensor $P_{\mu\nu}$ and Q-curvature
\begin{align}
P_{\mu\nu} &= \frac{1}{(d-2)}\Big(R_{\mu\nu}-Jg_{\mu\nu}\Big), \quad J = g^{\mu\nu}P_{\mu\nu}, \quad  Q = \frac{1}{4}\Big(d J^2 - 2\Box J-4 P_{\mu\nu}P^{\mu\nu}\Big). 
\end{align}
The Weyl invariance of \eqref{ctPanaction} is now apparent from the relation
\[\label{Pantr}
\tilde{\Delta}_{4}\tilde{\phi} = e^{-(d+4)\sigma/2}\Delta_4\phi,
\]
where $\tilde{\Delta}_4$ is evaluated on $\tilde{g}_{\mu\nu}=e^{2\sigma} g_{\mu\nu}$ and $\tilde{\phi} = e^{-(d-4)\sigma/2}\phi$. 
Indeed, we encountered the four-dimensional version of this action in \eqref{Panaction} of section \ref{TOOPaneitzsec} (see also \cite{Manvelyan:2006bk}).

Returning to the counterterm \eqref{Scov}, having enforced Weyl covariance, we now need to covariantise under gauge transformations.  Promoting $\phi\rightarrow\phi^I$ and $\p_\mu \rightarrow D_\mu^{IJ}$, we obtain the new counterterm
\begin{align}\label{S1cov}
S^{(\O\O)}_{\mathrm{ct}} 
&=\mathfrak{c}_{\O\O}\int\D^{4+2u\ep}\x\sqrt{g}\mu^{2v_2\ep}\Big[(D^2\phi^I)^2-\frac{2(1+v_2\ep)}{(1+u\ep)}R^{\mu\nu}D_\mu\phi^I D_\nu\phi^I
+\frac{(v_2-u)\ep}{(3+2u\ep)}R\phi^I D^2\phi^I
\nn\\[1ex]&\quad
+\frac{(2+u\ep)(1+v_2\ep)}{(3+2u\ep)(1+u\ep)}R(D\phi^I)^2
+\frac{(v_2-u)^2\ep^2}{4(3+2u\ep)^2}R^2(\phi^I)^2-\frac{(v_2-u)\ep(1+v_2\ep)}{4(1+u\ep)(1+2u\ep)}E_4(\phi^I)^2\Big].
\end{align}
On a flat background, this new counterterm reduces to the original $\mathfrak{c}_{\O\O}$ counterterm in \eqref{JOO_Sct_A}, but on a general metric has improved Weyl covariance properties: we will therefore use it in place of our original counterterm.
Notice however that in this last step of gauge-covariantising  we  effectively introduced the new cubic terms 
\begin{align}
&\mathfrak{c}_{\O\O}\int\D^{4+2u\ep}\x\sqrt{g}\mu^{2v_2\ep}ig(T^a_R)^{IJ}\Big[
-4A_\mu^a\Box\phi^I\p^\mu\phi^J+2\nabla_\mu A^{a\mu}\phi^I \Box\phi^J-
\frac{4(1+v_2\ep)}{(1+u\ep)}R^{\mu\nu} A_{\nu}^{a}\phi^I\p_\mu\phi^J\nn\\[1ex]&\quad\qquad\qquad\qquad\quad
-\frac{2}{(3+2u\ep)}\Big[(v_2-u)\ep-\frac{(2+u\ep)(1+v_2\ep)}{(1+u\ep)}\Big]R A^{a\mu}\phi^I\p_\mu\phi^J \Big].
\end{align}
These new cubic terms in fact break the full Weyl covariance we had in \eqref{Scov}: using
\begin{align}
\delta_\sigma (\nabla_\mu A^{a\mu}) &= 
-2\sigma\nabla_\mu A^{a\mu}+(d-2)A^{a\mu}\p_\mu\sigma,
\end{align}
after some calculation, we find that
\begin{align}\label{c1noncov}
&\int\D^{4+2u\ep}\x\sqrt{g}\sigma \mu\frac{\D}{\D\mu}\mathcal{L}^{(\O\O)}_{\mathrm{ct}} \nn\\&\qquad =\delta_\sigma S^{(\O\O)}_{\mathrm{ct}} 
- 4(1+v_2\ep)\mathfrak{c}_{\O\O}\int\D^{4+2u\ep}\x\sqrt{g}\mu^{2v_2\ep}ig(T^a_R)^{IJ}F^{\mu\nu a}\phi^I\p_\nu\phi^J\p_\mu\sigma.
\end{align}

For the combined counterterm and source action to satisfy the Weyl covariance relation  \eqref{Weylcov_A}, we now need to arrange for all the non-Weyl covariant terms proportional to $\p_\mu\sigma$ to cancel out.  Putting together our results \eqref{cJJmudep}, \eqref{Sct2mudep} and \eqref{c1noncov} above, we obtain the desired relation \eqref{covreln}.

\subsection{Evaluation of counterterm contributions}
\label{sec:evalctcontr}

\label{sec:TTOctcontr}

In this appendix, we compute the counterterm contributions to the form factors appearing in our transverse traceless decomposition of correlators.  
For this purpose, it suffices to work in a gauge where the {\it inverse} metric perturbation is transverse traceless, 
\[
g^{\mu\nu}=\delta_{\mu\nu}+\g_{\mu\nu}, \qquad \g_{\mu\mu}=0, \qquad \g_{\mu\nu,\nu}=0,
\]
since all other components 
are projected out in the calculation of form factors.

Writing $g_{\mu\nu}=\delta_{\mu\nu}+h_{\mu\nu}$, we then have
\[\label{deltagexp}
h_{\mu\nu} = -\g_{\mu\nu}+\g_{\mu\alpha}\g_{\alpha\nu}+O(\g^3)
\]
where
\[\label{newgauge}
h=h_{\mu\mu} = \g_{\mu\nu}\g_{\mu\nu}+O(\g^3), \qquad
h_{\mu\nu,\nu} = \g_{\mu\alpha,\nu}\g_{\alpha\nu}+O(\g^3).
\]
The Ricci curvature 
\begin{align}\label{Rexpressions}
R_{\mu\nu} &= -\frac{1}{2}\p^2h_{\mu\nu} -\frac{1}{2}h_{,\mu\nu}+h_{\alpha(\mu,\nu)\alpha} +\frac{1}{4}(h_{\alpha\beta}h_{\alpha\beta})_{,\mu\nu}-h_{\alpha\beta} \hat{S}_{\beta\mu\nu,\alpha}-\hat{S}_{\alpha\beta\mu}\hat{S}_{\beta\alpha\nu}+O(\g^3)
\end{align}
where
\[
\hat{S}_{\mu\nu\alpha} = \Gamma^{(1)\mu}_{\quad \nu\alpha} = \frac{1}{2}(h_{\mu\nu,\alpha}+h_{\mu\alpha,\nu}-h_{\nu\alpha,\mu})
\]
and we have used the fact that to $O(\g^3)$ we can treat $h_{\mu\nu}$ as transverse traceless where it appears quadratically.
We thus have
\begin{align}
R^{(1)}_{\mu\nu} &= \frac{1}{2}\p^2 \g_{\mu\nu}, \\
R^{(2)}_{\mu\nu} &= -\frac{1}{2}\p^2(\g_{\mu\alpha}\g_{\alpha\nu})+  (\g_{\alpha\beta}\g_{\alpha(\mu})_{,\nu)\beta}
-\frac{1}{4}(\g_{\alpha\beta}\g_{\alpha\beta})_{,\mu\nu}-\g_{\alpha\beta} S_{\beta\mu\nu,\alpha}-S_{\alpha\beta\mu}S_{\beta\alpha\nu}
\end{align}
where
\[
S_{\mu\nu\alpha} =  \frac{1}{2}(\g_{\mu\nu,\alpha}+\g_{\mu\alpha,\nu}-\g_{\nu\alpha,\mu}).
\]
The scalar curvature
\begin{align}
R^{(1)} &= 0,\\
R^{(2)} & = -\g_{\mu\nu}\p^2 \g_{\mu\nu}-\frac{5}{4}\g_{\mu\nu,\alpha}\g_{\mu\nu,\alpha}+\frac{1}{2}\g_{\mu\nu,\alpha}\g_{\mu\alpha,\nu},
\end{align}
while the Riemann curvature
\begin{align}\label{Riemannexp}
R^{(1)}_{\mu\nu\alpha\beta} &= -2 S_{\mu\nu[\alpha,\beta]} \\
R^{(2)}_{\mu\nu\alpha\beta} &= -2S_{\lambda\nu[\alpha}h_{\mu\lambda,\beta]}+2S_{\lambda \nu[\beta}S_{\mu\lambda \alpha]}.
\end{align}
We can now  evaluate the counterterms contributions  for  $\<T_{\mu_1\nu_1}T_{\mu_2\nu_2}\O\>$  with $d=\Delta_3=4$.
The first counterterm involves the  Weyl tensor (as defined in exactly four dimensions),
\begin{align}\label{Weyl2phi}
&\int\D^d\x\,\sqrt{g}\mu^{(u+v_3)\ep} \phi^I W_4^2
\nn\\&\quad =\int\D^d\x\,\mu^{(u+v_3)\ep}\phi^I \Big[\g_{\mu\nu,\alpha\beta}\g_{\mu\nu,\alpha\beta}+\g_{\mu\alpha,\nu\beta}\g_{\nu\beta,\mu\alpha}-2\g_{\mu\alpha,\nu\beta}\g_{\mu\nu,\alpha\beta}-\frac{1}{2}\p^2\g_{\mu\nu}\p^2\g_{\mu\nu}\Big].
\end{align}
Since $R^2$ vanishes to quadratic order in $\g_{\mu\nu}$, we then have
\begin{align}
\int\D^d\x \,\sqrt{g}\mu^{(u+v_3)\ep}\phi^IE_4 &=\int\D^d\x\,\mu^{(u+v_3)\ep} \phi^I \Big[W_4^2 -2R_{\mu\nu}^{(1)}R_{\mu\nu}^{(1)}\Big] \nn\\&= \int \D^d\x\,\mu^{(u+v_3)\ep}\phi^I\Big[W_4^2 - \frac{1}{2}\p^2\g_{\mu\nu}\p^2\g_{\mu\nu}\Big].
\end{align}
The counterterm action
\[
\int\D^d\x\, \sqrt{g}\mu^{(u+v_3)\ep}\phi^I\,(\mathfrak{a}^I E_4+\mathfrak{c}^I W_4^2)
\]
then generates the counterterm contributions
\begin{align}\label{TTOctcontr1}
A^{\mathrm{ct}\,I}_1 &= 8(\mathfrak{a}^I+\mathfrak{c}^I)\mu^{(u+v_3)\ep},\\[1ex]
A^{\mathrm{ct}\,I}_2 &= -8(\mathfrak{a}^I+\mathfrak{c}^I)(p_1^2+p_2^2-p_3^2)\mu^{(u+v_3)\ep},\\[1ex]
A^{\mathrm{ct}\,I}_3 &= \Big[4 \mathfrak{c}^I p_1^2 p_2^2 - 2(\mathfrak{a}^I+\mathfrak{c}^I)J^2 \Big]\mu^{(u+v_3)\ep}.
\label{TTOctcontr3}
\end{align}
In fact, the finite piece of the Weyl-squared counterterm \eqref{Weyl2phi} generates a stress tensor ambiguity of the form discussed in \cite{Erdmenger:1996yc}.  In four dimensions, to quadratic order in $\g_{\mu\nu}$ we have
\begin{align}\label{Osb}
\int\D^4\x\, \sqrt{g}\phi^I W_4^2 &= 2\int\D^4\x\,\phi^I W_{\mu\nu\alpha\beta}^{(1)}W_{\mu\nu\alpha\beta}^{(1)} \nn\\
&= 2\int\D^4\x\,\phi^I W_{\mu\nu\alpha\beta}^{(1)} \Big[
R^{(1)}_{\mu\nu\alpha\beta}-\delta_{\mu[\alpha}R^{(1)}_{\beta]\nu}+\delta_{\nu[\alpha}R^{(1)}_{\beta]\mu}+\frac{1}{3}R^{(1)}\delta_{\mu[\alpha}\delta_{\beta]\nu}\Big]\nn\\
&=2\int\D^4\x\,\phi^I W_{\mu\nu\alpha\beta}^{(1)}
R^{(1)}_{\mu\nu\alpha\beta}\nn\\
&=-4\int\D^4\x\,\phi^I W_{\mu\nu\alpha\beta}^{(1)}\g_{\nu\alpha,\mu\beta},
\end{align}
where the last two lines follow from the symmetries and tracelessness of the Weyl tensor, along with \eqref{Riemannexp}. To linear order in $\g_{\mu\nu}$, we then obtain a transverse-traceless contribution to the stress tensor of the form
\[
T_{\nu\alpha}(\x) = -\frac{2}{\sqrt{g}}\frac{\delta S}{\delta g^{\nu\alpha}(\x)} = 8\partial_\mu\partial_\beta\big[\phi^I W^{(1)}_{\mu\nu\alpha\beta}\big]. 
\]

Finally, for completeness, let us consider the counterterm
\[\label{Rboxphi}
\int\D^d\x\,\sqrt{g}\mu^{(u+v_3)\ep}R\Box\phi^I = \int \D^d\x\,\sqrt{g}\mu^{(u+v_3)\ep}\phi^I\Box R
=2 {\int}\D^d\x\,\sqrt{g}\mu^{(u+v_3)\ep} R_{\mu\nu}\nabla^\mu\nabla^\nu\phi^I, 
\]
where the last relation follows from the contracted Bianchi identities.  We find
\begin{align}
&\int\D^d\x\,\sqrt{g}\mu^{(u+v_3)\ep}R\Box \phi^I \nn\\[0ex]&\qquad = 
\int\D^d\x\, \mu^{(u+v_3)\ep}\p^2\phi^I \Big[-\g_{\mu\nu}\p^2 \g_{\mu\nu}-\frac{5}{4}\g_{\mu\nu,\alpha}\g_{\mu\nu,\alpha}+\frac{1}{2}\g_{\mu\nu,\alpha}\g_{\mu\alpha,\nu}\Big]
\end{align}
leading to
\begin{align}
A_2^{\mathrm{ct}\,I} &=-4p_3^2\mu^{(u+v_3)\ep},\\
A_3^{\mathrm{ct}\,I} &= \Big((p_1^2+p_2^2)p_3^2 -5p_3^4\Big)\mu^{(u+v_3)\ep}.
\end{align}
We will not use this counterterm, however, since it generates the mixed 2-point function
\[
\lla T_{\mu\nu}(\bs{p}) \O^I(-\bs{p})\rra_{\mathrm{ct}} = -2 p^4 \pi_{\mu\nu}(\bs{p}).
\]

\subsection{Shadow relations}

The transverse-(traceless) parts of the correlators satisfy the following shadow relations, where the scalar  $\O^I$ of dimension $\Delta$ is replaced with its shadow $\tilde{\O}^I$ of dimension $\tilde{\Delta}=d-\Delta$. 
\begin{align}\label{JOOshadow}
\lla j^{\mu a}(\bs{p}_1)\tilde{\O}^{I_2}(\bs{p}_2)\tilde{\O}^{I_3}(\bs{p}_3)\rra &\propto \frac{\lla j^{\mu a}(\bs{p}_1)\O^{I_2}(\bs{p}_2)\O^{I_3}(\bs{p}_3)\rra }{\lla \O^{K_2}(\bs{p}_2)\O^{K_2}(-\bs{p}_2)\rra
\lla \O^{K_3}(\bs{p}_3)\O^{K_3}(-\bs{p}_3)\rra}, 
\\[2ex]\label{TOOshadow}
\lla t_{\mu_1\nu_1}(\bs{p}_1)\tilde{\O}^{I_2}(\bs{p}_2)\tilde{\O}^{I_3}(\bs{p}_3)\rra &\propto \frac{\lla t_{\mu_1\nu_1}(\bs{p}_1)\O^{I_2}(\bs{p}_2)\O^{I_3}(\bs{p}_3)\rra }{\lla \O^{K_2}(\bs{p}_2)\O^{K_2}(-\bs{p}_2)\rra
\lla \O^{K_3}(\bs{p}_3)\O^{K_3}(-\bs{p}_3)\rra}, 
\\[2ex]\label{JJOshadow}
\lla j^{\mu_1a_1}(\bs{p}_1)j^{\mu_2a_2}(\bs{p}_2)\tilde{\O}^{I_3}(\bs{p}_3)\rra &\propto \frac{\lla j^{\mu_1a_1}(\bs{p}_1)j^{\mu_2a_2}(\bs{p}_2)\O^{I_3}(\bs{p}_3)\rra }{
\lla \O^{K_3}(\bs{p}_3)\O^{K_3}(-\bs{p}_3)\rra}, 
\\[2ex]\label{TTOshadow}
\lla t_{\mu_1\nu_1}(\bs{p}_1)t_{\mu_2\nu_2}(\bs{p}_2)\tilde{\O}^{I_3}(\bs{p}_3)\rra &\propto \frac{\lla t_{\mu_1\nu_1}(\bs{p}_1)t_{\mu_2\nu_2}(\bs{p}_2)\O^{I_3}(\bs{p}_3)\rra }{
\lla \O^{K_3}(\bs{p}_3)\O^{K_3}(-\bs{p}_3)\rra}.
\end{align}
These relations are valid for generic $d$ and $\Delta$, but not in cases where renormalisation is required.    
(The shadow correlators have singularities of  the opposite type, in which all $\pm$ signs are reversed; this alters the renormalisation procedure as discussed in \cite{Bzowski:2015pba}.)
Moreover, these relations do not always extend to the longitudinal parts of the correlators, as can be seen from examining the corresponding transverse Ward identities ({\it e.g.,} for \eqref{TOOshadow}).  

Their proof follows from the fact that 
the modified Bessel function is an even function of its index. 
Since 
$K_{\beta}(x)=K_{-\beta}(x)$, it follows that
\[
I_{\alpha\{\beta_1,\beta_2,\beta_3\}} = p_3^{2\beta_3}I_{\alpha\{\beta_1,\beta_2,-\beta_3\}}=p_2^{2\beta_2}p_3^{2\beta_3}I_{\alpha\{\beta_1,-\beta_2,-\beta_3\}}.
\]
The relations \eqref{JOOshadow} and \eqref{TOOshadow} then follow from our solutions \eqref{e:pri1JOO} and \eqref{e:priTOO1}, using \eqref{e:2ptOO}.
For \eqref{JJOshadow}, we instead use equation (3.12) of \cite{Bzowski:2015yxv} to rewrite the form factors  for $\lla j^{\mu_1a_1}j^{\mu_2a_2}\O^{I}\rra$ , given in  \eqref{a:JJO1}\,-\,\eqref{a:JJOlast} and \eqref{secCWI_JJO_constr}, as
\begin{align}
A_1^{a_1a_2I} &=C_1^{a_1a_2I}J_{2\{000\}},\\
A_2^{a_1a_2I} &=C_1^{a_1a_2I} p_3^{\Delta_3}\Big[\frac{1}{2}(\Delta_3-2)(\tilde{\Delta}_3-2)-\Big(2+p_3\frac{\p}{\p p_3}\Big)\Big]\Big(p_3^{-\Delta_3}J_{0\{000\}}\Big), 
\end{align}
where $\tilde{\Delta}_3=d-\Delta_3$.  
Since
\begin{align}
p_3^{-\Delta_3}J_{N\{000\}} &= p_3^{-\Delta_3}I_{\frac{d}{2}-1+N\{\frac{d}{2}-1,\frac{d}{2}-1,\Delta_3-\frac{d}{2}\}}\nn\\& =p_3^{-\tilde{\Delta}_3}I_{\frac{d}{2}-1+N\{\frac{d}{2}-1,\frac{d}{2}-1,\tilde{\Delta}_3-\frac{d}{2}\}}=p_3^{-\tilde{\Delta}_3}J_{N\{00\tilde{0}\}},
\end{align}
it is then straightforward to see that 
\begin{align}
p_3^{-\tilde{\Delta}_3}\tilde{A}_j^{a_1a_2I} = p_3^{-\Delta_3}A_j^{a_1a_2I},\qquad 
j=1,2,
\end{align}
where the tilded form factors are those of the shadow correlator $\lla j^{\mu_1a_1}j^{\mu_2a_2}\tilde{\O}^{I}\rra$.  The relation \eqref{JJOshadow} immediately follows.

The proof of \eqref{TTOshadow} is analogous, and follows by writing the form factors \eqref{a:TTO1}\,-\,\eqref{a:TTO3}  for $\lla t_{\mu_1\nu_1}t_{\mu_2\nu_2}\O^I\rra$, with \eqref{secsolTOO1}\,-\,\eqref{secsolTOO2}, in the form
\begin{align}
A_1^I &= C_1^I J_{0\{000\}},\\
A_2^I &= C_1^I p_3^{\Delta_3}\Big[(\Delta_3-2)(\tilde{\Delta}_3-2)-4\Big(2+p_3\frac{\p}{\p p_3}\Big)\Big]\Big(p_3^{-\Delta_3}J_{2\{000\}}\Big),\\
A_3^I &= C_1^Ip_3^{-\Delta_3}\Big[(\Delta_3-2)(\tilde{\Delta}_3-2)\Big(\frac{\Delta_3\tilde{\Delta}_3}{4} - p_3\frac{\p}{\p p_3}\Big)\nn\\&\qquad\qquad\qquad +2\Big(2+p_3\frac{\p}{\p p_3}\Big)p_3\frac{\p}{\p p_3}\Big]\Big(p_3^{-\Delta_3}J_{0\{000\}}\Big).
\end{align}

\bibliographystyle{JHEP}
\bibliography{cwis2017}

\end{document}